\documentclass[usenatbib]{mn2e}

\usepackage[toc,page]{appendix}
\usepackage{amsmath}
\usepackage[dvips]{graphicx}
\usepackage{pdflscape}
\usepackage{afterpage}
\usepackage{rotating}
\usepackage{epsfig}
\usepackage{lscape}
\usepackage{natbib}
\usepackage{amssymb}
\usepackage{amsmath}
\usepackage{tabularx}
\usepackage{longtable}

\newcommand{\RE}{R_{\rm E}}
\newcommand{\Reff}{R_{\rm eff}}

\newcommand{\kms}{{\rm \,km\,s^{-1}}}
\newcommand{\Mpc}{\,{\rm Mpc}}
\newcommand{\kpc}{\,{\rm kpc}}

\newcommand{\Dd}{D_{\rm d}}
\newcommand{\Ds}{D_{\rm s}}
\newcommand{\Dds}{D_{\rm ds}}

\title[Illustris early-type galaxies] {The inner structure of
  early-type galaxies in the Illustris simulation} \author[Xu et al.]
      {Dandan Xu$^{1}$\thanks{E-mail: Dandan.Xu@h-its.org}, Volker
        Springel$^{1,2}$, Dominique Sluse$^{3}$, Peter
        Schneider$^{4}$, \and Alessandro Sonnenfeld$^{5}$, Dylan
        Nelson$^{6}$, Mark Vogelsberger$^{7}$, Lars Hernquist$^{8}$
        \vspace*{0.2cm}\\ $^{1}$ Heidelberg Institute for Theoretical
        Studies, Schloss-Wolfsbrunnenweg 35, 69118 Heidelberg, Germany
        \\ $^{2}$ Zentrum f\"ur Astronomie der Universit\"at
        Heidelberg, Astronomisches Recheninstitut,
        M\"{o}nchhofstr. 12-14, 69120 Heidelberg, Germany \\ $^{3}$
        STAR Institute, Quartier Agora - All\'ee du six Ao\^ut, 19c
        B-4000 Li\`ege, Belgium \\ $^{4}$ Argelander-Institut
        f$\ddot{u}$r Astronomie, Universit$\ddot{a}$t Bonn, Auf dem
        H$\ddot{u}$gel 71, 53121 Bonn, Germany \\ $^{5}$ Kavli
        Institute for the Physics and Mathematics of the Universe of
        Tokyo, 5-1-5 Kashiwanoha Kashiwa, 277-8583 Japan \\ $^{6}$ Max
        Planck Institute for Astrophysics, Karl-Schwarzschild-Str. 1,
        Postfach 1317, D-85741 Garching, Germany \\ $^{7}$ Department
        of Physics, Massachusetts Institute of Technology, 77
        Massachusetts Avenue, Cambridge, United States \\ $^{8}$
        Harvard Astronomy Department, 60 Garden Street MS 46,
        Cambridge, MA 02138, United States }

\begin{document}
\pagerange{\pageref{firstpage}--\pageref{lastpage}} \pubyear{2016}
\maketitle
\label{firstpage}

\begin{abstract}
  Early-type galaxies provide unique tests for the predictions of the
  cold dark matter cosmology and the baryonic physics assumptions
  entering models for galaxy formation.  In this work, we use the
  Illustris simulation to study correlations of three main properties
  of early-type galaxies, namely, the stellar orbital anisotropies,
  the central dark matter fractions and the central radial density
  slopes, as well as their redshift evolution since $z=1.0$. We find
  that lower-mass galaxies or galaxies at higher redshift tend to be
  bluer in rest-frame colour, have higher central gas fractions, and
  feature more tangentially anisotropic orbits and steeper central
  density slopes than their higher-mass or lower-redshift
  counterparts, respectively. The projected central dark matter
  fraction within the effective radius shows a very mild mass
  dependence but positively correlates with galaxy effective radii due
  to the aperture effect. The central density slopes obtained by
  combining strong lensing measurements with single aperture
  kinematics are found to differ from the true density slopes. We
  identify systematic biases in this measurement to be due to two
  common modelling assumptions, isotropic stellar orbital
  distributions and power-law density profiles. We also compare the
  properties of early-type galaxies in Illustris to those from
  existing galaxy and strong lensing surveys, we find in general broad
  agreement but also some tension, which poses a potential challenge
  to the stellar formation and feedback models adopted by the
  simulation.
\end{abstract}


\section{Introduction}

Cosmic structures grow in a hierarchical fashion whereby small halos
merge to form larger ones. This prevailing picture of structure
assembly is predicted by the cold dark matter (CDM) cosmological
theory.  Early-type galaxies are in some sense the end products of the
corresponding galaxy merging and accretion processes (e.g.,
\citealt{WhiteRees1978, Davis1985, KWG1993, Cole1994, Kauffmann1996}),
and thus provide an interesting testing ground of the CDM cosmological
model.

Due to their association with mergers, early-type galaxies tend to
live in high-density environments and have quite old stellar
populations. Though traditionally thought to be structureless and
``red and dead'', mounting evidence has shown over the past decades
that there is considerable richness and complexity in the origin and
evolution of their mass-size relations, star formation activities, and
central density profiles, etc. Utilizing for example their fundamental
plane relations and/or their gravitational lensing effects, early-type
galaxies are also widely used as probes of the high-redshift Universe,
making them a powerful and important tool in modern astrophysics and
cosmology.

Among the various properties of early-type galaxies, their central
dark matter fraction and central density profiles are particularly
closely tied to their formation and evolution histories. The CDM model
predicts universal NFW (\citealt{NFW1997}) profiles for the dark
matter distribution in halos over a wide range of mass scales. However,
on galaxy-scales, baryonic matter strongly dominates the central
regions of dark matter halos. As a result of dissipation, baryons
follow more centrally concentrated density distributions, which in
turn changes the central mass fraction of dark matter (e.g.,
\citealt{TK2004, Koopmans2006, Napolitano2010, Barnabe2011III,
  Ruff2011SL2S, Cappellari2013XV, Sonnenfeld2015SL2SV, Oguri2014a})
and also modifies the inner dark matter slopes, making them steeper
than the NFW prediction (e.g., \citealt{Sonnenfeld2012, Grillo2012,
  Johansson2012, Remus2013, Cappellari2013XV, Oguri2014a}).

Regarding the total density profiles in the central regions, the most
intriguing fact is that the sum of dark and baryonic matter
approximately follows an isothermal profile,
i.e.~$\rho(r)\propto r^{-2}$, even though neither the dark nor the
baryonic matter exhibit an isothermal distribution individually.
Evidence for such a profile comes from stellar kinematical studies
(e.g., \citealt{BinneyTremaineBook, Cappellari2015}), strong and weak
lensing observations (e.g., \citealt{Rusin2003, Koopmans2006,
  Koopmans2009, Gavazzi2007, Barnabe2009II, Barnabe2011III};
\citeauthor{Auger2010SLACSX} 2010b; \citealt{Ruff2011SL2S,
  Bolton2012BELLS, Sonnenfeld2013SL2SIV}), as well as X-ray studies of
early-type galaxies (e.g., \citealt{Humphrey2006, HumphreyBuote2010}).
Theoretically, the formation of such a total density distribution is
speculated to occur through a two-phase process, where active
(central) star formation and adiabatic contraction in an early stage
is followed by dissipationless mergers and accretion later on. The
former steepens the central density slopes while the latter in general
makes them shallower. The observed central density slopes and their
evolutionary trend, therefore, put stringent constraints on both the
CDM structure formation model and models for baryonic physics
processes.

Galaxy-scale strong gravitational lensing is among the major tools to
probe galaxies out to high redshifts. It robustly measures the
projected mass within the Einstein radius, typically at within a few
kpc from the centre of a lensing galaxy. Traditionally, the strong
lensing technique has also been combined with stellar kinematics,
which provides the mass measured within some different aperture radius
(\citealt{RomanowskyKochanek1999SLD}). The combination thus allows
measurements of the radial density slopes and dark matter fractions.
The method has been put into good use for many existing strong lensing
surveys, e.g., the Lenses Structure and Dynamics Survey (LSD; e.g.,
Treu \& Koopmans 2004); the Sloan Lens ACS Survey (SLACS; Koopmans et
al. 2006, 2009; \citeauthor{Bolton2008SLACSV} 2008a; Auger et
al. 2010b), the BOSS Emission-Line Lens Survey (BELLS;
\citealt{Brownstein2012BELLS}; Bolton et al. 2012), and the Strong
Lensing Legacy Survey (SL2S; Ruff et al. 2011;
\citealt{Gavazzi2012SL2S}; Sonnenfeld et al. 2013, 2015). To date,
hundreds of strong-lensing early-type galaxies have been well studied
out to redshift $z=1.0$.  The average central density slopes have been
found to be approximately consistent with isothermal radial profiles
with a small intrinsic scatter. This isothermal behaviour seems to
have evolved very little in the past 7 Gyrs (i.e.~since $z=1.0$).

To date various theoretical approaches, including semi-analytical
models and N-body simulations have been exploited in order to address
open issues regarding the mass-size relations, the central dark matter
fractions, as well as the total density slopes of early-type galaxies
(e.g., \citeauthor{Nipoti2009Size} 2009a,\,b; \citealt{Johansson2012,
  Remus2013, Dubois2013, SNT2014}).  These studies have made important
progress in finding the missing links in the framework of the
formation and evolution theories and in understanding potential
systematic biases of the observational techniques. However, these
modelling techniques often lacked a self-consistent treatment of
baryonic physics in a cosmological context, which limited their
predictive power.

In this regard, the latest generation of cosmological hydrodynamical
simulations of galaxy formation represents a significant step forward
and enables a much closer comparison between theoretical predictions
and observations (e.g., \citealt{Wellons2015, Wellons2016,
  Remus2016}).  Among these new simulations is the Illustris Project
(\citeauthor{Illustris2014Nat} 2014a,\,b; \citealt{Genel2014Illustris,
  Sijacki2015Illustris, Nelson2015IllustrisDataRelease}), which
provides an ideal tool for such purposes. Run with the accurate
moving-mesh hydro solver {\small AREPO} (\citealt{Springel2010Arepo}),
the Illustris simulation took into account a wide range of baryonic
processes, resolved the formation of 40\,000 galaxies of different
morphology types, and managed to reproduce many fundamental properties
of observed galaxies.

In this paper, we report a variety of properties of early-type
galaxies in the highest resolution simulation of the Illustris
project\footnote{The highest resolution Illustris run covers a
  cosmological volume of (106.5 Mpc)$^3$ and has a dark matter mass
  resolution of $6.26\times10^6{\rm M}_{\odot}$ and an initial
  baryonic mass resolution of $1.26\times10^6{\rm M}_{\odot}$,
  resolving gravitational dynamics down to a physical scale of
  $\epsilon=710$ pc.}. In particular, we investigate the dependencies
and the redshift evolution since $z=1.0$ of (1) the stellar orbital
anisotropies, (2) the central dark matter mass fractions, and (3) the
central radial density slopes over the past $\sim7$ Gyrs. The main aim
of this work is to unveil correlations between these galaxy properties
and to link them to underlying physical processes. Also, we are
interested in identifying differences between simulation predictions
and observations, through which one can establish systematic biases of
observational techniques and the interpretations of the measurements.

The paper is organized as follows. In Sect.\,2, we describe in detail
how the light distributions of the simulated galaxies were determined
and how the observed galaxy properties were measured. In Sect.\,3, we
report general properties of the selected galaxies, including galaxy
and total matter morphologies (Sect.\,3.1), the mass-size-velocity
dispersion relations (Sect.\,3.2), and fundamental plane properties
(Sect.\,3.3). We then present the dependencies and the redshift
evolution of the stellar orbital anisotropies in Sect.\,4, those of
the central dark matter fraction in Sect.\,5 and those of the central
density slopes in Sect.\,6. Finally, a discussion and our conclusions
are given in Sect.\,7. We note that all the galaxy properties reported
in this paper have been made publicly available from the Illustris
website. In the Appendix, we give a brief summary of the content of
this online catalogue.

In this work, we adopted the same cosmology as used in the Illustris
simulation, i.e., a matter density of $\Omega_{\rm m}$ = 0.27, a
cosmological constant of $\Omega_{\Lambda}$ = 0.73, a Hubble constant
$h=H_0/(100\kms\Mpc^{-1})=0.70$ and a linear fluctuation amplitude
$\sigma_8=0.81$. These values are consistent with the Wilkinson
Microwave Anisotropy Probe (WMAP)-9 measurements
(\citealt{Hinshaw2013WMAP9}).

\section{Illuminating galaxies and measuring galaxy properties}

The galaxies simulated in Illustris are identified as gravitationally
bound structures of gas cells, dark-matter particles, and stellar
particles using the {\sc subfind} algorithm
(\citealt{Springel2001subfind, Dolag2009HydroSubfind}). In this
section, we describe how we calculated in a post-processing procedure
observational properties for the simulated galaxies.

It is noteworthy to recall that when a smaller halo is accreted onto a
bigger structure, its dark and baryonic matter at the outskirt can be
tidally stripped while sinking to the centre of the host. As a result,
a (galaxy) halo in a group or cluster environment can be composed of a
tightly-bound central region and a loosely-bound outskirt that extends
to large radii. Observationally, the measured ``galaxy'' properties
are only accounting for the former. However, the latter component,
which in the context of galaxy clusters is also known as
``intracluster light'', can make up for $\la 50\%$ of the total
stellar mass and luminosity (e.g., \citealt{LinMohr2004ICL,
  Zibetti2005ICL, Puchwein2010ICL}). This stellar mass is still bound
gravitationally to the galaxy and hence normally included in the raw
measurement of {\small SUBFIND}.

Often, a simple radial cut-off radius has been used in numerical
simulations to deal with this problem and to calculate properties that
are associated with the tightly-bound galaxy component. For example,
in Puchwein et al. (2010, see their eq.\,1), a halo mass-dependent
radial cut was applied to the central brightest cluster galaxies
(BCGs) more massive than
$2\times10^{13}M_{\odot}$. \citet{Schaye2015EAGLE} applied a
three-dimensional (3-D) radial cut of 30 physical kpc to all galaxies
from the {\small EAGLE} project. Such a choice was found to
significantly affect the measured properties of the tightly-bound
galaxy component that has a stellar mass more than
$10^{11}M_{\odot}$. Such a radial cut was also able to reproduce the
observed galaxy stellar mass function that was derived using the
frequently applied Petrosian apertures (see $\S5.1.1$ in
\citet{Schaye2015EAGLE} for detailed discussion). In this work, we
used a similar strategy. In calculations of projection-independent
intrinsic galaxy properties, such as the stellar mass, a 3-D radial
cut of 30 kpc is adopted. For the luminosity-based and thus
projection-dependent properties, we used a 2-D radial cut of 30 kpc
from the centre of light for a given galaxy projection. Any resolution
element still bound to the system but located or projected outside
this radius is excluded from the corresponding calculations.

\subsection{From stellar particles to galaxy light}

We calculated the emission properties of individual simulated galaxies
from the luminosities of their constituent stellar particles. Each
stellar particle of $\sim10^6h^{-1}{\rm M}_{\odot}$ is treated as a coeval
single stellar population that has the \citet{Chabrier2003IMF} initial
mass function (IMF). Given the star formation time and metallicity of
each stellar particle, a ``raw'' luminosity $L_{\rm raw}$ in a given
bandpass can thus be derived using the \citet{BC2003GALAXEV} stellar
population synthesis (SPS) model {\sc galaxev}.

This raw galaxy light can be processed by dust through absorption and
scattering at shorter wavelengths and re-emission at longer
wavelengths. We implemented such a dust attenuation process through a
simple semi-analytical approach as follows.  Assuming that a galaxy
can be approximated as a uniformly mixed slab of stars, gas and dust,
the amount of extinction of each stellar particle is given by
\begin{equation}
  \label{eq:DMextinction}
  \frac{L_{\rm obs}}{L_{\rm raw}}=
    \frac{1-\exp(-\tau_\lambda)}{\tau_\lambda},
\end{equation}
where $L_{\rm obs}$ and $L_{\rm raw}$ are the ``observed''
(dust-attenuated) and the raw (dust-free) luminosities, respectively.
$\tau_{\lambda}$ is the optical depth, which depends on the wavelength
$\lambda$ and is caused by both dust absorption and scattering along
the line of sight.

In order to predict the amount of dust extinction for each stellar
particle, a $100\times100$ mesh that covers the (projected) central
region of a simulated galaxy, from $-3$ to $+3$ times the half-stellar
mass radius in either dimension, was used to tabulate the distribution
of $\tau_{\lambda}$. The $\tau_\lambda$-mesh can be derived from the
neutral hydrogen (HI) distribution of the gas cells using a
semi-analytical prescription as follows.

A redshift- and metallicity-dependent optical depth
$\tau_\lambda^{\rm a}$ due to the dust absorption process is
traditionally modelled as (e.g., \citealt{GR1987GSED,
  Devriendt1999GSED, DG2000GSED})
\begin{equation}
\label{eq:KW07}
\tau_\lambda^{\rm a}=\bigg(\frac{A_\lambda}{A_{\rm
    v}}\bigg)_{Z_{\odot}} (1+z)^{\beta}\bigg(\frac{Z_{\rm
    g}}{Z_{\odot}}\bigg)^s\frac{\left<N_{\rm H}\right>}{2.1\times10^{21}{\rm
    cm}^{-2}},
\end{equation}
where $(A_\lambda/A_{\rm v})_{Z_{\odot}}$ is the solar-neighbourhood
extinction curve (\citealt{Cardelli1989ExtCurve}), $Z_{\rm g}$ is the
gas metallicity (of the galaxy) and $Z_{\odot}=0.02$ is the measured
value for the Sun. $\left<N_{\rm H}\right>$ is the average neutral
hydrogen column density. A power-law index $s=1.35$ for
$\lambda<2000\,$\AA\, and $s=1.6$ for $\lambda>2000\,$\AA\, was found
by \citet{GR1987GSED} for the metallicity dependence. In addition,
\citet{KW2007GFM} found that $\beta=-0.5$ can reproduce measurements
of Lyman-break galaxies at $z\sim3$. We also adopted this value for
$\beta$.

To also account for dust scattering, we used an approximate solution
from \citet{Calzetti1994Dust}, in which the total effective optical
depth $\tau_\lambda$ is given by
\begin{equation}
\tau_\lambda=h_{\lambda}\sqrt{1-\omega_{\lambda}}\tau_\lambda^{\rm
  a}+(1-h_{\lambda})(1-\omega_{\lambda})\tau_\lambda^{\rm a},
\label{eq:scatslabtau}
\end{equation}
where $\omega_{\lambda}$ is the albedo, defined as the ratio between
the scattering and the extinction coefficients, and $h_{\lambda}$ and
$1-h_\lambda$ are the weighing factors for the isotropic and
the forward-only scattering, respectively.

For each cell $(i,j)$ of the $\tau_\lambda$-mesh, a mean neutral
hydrogen column density $\left<N_{\rm H}^{(i,j)}\right>$ was first
calculated by scattering the (fractional) cold hydrogen masses of each
gas cell onto the mesh using a SPH smoothing technique.  $\left<N_{\rm
  H}^{(i,j)}\right>$ was then converted into $\tau_\lambda^{(i,j)}$
via Eqs.\,(\ref{eq:KW07}) and (\ref{eq:scatslabtau}), assuming
$\lambda$ being the effective wavelength of a given bandpass. For
stellar particles that are projected within the mesh coverage, the
exact $\tau_\lambda$ (at the position of a given stellar particle) was
then interpolated from the $\tau_\lambda$-mesh. For those outside,
$\tau_\lambda=0$ was assumed. The total light distribution of a
simulated galaxy in its viewing direction in a given bandpass was then
computed from $L_{\rm obs}$, the ``observed'' (dust-attenuated)
luminosities of the constituent stellar particles.

\subsection{Measurement of galaxy centres, ellipticities and
  orientation angles}

Each simulated galaxy (together with its dark matter halo) has a
``centre'' that was calculated as the position of the particle with
the minimum gravitational potential found using {\sc subfind}. The
observed galaxy centre is light-based and projection-dependent, we
therefore defined a two-dimensional (2-D) galaxy centre ($x_{\rm gc}$,
$y_{\rm gc}$) as:
\begin{equation}
\begin{array}{l}
x_{\rm gc}=\sum\limits_{i}L_i x_i\bigg(\sum\limits_{i}L_i\bigg)^{-1},\\
y_{\rm gc}=\sum\limits_{i}L_i y_i\bigg(\sum\limits_{i}L_i\bigg)^{-1},
\end{array}
\label{eq:GalCent}
\end{equation}
where $x_i$ and $y_i$ are the $x$- and $y$-coordinates of the $i$-th
stellar particle in the plane of a given galaxy projection. $L_i$ is
the ``observed'' (dust-attenuated) luminosity, and $\sum\limits_{i}L_i$ is
the total luminosity within a given aperture.

The orientation and ellipticity (or axis ratio) of a galaxy were
measured through luminosity-weighted second moments, which are defined
as:
\begin{equation}
\begin{array}{l}
M_{xx}=\sum\limits_{i}L_i (x_i-x_{\rm gc})^2\bigg(\sum\limits_{i}L_i\bigg)^{-1},\\
M_{yy}=\sum\limits_{i}L_i (y_i-y_{\rm gc})^2\bigg(\sum\limits_{i}L_i\bigg)^{-1},\\
M_{xy}=\sum\limits_{i}L_i (x_i-x_{\rm gc})(y_i-y_{\rm gc})\bigg(\sum\limits_{i}L_i\bigg)^{-1}.
\end{array}
\label{eq:SecondMoment}
\end{equation}
The axis ratio $b/a$ of a galaxy projection is then given by:
\begin{equation}
b/a=\bigg(\frac{M_{xx}+M_{yy}-\sqrt{(M_{xx}-M_{yy})^2+4M_{xy}^2}}
{M_{xx}+M_{yy} +\sqrt{(M_{xx}-M_{yy})^2 +4M_{xy}^2}}\bigg)^{1/2}.
\label{eq:b2a}
\end{equation}
The orientation angle $\phi_{\rm PA}$ is given by:
\begin{equation}
\phi_{\rm PA}=\frac{1}{2}\tan^{-1}\bigg(\frac{2M_{xy}}{M_{xx}-M_{yy}}\bigg).
\label{eq:PA}
\end{equation}

\subsection{Luminosity and effective radius measurement}

In order to calculate the galaxy luminosity and effective radius
within non-circular apertures, we first assumed that the 2-D surface
brightness distribution in a given viewing projection follows a series
of elliptical isophotes that are well described by $b/a$ and
$\phi_{\rm PA}$ measured using Eqs.\,(\ref{eq:b2a}) and (\ref{eq:PA})
within three-times the half stellar-mass radius from the centre of the
galaxy. A ``summed'' galaxy luminosity $L^{\rm sum}$ is calculated by
directly summing up $L_{\rm obs}$ for all constituent stellar
particles that are projected within 30 kpc.

A ``direct'' effective radius $R^{\rm dir}_{\rm eff}$ is determined as
the geometric mean of the semi-major and semi-minor radii of the
elliptical isophote which encloses half of $L^{\rm sum}$. We used
$L^{\rm sum}$ and $R^{\rm dir}_{\rm eff}$ as approximate estimates of
the intrinsic luminosity and size of a simulated galaxy. They are,
however, different from those used in observations. In order to make a
fair comparison, we followed one of the observational conventions to
derive a ``model'' luminosity $L^{\rm mod}$ and a ``model'' effective
radius $R^{\rm mod}_{\rm eff}$, i.e., by fitting a Sersic profile
(\citealt{SersicProfile1963}) to the radial distribution of the
elliptical isophotes that are assumed to closely trace the 2-D surface
brightness distribution of a given galaxy projection.

The surface brightness $I(R)$ of a Sersic profile at radius $R$ is
given by:
\begin{equation}
I(R)=I(R_{\rm eff})\exp\{-b_m[(R/R_{\rm eff})^{1/m}-1]\},
\end{equation}
where $R_{\rm eff}$ is the effective radius of the Sersic profile, and
$m$ is the Sersic index. The factor $b_m$ can be determined by
satisfying $\int_0^{\Reff}I(R)R\;{\rm d}R
=\frac{1}{2}\int_0^{\infty}I(R)R\;{\rm d}R$. We adopted the values of
$b_m$ from \citet{CiottiBertin1999Sersicm} for $m\geqslant0.36$ and
those of \citet{MacArthur2003Sersicm} for $m<0.36$.  The luminosity
$L(R)$ of a Sersic distribution enclosed within a radius of $R$ is
given by:
\begin{equation}
L(R)=I(R_{\rm eff})\exp(b_m)R_{\rm eff}^2\frac{2\pi
  m}{b_m^{2m}}\gamma(2m,\,b_m(R/R_{\rm eff})^{1/m}),
\end{equation}
where $\gamma(a,x)$ is the incomplete gamma function,
$\gamma(a,x) \equiv \int_0^{x}\exp(-t)t^{a-1}\;{\rm d}t$. Note that in
the formulae above, the radius $R$ is defined as the geometric mean of
the semi-major and semi-minor radii of the elliptical isophotes.

For each galaxy projection, the binned radial profile between
0.05$\,R^{\rm dir}_{\rm eff}$ and 3.0$\,R^{\rm dir}_{\rm eff}$ was
fitted by the Sersic model using a minimum $\chi^2$ fitting
approach. In general, the radial surface brightness distributions of
the simulated galaxies well follow Sersic profiles. The model
effective radius $R^{\rm mod}_{\rm eff}$ of a given galaxy projection
was set to be the best-fitting Sersic effective radius.  The model
luminosity $L^{\rm mod}$ was then given by the best-fitting Sersic
luminosity within $7\,R^{\rm mod}_{\rm eff}$.

We measured the light distribution of each simulated galaxy in a
variety of optical filter bandpasses in order to compare with latest
observations. Note that galaxy luminosities (magnitudes) and effective
radii are band-dependent (also see \citealt{Barbera2009}). Hereafter,
the effective radii $\Reff$ specifically refer to the Sersic model
effective radius that was derived in the rest-frame Johnson
$V$-band. For galaxies at $z=0.3$ measured in this band, the
luminosity ratios have a (stellar-mass weighted) mean of $\langle
L^{\rm mod}/L^{\rm sum} \rangle=1.06$ with a standard deviation of
0.16, and the effective radii ratios have a (stellar-mass weighted)
mean of $\langle \Reff^{\rm mod}/\Reff^{\rm dir} \rangle=1.13$ with a
0.33 scatter. These ratios vary little across the investigated
redshift range.

\begin{figure}
\centering
\includegraphics[width=8cm]{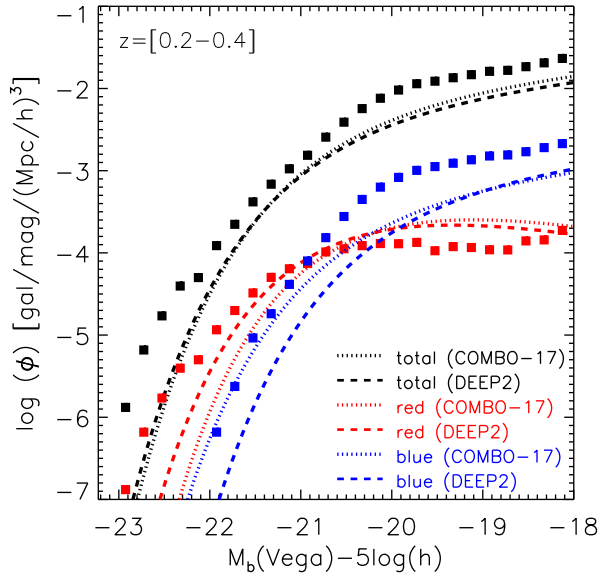}
\caption{Galaxy luminosity functions at intermediate redshifts
  ($0.2<z<0.4$), measured in Johnson-$B$ band. Simulation data are
  given by square symbols, where black, red and blue colours represent
  the total, ``red'', and ``blue'' galaxy samples, respectively. The
  latter two are rescaled by a factor of 0.1 for graphical
  clarity. The red and blue galaxy samples were selected from
  the simulation according to eq.\,1 in \citet{Faber2007LF}. The
  dotted and the dashed lines give single power-law Schechter function
  fits to the observed galaxy luminosity functions from the COMBO-17
  and DEEP2 surveys, respectively.}
\label{fig:LFs}
\end{figure}

In Fig.\,\ref{fig:LFs} we present the dust-corrected luminosity
functions (LF) of the Illustris galaxies at intermediate redshifts
($0.2<z<0.4$), measured in the Johnson-$B$ band. Plotted in the same
panel are the Schechter function fits to the observed LFs from DEEP2
and COMBO-17 surveys (\citealt{Faber2007LF}). The red and blue galaxy
samples were selected from the simulation according to eq.\,1 in
\citet{Faber2007LF}. The Illustris galaxy LFs have been found to
roughly match observational results in various bands and within a wide
range of redshifts (also see Vogelsberger et al. 2014b;
\citealt{Hilbert2016IllustrisIA}). We note, in particular, that the
effect of dust attenuation is significant especially for blue galaxy
sample, which, without taking dust into account, would contribute to
at least half of the total galaxy count at any given luminosity up to
$M\sim-22$. In comparison, the observed LFs have this transition at
two magnitudes higher, where the blue and red galaxy samples have
equal contributions to the total LF and below which (in the brighter
end), red galaxies would dominate the total LF. As can be seen,
applying dust attenuation to the simulated galaxies has pushed this
transition magnitude much closer to the observation. At the bright
end, there are larger measurement uncertainties, and the derived
magnitudes can differ strongly depending on the assumed light profiles
(e.g., \citealt{Bernardi2013LFprofile}).

\subsection{Galaxy type classification}

The method that we used for galaxy classification is similar to the
practice of the Sloan Digital Sky Survey (SDSS). Using the rest-frame
SDSS $g$, $r$ and $i$ filters simultaneously, we fitted both de
Vaucouleurs profiles (\citealt{DeVaucouleurs1948}) and exponential
profiles to the radial surface brightness distributions of the
elliptical isophotes (between 0.05$\,R^{\rm dir}_{\rm eff}$ to
3.0$\,R^{\rm dir}_{\rm eff}$). If the former provides a better fit,
then the galaxy is classified as an early-type (elliptical) galaxy,
whereas if the latter fits better, then it is considered a late-type
(disk) galaxy.

\begin{figure}
\centering
\includegraphics[width=8cm]{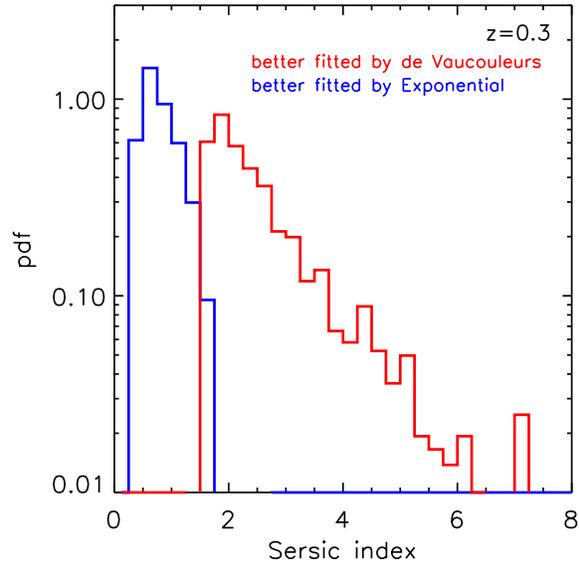}
\caption{Histograms of the best-fitting Sersic indices of galaxies
  whose radial surface brightness distributions are better fitted by
  de Vaucouleurs profiles (red) and of those whose radial
  distributions are better fitted by exponential profiles (blue).}
\label{fig:SersicHisto}
\end{figure}

\begin{figure}
\centering
\includegraphics[width=4cm]{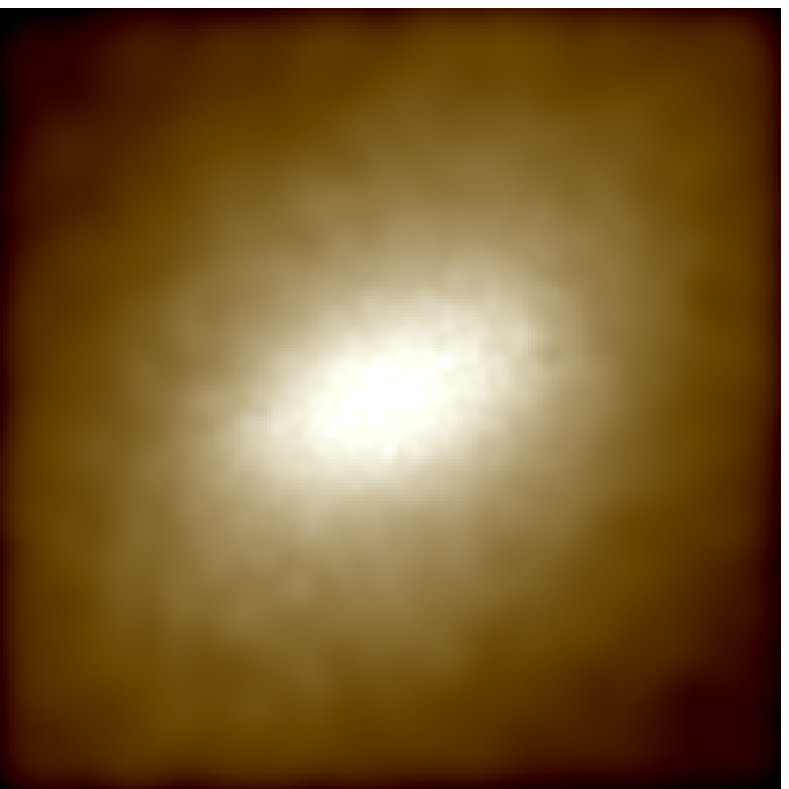}
\includegraphics[width=4cm]{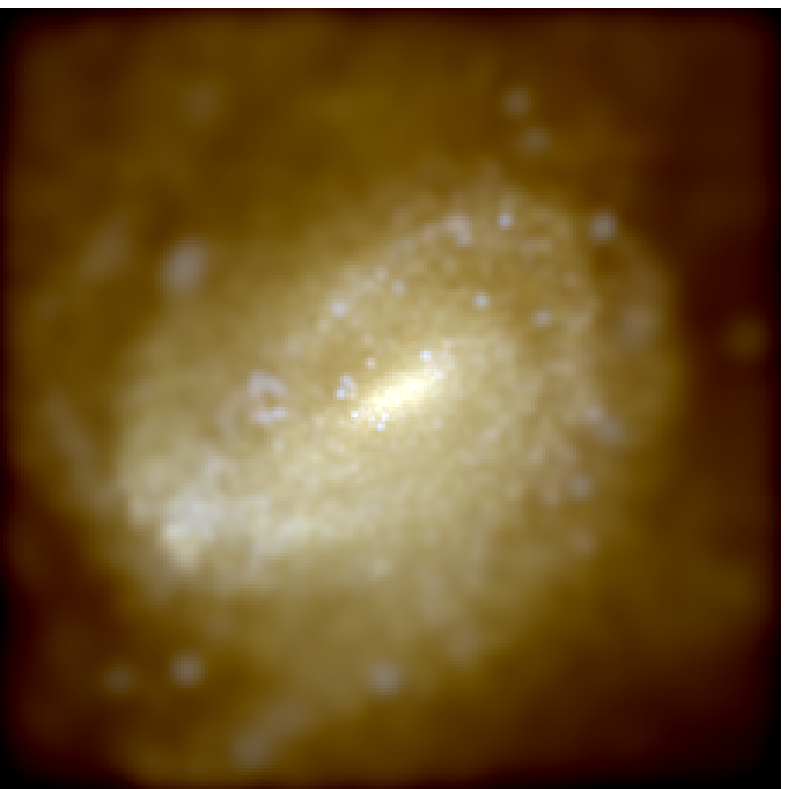}
\includegraphics[width=8cm]{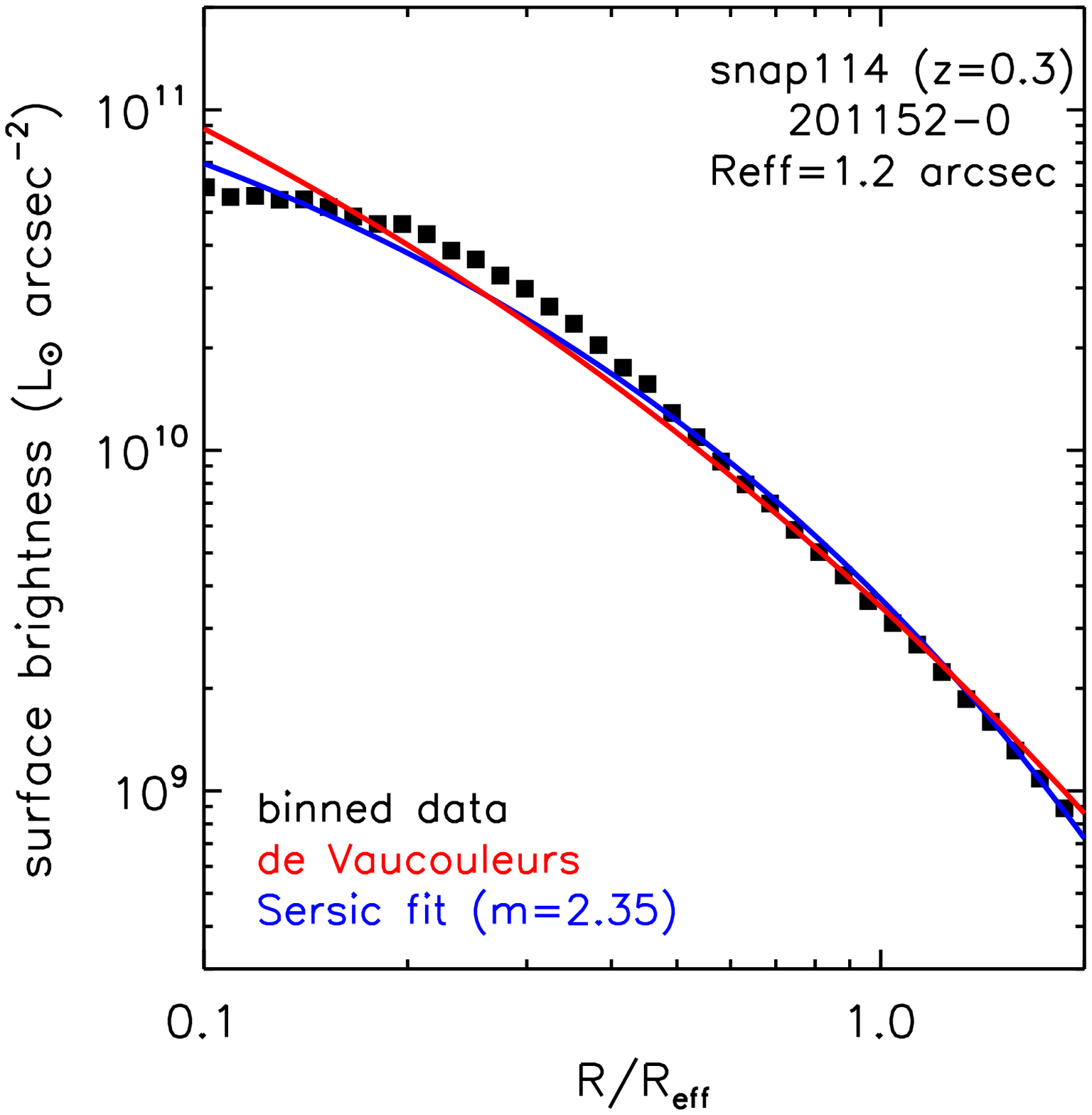}
\includegraphics[width=8cm]{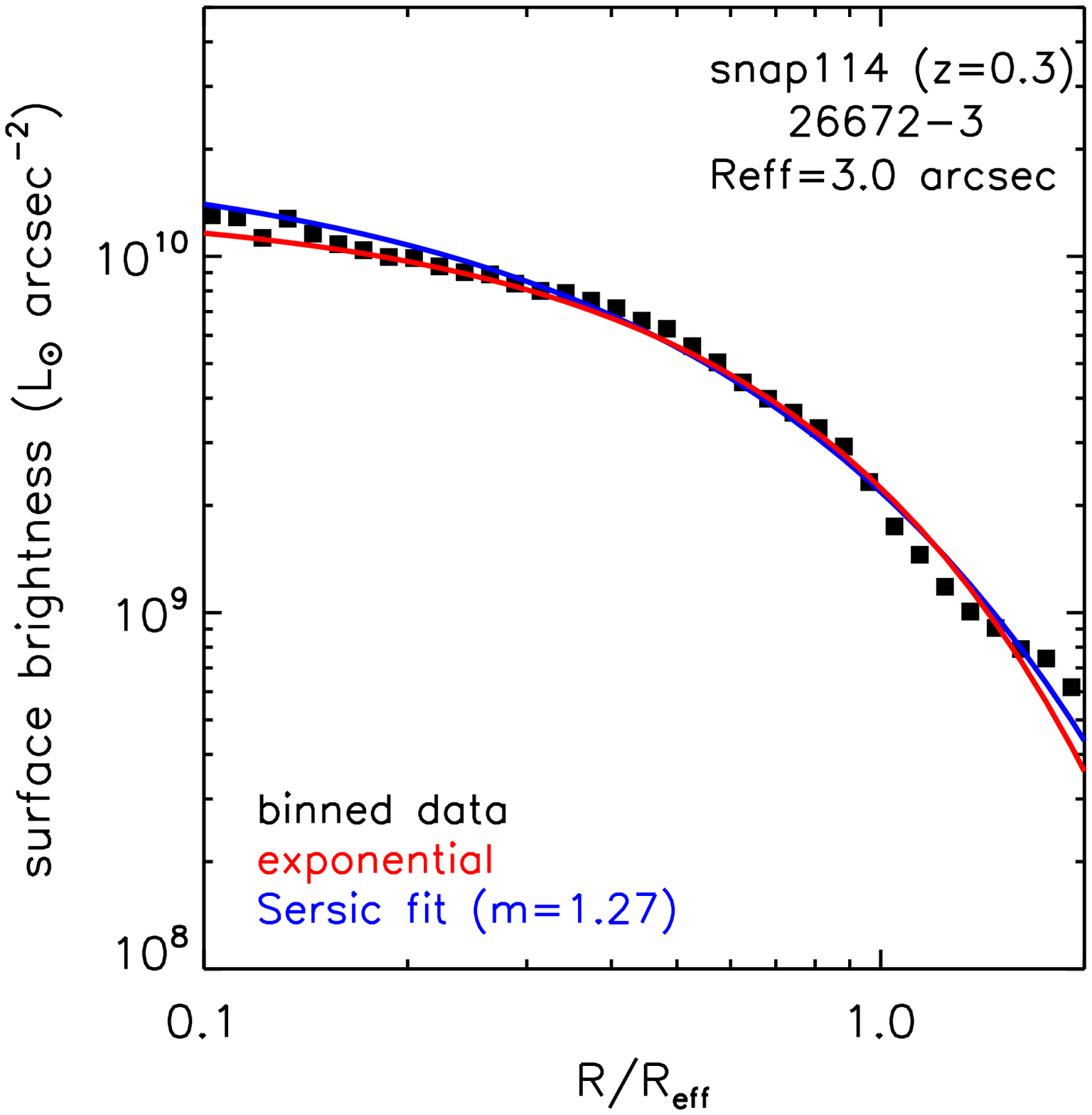}
\caption{The top row presents synthesized images of an elliptical
  (left panel) and a disk (right panel) galaxy from the simulation at
  redshift $z=0.3$. The image was made by combining the surface
  brightness distributions in the rest-frame SDSS $g$, $r$ and $i$
  filter bandpasses, within a $3R_{\rm eff}\times 3R_{\rm eff}$ region
  from the galaxy centres. The middle and bottom panels show radial
  surface brightness distributions of the early- and late-type
  galaxies, respectively. In these panels, black dots are the binned
  data measured in the SDSS $r$ band, blue lines give the best-fitting
  Sersic profiles (Sersic indices of 2.35 and 1.27, respectively), and
  red lines show the best-fitting de Vaucouleurs and exponential
  profiles of the two galaxies, respectively.}
\label{fig:EllDisk-example}
\end{figure}

Fig.\,\ref{fig:SersicHisto} shows histograms of the best-fitting
Sersic indices of the simulated early- and late-type galaxies at
redshift $z=0.3$. As expected, the former have larger ($m\ga2$) Sersic
indices, while the latter have $m<2$. As an illustration, we select
two typical galaxies with different type classifications and show
their synthesized light distributions in the top panel of
Fig.\,\ref{fig:EllDisk-example}, where the left and right sub-panels
display an elliptical and a disk galaxy at $z=0.3$, respectively. The
image was made by combining the surface brightness distributions in
the rest-frame SDSS $g$, $r$ and $i$ filter bandpasses, within a
$3R_{\rm eff}\times 3R_{\rm eff}$ region from the light centres of the
corresponding galaxy projections.

The lower panels of Fig.\,\ref{fig:EllDisk-example} show the radial
surface brightness distributions of the early- (middle) and late-type
(bottom) galaxies above. In these panels, black dots are the binned
data measured in the SDSS $r$ band, blue lines give the best-fitting
Sersic profiles (Sersic indices of 2.35 and 1.27, respectively), while
red lines show the best-fitting de Vaucouleurs and exponential
profiles for the two galaxies, respectively.

\section{General properties of the galaxy catalogue}

Apart from the above mentioned luminous properties, we derived a wide
range of projection-dependent lensing and dynamical properties, such
as the Einstein radii $\RE$, dark matter fractions $f_{\rm dm}$,
stellar velocity dispersions $\sigma$, orbital anisotropies $\beta$,
and various mass density slope estimators $\gamma$. All these
calculations were carried out for galaxies (regardless of galaxy
types) with stellar masses $M_{*}\ga10^{10}{\rm M}_\odot$
(corresponding to more than $\sim$ 10\,000 stellar particles) at
redshifts in the range $z\in[0.1,~1.0]$ with an interval spacing
$\Delta z=0.1$. We defined artificial source redshifts at $z_{\rm
  s}=[0.5, ~0.6, ~0.7, ~0.9, ~2.0, ~2.0, ~2.0, ~2.0, ~2.0, ~2.0]$
accordingly, in order to calculate the Einstein radii\footnote{The
  Einstein radius $\RE$ is found as the radius within which the mean
  surface density is equal to the lensing critical density
  $\Sigma_{\rm cr}=\left(\frac{c^2}{4\pi G}\right)
  \left(\frac{\Ds}{\Dds\Dd}\right)$, where $\Dd$, $\Ds$ and $\Dds$ are
  the angular diameter distances to the lens, to the source, and from
  the lens to the source, respectively.}. These choices for $z_{\rm
  s}$ are motived by the observed lens-source redshift
distributions. All the calculated properties have been catalogued and
are publicly available from the Illustris website
(www.illustris-project.org), and a detailed description of the
different catalogue fields can be found in the Appendix.

In the following part of the paper, we aim at presenting the
statistical properties of the simulated early-type galaxies and
comparing them to those resulted from the SDSS early-type galaxy
survey (\citealt{HB2009SDSSETG}), and to those from recent strong
lensing surveys, i.e., the SLACS and SL2S surveys, as well as those
from the Cosmological Monitoring of Gravitational Lenses projects
(COSMOGRAIL; \citealt{Sluse2012COSMOGRAIL}). These surveys provide
comprehensive observational samples predominantly composed of isolated
early-type galaxies within a wide redshift range, similar to the one
studied here.

We are aware that the lensing selection effect has always been a
complication when comparing simulation samples to observations (or
using observed samples to interpret physical properties of
galaxies). Detailed work in this regard can be found in, e.g.,
\citet{MVK2009LensingSelection}. For the present study, we do not
apply any sophisticated selection criteria. We selected Illustris
galaxies at each given redshift according to the following two simple
criteria: (1) central and early-type galaxies, and (2) the stellar
line-of-sight velocity dispersions $\sigma_{e/2}$ (measured within
0.5$\,\Reff$) satisfies $\sigma_{e/2} \in[160,400]\kms$. These
criteria are mainly observationally motivated by the strong lensing
galaxy surveys.  As will be seen in Sect.\,3.2, the criteria above
resulted in galaxy samples that roughly reproduced the observed
mass-size-velocity dispersion relations.

It is worth noting that when calculating properties of a central
galaxy, any self-gravitationally-bound substructures (satellite
galaxies) identified by {\sc subfind} were excluded. To increase the
sample size, we treated galaxies that are viewed along the three
principal directions of the simulation box as independent. The
selection criteria resulted in $\sim600$ independent galaxy
projections at each of the simulation redshifts investigated. In this
section, we present the measured shapes of the luminous and dark
matter distributions (Sect.\,3.1), mass-size-velocity dispersion
relations (Sect.\,3.2) and fundamental plane relations (Sect.\,3.3) of
the selected galaxy samples at various redshifts.

\subsection{{Shape of luminous and total matter}}

\begin{figure}
\centering
\includegraphics[width=8cm]{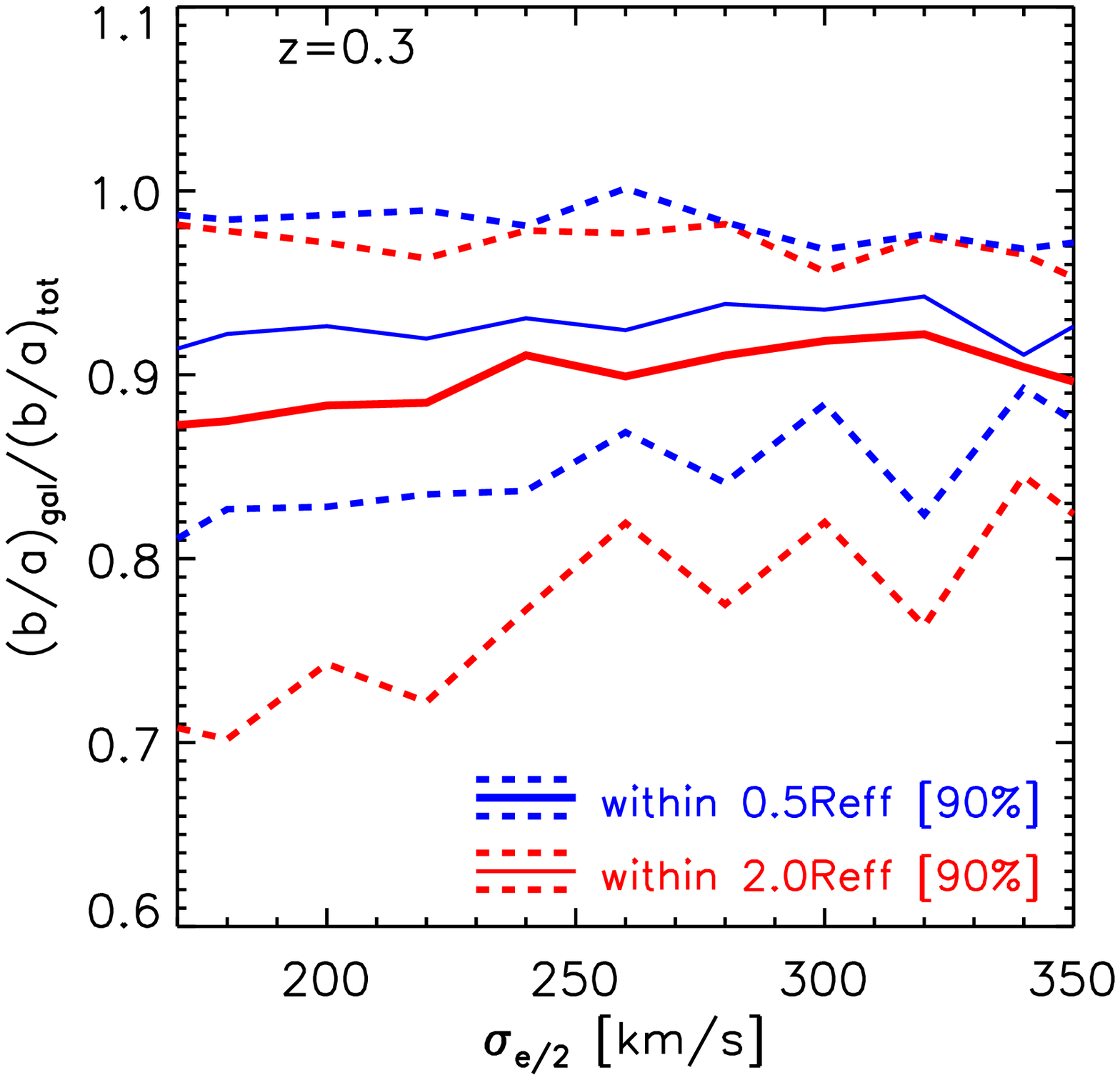}
\includegraphics[width=8cm]{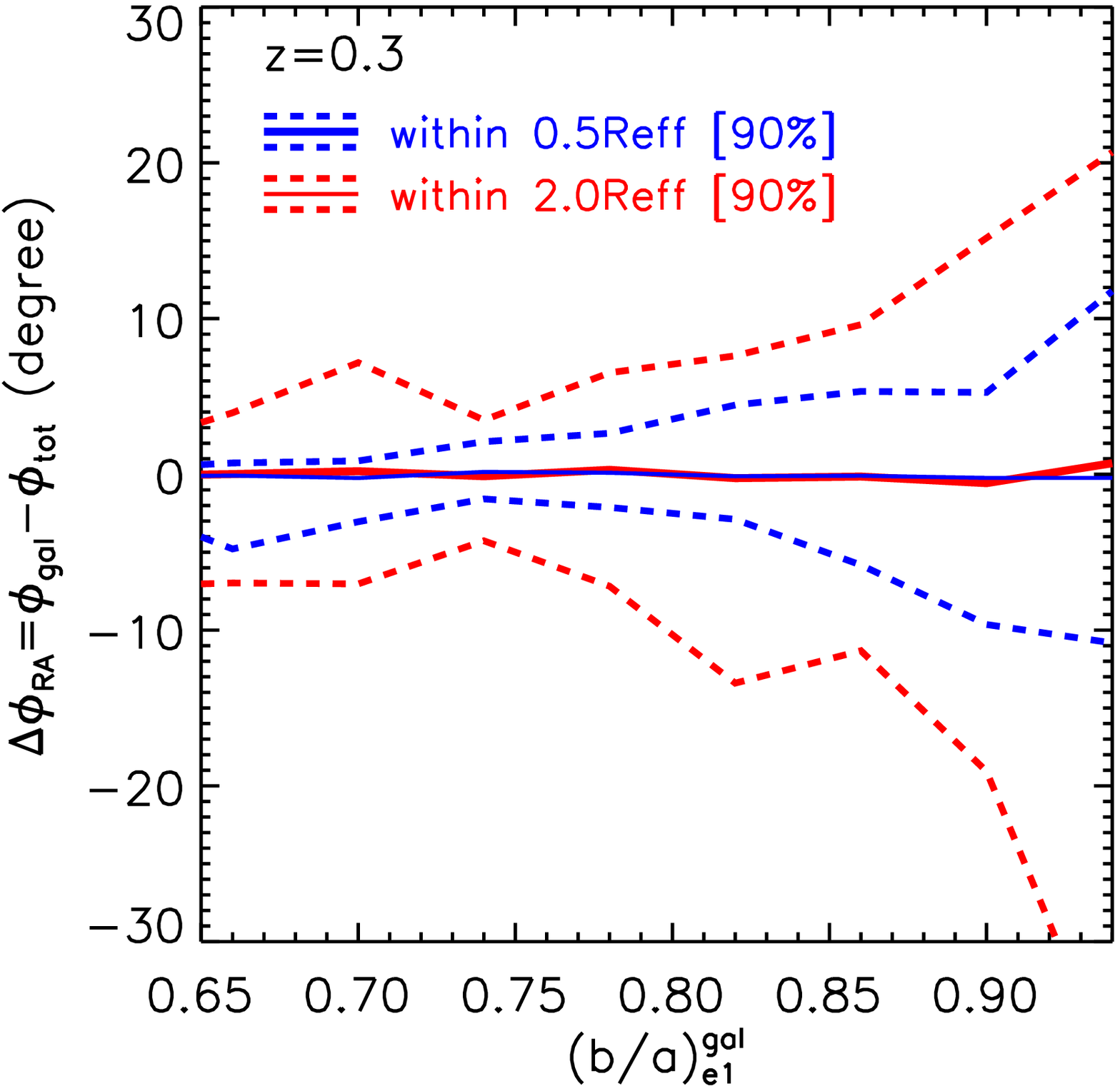}
\caption{{\em Top panel:} the ratio between the axis ratio $b/a$
  measured for the light distribution and that for the total matter
  distribution, as a function of the central stellar velocity
  dispersion $\sigma_{e/2}$, for the selected early-type galaxy sample
  at $z=0.3$. {\em Bottom panel:} the difference between the
  orientation angle of the light distribution and that of the total
  matter distribution, as a function of the galaxy axis ratio
  $b/a^{\rm gal}_{e1}$ measured within $R_{\rm eff}$, for the same
  galaxy sample. In both panels, blue and red represent the
  distributions measured within $0.5\,R_{\rm eff}$ and $2.0\,R_{\rm
    eff}$, respectively. The solid and the dashed lines indicate the
  median and the 90\% boundaries of the distributions, respectively.}
\label{fig:b2a_ratioRA_diff}
\end{figure}

It has always been a critical question ``how well the light follows
the matter''. In central regions of early-type galaxies, the projected
total matter distributions are in general rounder\footnote{The
  situation is the opposite at larger radii for which weak lensing
  technique can be applied. At those radii, dark matter dominates the
  total matter distribution, which appears to be flatter than the
  light profile (e.g., \citealt{Hoekstra2004WLGDMH}).} than the
luminous (stellar) distributions. This can be seen from the top panel
of Fig.\,\ref{fig:b2a_ratioRA_diff}, which shows the ratio between the
luminous axis ratio $(b/a)_{\rm gal}$ and the total axis ratio
$(b/a)_{\rm tot}$, as a function of the central stellar velocity
dispersion $\sigma_{e/2}$: the solid blue and red line indicates the
median of the distribution measured within $0.5\,R_{\rm eff}$ and
$2.0\,R_{\rm eff}$, respectively; the dashed lines show the 90\%
boundaries of the distributions. Note that the ellipticity ratio
decreases as the aperture size increases from $0.5\,R_{\rm eff}$ to
$2.0\,R_{\rm eff}$.  This is mainly attributed to the fact that as the
aperture size increases the total matter distribution becomes rounder,
while the ellipticity of the stellar distribution only varies
mildly. The ellipticity ratio distribution does not seem to evolve
strongly with redshift, at higher redshift the scatter increases
marginally.

Observationally, the shape of the total matter distribution of a
(lensing) galaxy can be inferred via a lens modelling technique, while
that of the stellar distribution can be obtained from direct
imaging. In this regard, previous studies based on singular isothermal
ellipsoidal (SIE) lens models found that the median ellipticity ratios
of the galaxy samples from the SLACS, SL2S and COSMOGRAIL (see
Koopmans et al. 2006; Gavazzi et al. 2012;
\citealt{Sluse2012COSMOGRAIL}; \citealt{Shu2015SLACSXII}) are very
close to 1.0, which lies above both the median and the 90\% upper
boundary of the simulation distribution. Regarding this disagreement,
we note that, as will be seen in Sect.\,5, the central dark matter
fractions in the Illustris early-type galaxies were found to be
systematically higher compared to the observational results. This may
also explain the systematically lower ellipticity ratios between the
luminous and the total matter distributions of the simulated
galaxies. However, it is also noteworthy to point out that recent
lensing studies that have either applied non-parametric lens modelling
(e.g., \citealt{Bruderer2016NonParamLensShapeDMP}) or adopted novel
techniques to extract galaxy morphology data from adaptive optics
observations (e.g., \citealt{Rusu2016SubaruSDSS}) found that the total
matter distributions are systematically rounder than the stellar
distributions in central regions of early-type galaxies, in agreement
with our simulation results.

The orientation of the projected total matter distribution follows in
general that of the light distribution very well. The bottom panel of
Fig.\,\ref{fig:b2a_ratioRA_diff} shows the misalignment angle
$\Delta\phi_{\rm RA}$ between the orientation angles $\phi_{\rm gal}$
of the light distribution and $\phi_{\rm tot}$ of the total matter
distribution, as a function of the galaxy axis ratio $(b/a)^{\rm
  gal}_{e1}$ that was measured within $R_{\rm eff}$.  Larger scatter
of $\Delta\phi_{\rm RA}$ for rounder galaxies is due to less clean
measurements of their orientation angles. A marked evolution exists
for the selected early-type galaxy samples at different redshifts,
e.g., for a sub-sample of galaxies that have $(b/a)^{\rm gal}_{e1} \in
[0.75,0.85]$, the standard deviation of $\Delta\phi_{\rm RA}$ measured
within $R_{\rm eff}$ decreases from $\sim12^{\circ}$ at $z=1.0$ to
$\sim5^{\circ}$ at $z=0.6$, and finally to $\la2^{\circ}$ at
$z\leqslant0.3$.  This evolution coincides with a mild increase of the
(projected) central baryonic fraction towards lower redshifts (see
Fig.\,\ref{fig:slpZev}). In the intervening period, stars and dark
matter become more mixed and thus better aligned with cosmic
time. These results are consistent with strong lensing observations
(e.g., \citealt{Sluse2012COSMOGRAIL, Dye2014HerschelATLAS}, Rusu et
al. 2016). In particular, for galaxies that have stellar axis ratios
falling within similar ranges as above, Gavazzi et al. (2012) measured
an rms scatter of $18^{\circ}$ for the SL2S sample ($0.3<z<1.0$) and
Koopmans et al. (2006) reported an rms deviation of $3^\circ$ for the
SLACS galaxy sample ($z<0.3$).

\subsection{Galaxy mass, size, and velocity dispersion}

\begin{figure}
\centering
\includegraphics[width=8cm]{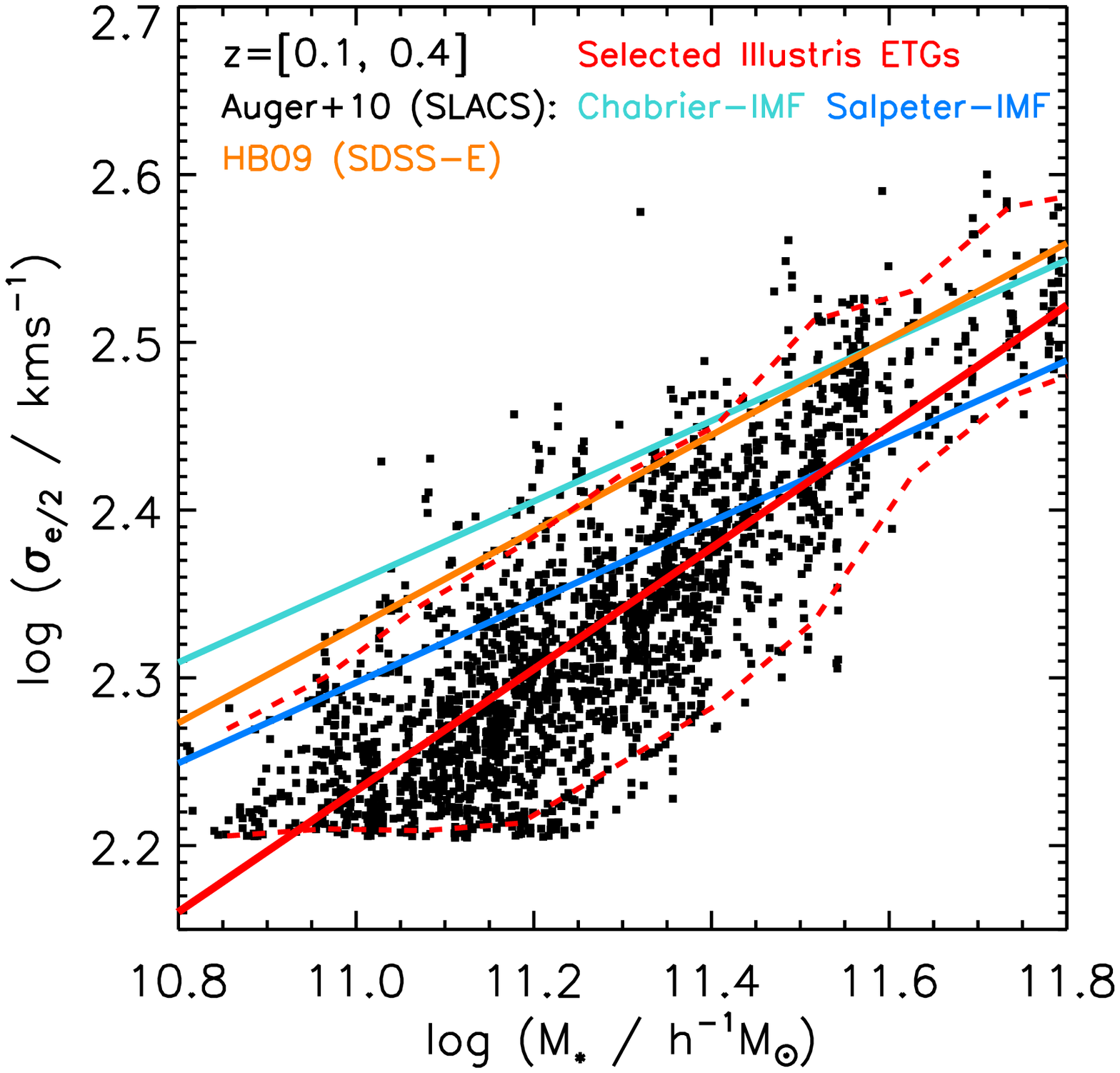}
\includegraphics[width=8cm]{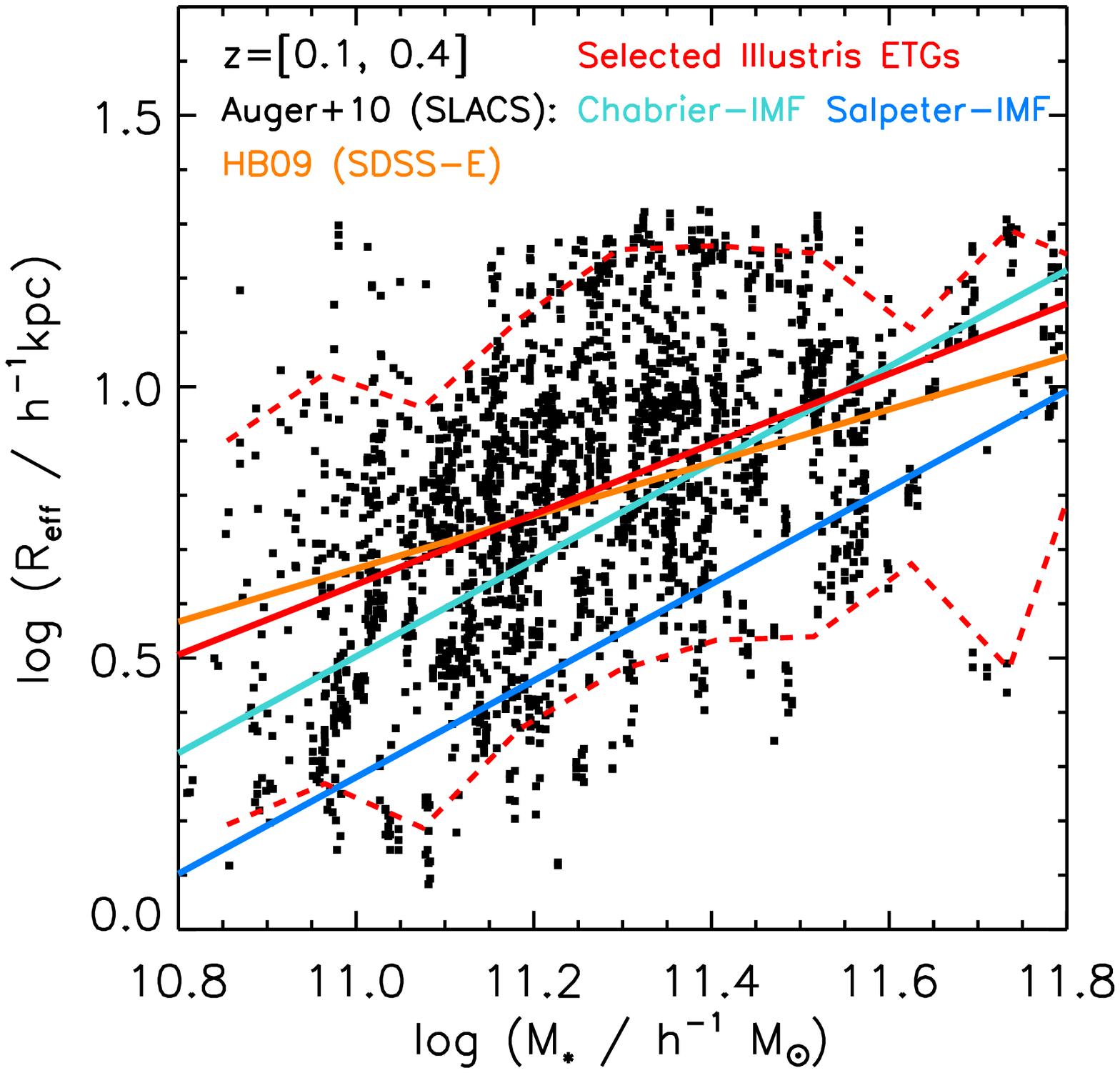}
\caption{The $\sigma_{e/2}-M_{*}$ (top panel) and $R_{\rm eff}-M_{*}$
  (bottom panel) relations. The black squares mark the simulation data
  of the selected early-type galaxies at $z=0.1-0.4$. The lower cut in
  stellar velocity distribution (in the top panel) is due to our
  selection criteria. The solid and the dashed red lines indicate the
  best linear fit to the data and the 90\% boundaries of the
  distribution, respectively. For comparison, the solid blue and cyans
  lines represent the best linearly fitted relations for the SLACS
  galaxies (Auger et al. 2010b), assuming either Salpeter IMFs or
  Chabrier IMFs, respectively. The solid orange lines indicate the
  linearly fitted relations for the observed early-type galaxies from
  the SDSS survey (\citealt{HB2009SDSSETG}).}
\label{fig:SigmaReffMass}
\end{figure}

The relations between galaxy size, velocity dispersion and stellar
mass are among the most basic scaling relations.
Fig.\,\ref{fig:SigmaReffMass} shows the $\sigma_{e/2}-M_{*}$ and
$R_{\rm eff}-M_{*}$ relations measured for the selected early-type
galaxies at $z=0.1-0.4$. In either panel, the black squares mark the
simulation data, and the solid and the dashed red lines indicate the
best linear fit to the data and the 90\% boundaries of the
distribution, respectively. Note that the lower cut in the stellar
velocity distribution (in the upper panel) is due to our sample
selection criteria.

For comparison, the solid blue and cyans lines represent the best
linearly fitted relations for the SLACS galaxies (Auger et al. 2010b),
assuming either Salpeter IMFs (\citealt{Salpeter1955IMF}) or Chabrier
IMFs (\citealt{Chabrier2003IMF}), respectively. Recent studies suggest
that a Salpeter IMF is more compatible with observational inferences
for massive early-type galaxies (e.g., \citeauthor{Auger2010IMF}
2010a; \citealt{Treu2010IMF, GrilloGobat2010ETGIMF, Barnabe2011III,
  Spiniello2011DMFSIMF, Oguri2014a, Sonnenfeld2015SL2SV}), while a
Chabrier IMF is more suitable for those at the lower end of the
spectrum (e.g., \citealt{Shu2015SLACSXII}). In addition, the linearly
fitted relations for the observed early-type galaxies from the SDSS
survey (\citealt{HB2009SDSSETG}) are also given, as indicated by the
solid orange lines.

As can be seen, there exists some slight underestimation of the
central velocity dispersions for galaxies at the lower stellar-mass
end. However, we note that the level of the disagreement between the
simulation and the observations is comparable to that the systematic
divisions among the observational relations themselves, which can be
attributed to the uncertainty in the IMF and/or different sample
biasing etc. Overall, the simulation broadly reproduces the observed
$\sigma_{e/2}-M_{*}$ and $R_{\rm eff}-M_{*}$ relations.

\subsection{The fundamental plane and the mass planes}

Elliptical galaxies are observed to tightly follow the so-called
Fundamental Plane (hereafter FP; \citealt{Faber1987FP, DD1987FP,
  Dressler1987FP}) and its mass counterparts -- the Stellar Mass Plane
(hereafter M$_{*}$P; \citealt{HB2009StellarMP}) and Mass Plane
(hereafter MP; \citealt{Bolton2007MP}). We examined how well the
simulated galaxies trace these observed planes. To this end we adopted
the same definitions as Auger et al. (2010b) and fitted the plane
relations with the following form:
\begin{eqnarray}
&& \log \frac{R_{\rm eff}}{\rm kpc} = a + b
  \log\frac{\sigma_{e/2}}{100 \kms} + c \log\Lambda, \\ &&
  \Lambda=\frac{1}{2\pi} \frac{L_V}{10^9 L_{\odot}}
  \bigg(\frac{\Reff}{\kpc}\bigg)^{-2} (\mbox{for FP}),
  \\ && \Lambda=\frac{1}{2\pi} \frac{M_{*}}{10^9 M_{\odot}}
  \bigg(\frac{\Reff}{\kpc}\bigg)^{-2} (\mbox{for M$_{*}$P}),
  \\ && \Lambda=\frac{4}{\pi} \frac{M_{\rm tot}(\leqslant
    0.5\,\Reff)}{10^{10} M_{\odot}} \bigg( \frac{\Reff}{\kpc}
  \bigg)^{-2} (\mbox{for MP}).
  \label{eq:FPMP}
\end{eqnarray}
In Eq.\,(11), $L_V$ is the Sersic luminosity measured in the
rest-frame Johnson-$V$ band.  A measurement uncertainty of $5\%$ was
assumed for each of the five quantities of the simulated galaxies. The
fitting of the coefficients $a$, $b$ and $c$ was done using of the
{\sc lts\_planefit} program described in \citet{Cappellari2013XV},
which combines the Least Trimmed Squares robust technique of
\citet{RD2006PlaneFit} with a least-squares fitting algorithm which
allows for errors in all variables as well as intrinsic scatter.

\begin{figure}
\centering
\includegraphics[width=8cm]{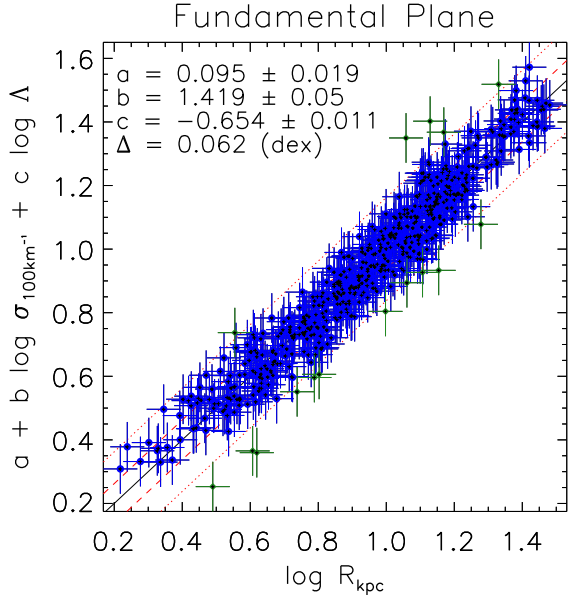}
\caption{The fundamental plane relation of the selected early-type
  galaxy sample at $z=0.3$ from the simulation. $R_{\rm kpc}$ is
  $R_{\rm eff}/{\rm kpc}$ and $\sigma_{100\kms}$ is
  $\sigma_{e/2}/(100\kms)$. A measurement uncertainty of $5\%$ was
  assumed for each ``observed'' quantity. The blue circles filled with
  black dots indicate individual simulated galaxies. The red solid and
  dashed lines give the $2.6\sigma$ and $1\sigma$ bounds,
  respectively, from the best-fitting relation given by the black
  solid line. Galaxies indicated by green symbols are outside
  $3\sigma$ of the fitted relation. The fitting and plotting software
  {\sc lts\_planefit} used here was provided by Cappellari et
  al. (2013). }
\label{fig:FPSMP}
\end{figure}

Fig.\,\ref{fig:FPSMP} shows the FP relation of the selected early-type
galaxy sample at $z=0.3$ from the simulation. The blue circles filled
with black dots indicate individual galaxies. The red solid and dashed
lines show $2.6\,\sigma$ and $1\,\sigma$ scatter, respectively, around
the best-fitting relation, which is given by the black solid
line. Galaxies indicated by green symbols are outside $3\,\sigma$ of
the fitted relation.  For the FP relation, we found best-fitting
coefficients equal to $a=0.10\pm0.02$, $b=1.42\pm0.05$ and
$c=-0.65\pm0.01$. For the M$_{*}$P, we obtained $a=0.43\pm0.01$,
$b=1.11\pm0.04$ and $c=-0.54\pm0.01$; and for MP, $a=-0.04\pm0.02$,
$b=1.63\pm0.06$ and $c=-0.96\pm0.02$.

The fitting coefficient $a$ represents the normalizations of the plane
relations; while $b$ and $c$ depict the slopes of the plane
relations. In particular, coefficients $b$ and $c$ depend on the
sample selection, observational bands, and fitting methods (see
\citealt{Bernardi2003SDSSFP}). Observationally, $b$ ranges from 0.99
to 1.52, and $c$ ranges from $-0.88$ to $-0.74$ for the FP; for the
MP, $b$ varies from 1.77 to 1.86, and $c$ ranges from $-1.30$ to
$-0.83$ (\citealt{Bernardi2003SDSSFP}; \citeauthor{Bolton2007MP,
  Bolton2008SLACSVIIFP} 2008b; Auger et al. 2010b; Cappellari et
al. 2013). Within the fitting uncertainties, our best-fitting values
for $a$, $b$ and $c$ are not in stark disagreement with observational
results. A more detailed study of the FP relation of the Illustris
early-type galaxies will be presented in a separate paper (Li et
al. in preparation).


\section{Stellar orbital anisotropies}

One of the major uncertainties in interpreting kinematical data is the
stellar orbital anisotropy, which cannot be directly measured or
constrained with high accuracy (e.g., \citealt{Li2016IllustrisJAM}) as
it is degenerate with galaxy density slopes, unless the latter can be
determined independently. When combined with single-aperture stellar
kinematics, strong lensing studies often have to assume zero orbital
anisotropies (but see e.g., \citealt{Barnabe2009II, Barnabe2011III}
for 2-D kinematics studies of the SLACS lenses). In this section, we
thus investigate the dependencies of the anisotropy parameter $\beta$
and its redshift evolution since $z=1.0$ to check the validity of this
assumption.

For spherically symmetric systems, the anisotropy parameter $\beta$
can be written as (\citealt{BinneyTremaineBook})
\begin{equation}
\beta=1-\frac{\overline{V_{\phi}^2}+\overline{V_{\theta}^2}}{2\overline{V_{r}^2}}=1-\frac{\sigma_{\phi}^2+\sigma_{\theta}^2}{2\sigma_{r}^2},
\label{eq:beta}
\end{equation}
where $\overline{V_{\phi}^2}$, $\overline{V_{\theta}^2}$ and
$\overline{V_{r}^2}$ are the second velocity moments, and
$\sigma_{\phi}^2$, $\sigma_{\theta}^2$ and $\sigma_{r}^2$ are the
velocity dispersions measured in the azimuthal $\hat{\phi}$, polar
$\hat{\theta}$ and radial $\hat{r}$ directions of a spherical
coordinate system. By definition,
$\sigma^2\equiv\overline{V^2}-\overline{V}^2$, where $\overline{V}$ is
the mean velocity. The second equality sign in Eq.\,(\ref{eq:beta}) is
valid for stationary non-rotating systems, where
$\overline{V_{\phi}}$, $\overline{V_{\theta}}$ and $\overline{V_{r}}$
vanish. Note that observationally, as the measurements are carried out
for the light components, $\overline{V_{\phi}}=0$,
$\overline{V_{\theta}}=0$ and $\overline{V_{r}}=0$ by construction.
We followed the same convention, measuring $\beta$ through $\sigma^2$
instead of $\overline{V^2}$. In this sense, $\beta$ is constructed to
measure the anisotropy of the velocity dispersion. $\beta=0$
corresponds to the ``isotropic'' case, and $\beta>0$ ($\beta<0$)
describes a radially (tangentially) anisotropic orbital distribution.

\begin{figure}
  \centering
  \includegraphics[width=8cm]{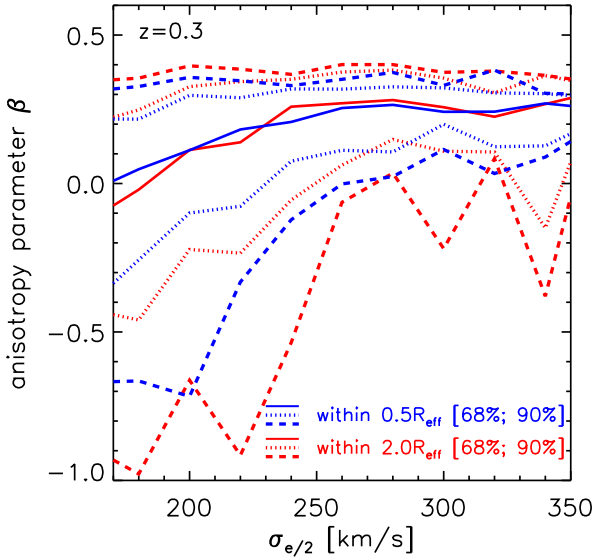}
  \caption{The anisotropy parameter $\beta$ as a function of the
    central stellar velocity dispersion $\sigma_{e/2}$, measured for
    the selected early-type galaxies at $z=0.3$. The blue and red
    curves present the distributions of $\beta(r\leqslant 0.5\,R_{\rm
      eff})$ and $\beta(r\leqslant 2.0\,R_{\rm eff})$,
    respectively. The solid, dotted and dashed lines indicate the
    medians, the 68\% and 90\% boundaries of the distributions,
    respectively. }
  \label{fig:betadependence}
\end{figure}

For each simulated galaxy, measurements of $\beta$ were made (for
stellar particles) within 3-D radii of $0.5\,R_{\rm eff}$ and
$2.0\,R_{\rm eff}$ from the centres of galaxies.
Fig.\,\ref{fig:betadependence} shows the dependence of $\beta$ on
$\sigma_{e/2}$ for the selected galaxies at $z=0.3$. The blue and red
curves present the distributions of $\beta(r\leqslant 0.5\,R_{\rm
  eff})$ and $\beta(r\leqslant 2.0\,R_{\rm eff})$, respectively. The
solid and dashed lines indicate the medians and the 90\% percentiles,
respectively. The stellar orbits of more massive galaxies tend to be
more radially anisotropic than those of their lower-mass
counterparts. Observationally, Koopmans et al. (2009) applied two
independent techniques to measure the logarithmic density slopes for
SLACS early-type galaxies\footnote{The majority of the SLACS galaxies
  are located at lower redshifts ($z<0.3$) and have $\sigma_{e/2}$
  peaks around $\sim250\pm40\kms$.}. The combination of the two
measurements provided a (weak) constraint on the orbital anisotropies,
$\langle\beta\rangle=0.45\pm0.25$, consistent with the values we
measured for their counterparts in the Illustris simulation.

\begin{figure}
  \centering
  \includegraphics[width=8cm]{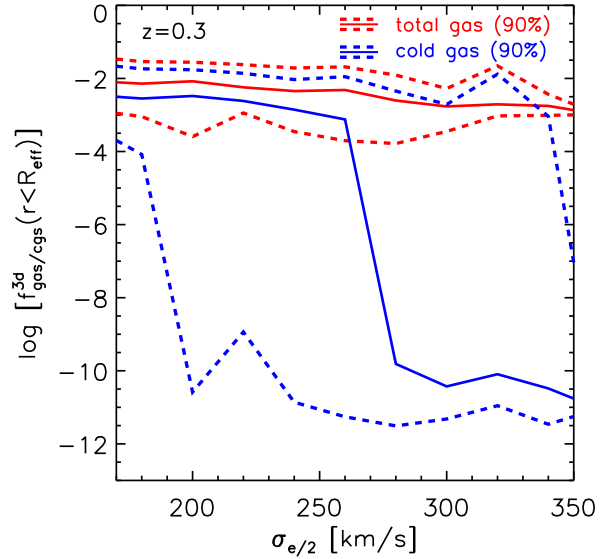}
  \caption{
    The central (cold) gas fractions versus $\sigma_{e/2}$ for the
    early-type galaxy sample at $z=0.3$. The median and 90\%
    boundaries of the distribution are given by the solid and the
    dashed lines, respectively. The blue and red curves show the
    distributions of the cold (HI) and of the total gas fractions
    within a 3-D radius of $\Reff$, respectively. }
  \label{fig:ColourGasdependence}
\end{figure}

It is interesting to note that the different behaviour of the average
$\beta$ in low- and high-mass galaxies is possibly related to the
recent star formation histories. To demonstrate this, we consider in
Fig.\,\ref{fig:ColourGasdependence} the central (cold) gas fractions
versus $\sigma_{e/2}$ for the same galaxy sample at $z=0.3$. The
median and 90\% boundaries of the distribution are given by the solid
and dashed lines, respectively. The blue and red curves show the
distributions of the cold (HI) and total gas fractions, respectively,
measured within a 3-D radius of $\Reff$ from the galaxy centres. We
can see that more massive galaxies that tend to have higher radial
anisotropies also contain less cold gas in their central regions,
while the less massive galaxies with higher tangential velocity
contributions have on average higher central gas fractions. The former
were also seen to have Johnson $B-V$ colours redder than the
latter. We also note that galaxies at higher redshifts are markedly
bluer and contain higher fractions of central cold gas than their
lower redshift counterparts.

These correlations provide a consistent picture. As the cold gas is
channelled down to the centre, star-formation activity preferably
happens on tangential orbits as a consequence of gas accretion and
rotational support. When the system (passively) evolves, more radial
anisotropies emerge.  In this case, one would expect that the stellar
orbits at higher redshifts are more tangentially dominated than their
lower-redshift counterparts. 

\begin{figure}
\centering
  \includegraphics[width=8cm]{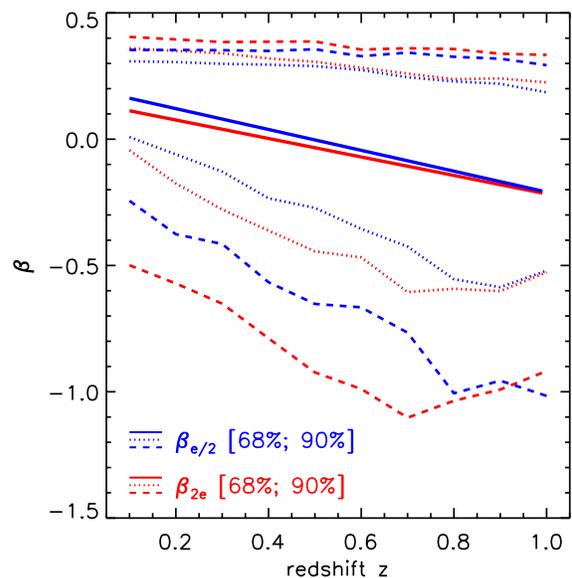}
  \caption{The redshift evolution of the anisotropy parameter $\beta$
    measured for the early-type galaxy samples at $z=0.1-1.0$ from the
    simulation. The blue and red curves represent the distributions of
    $\beta(\leqslant 0.5\,R_{\rm eff})$ and $\beta(\leqslant
    2.0\,R_{\rm eff})$, respectively. The solid, dotted and dashed
    lines indicate the best linear fit to the data, the 68\% and 90\%
    boundaries of either distribution, respectively. Linear regression
    resulted in $\partial \beta(r\leqslant 0.5\,R_{\rm eff}) /
    \partial z = -0.41 \pm 0.02 $ with a linear correlation
    coefficient $r=-0.33$ and $\partial \beta(r\leqslant 2.0\,R_{\rm
      eff}) / \partial z = -0.37 \pm 0.02 $ with $r=-0.26$. }
  \label{fig:betaevolution}
\end{figure}

Indeed, this can be clearly seen in Fig.\,\ref{fig:betaevolution},
which shows the redshift evolution of $\beta$ since $z=1.0$. In order
to quantify this redshift evolution, we randomly assigned a galaxy
from a given snapshot at $z_0$ with a redshift of $z=z_0\pm\Delta
z/2$, where $\Delta z\leqslant 0.05$. We then fit $\beta$ versus $z$
using a linear regression approach, which resulted in $\partial
\beta(r\leqslant 0.5\,R_{\rm eff}) / \partial z =-0.41 \pm 0.02 $ with
a linear correlation coefficient $r=-0.33$ and $\partial
\beta(r\leqslant 2.0\,R_{\rm eff}) / \partial z = -0.37 \pm 0.02 $
with $r=-0.26$. We verify that changing the range of $\Delta z$ from
0.01 to 0.1 makes no difference in the linear regression results.

We mention in passing that the observed correlation between $\beta$
and $\sigma_{e/2}$ could strongly depend on the details of the adopted
galactic wind and AGN feedback models, which, as shown in
\citet{Genel2015}, efficiently affect the gas distribution and
determine the stellar angular momentum and thus orbital
anisotropies. Observational constraints on the distribution of $\beta$
are crucial in establishing the validity of various feedback models.

\section{Projected central dark matter fractions}

The projected central dark matter fraction $f_{\rm dm}$ of observed
galaxies has often been constrained through combined measurements of the
stellar and the total masses. A galaxy's stellar mass can be obtained
using the SPS method applied to multi-band photometric data, provided
that the IMF is independently constrained. Constraints on the total
mass may come from stellar kinematics and/or strong lensing
measurements. As already discussed in Sect.\,4, this is often
complicated due to the lack of knowledge about stellar orbital
anisotropies. In addition, it also needs to assume parameterized
density profiles either of the total matter distribution (e.g.,
Koopmans et al. 2006; Auger et al. 2010b) or of the individual dark
and luminous components (e.g., Cappellari et al. 2013; Sonnenfeld et
al. 2015; Li et al. 2016).

In this section, we first compare the {\it projected} central dark
matter fraction of the early-type Illustris galaxies to the ones
derived for the SLACS (Auger et al. 2010b) and SL2S (Sonnenfeld et
al. 2015) galaxies. To this end, we use the quantity $f_{\rm
  dm}(R\leqslant 5\kpc)$, which is defined as the projected dark
matter fraction within a fixed aperture of $5\kpc$. The choice of a
fixed aperture for comparison purposes instead of, say, $\Reff$, is to
eliminate systematic differences due to possible sample bias.

\begin{figure}
\centering
\includegraphics[width=8cm]{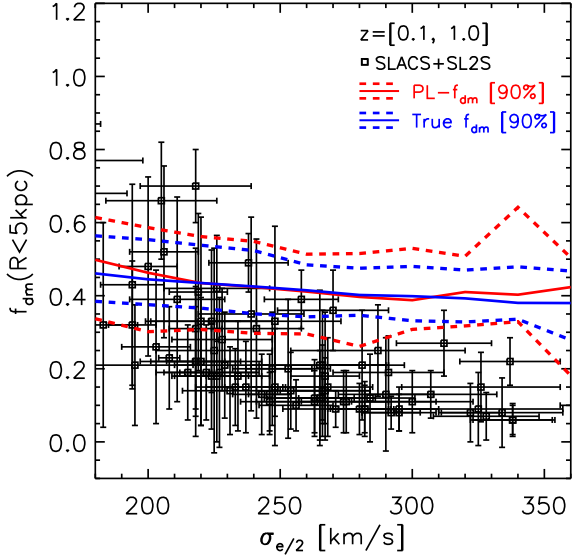}
\caption{The projected central dark matter fraction $f_{\rm
    dm}(R\leqslant 5\kpc)$ as a function of the central velocity
  dispersion $\sigma_{e/2}$.  The blue curves present the distribution
  of the directly measured $f_{\rm dm}(R\leqslant 5\kpc)$ for the
  selected early-type galaxies at $z=0.1-1.0$. The red curves indicate
  the distribution of the measurements for the same galaxies but
  obtained by assuming the total matter density distributions of
  galaxies to be power laws. The solid and the dashed lines give the
  median and the 90\% boundaries of the simulation distributions. The
  black squares show measurements for the SLACS and SL2S galaxy
  samples, where the error bars indicate 1\,$\sigma$ error of the
  data. For the observed galaxy sample, a dark halo component was
  modelled by a NFW profile, and a de-projected best-fitting de
  Vaucouleur distribution was adopted to model the stellar
  distribution. The sum of the two was then used to fit both strong
  lensing and kinematics data (see Sonnenfeld et al. 2015 for
  details). }
\label{fig:Fdm5kpcdependence}
\end{figure}

Fig.\,\ref{fig:Fdm5kpcdependence} shows $f_{\rm dm}(R\leqslant 5\kpc)$
as a function of the central velocity dispersion $\sigma_{e/2}$.  The
blue curves give the distribution of the directly measured $f_{\rm
  dm}(R\leqslant 5\kpc)$ for the selected early-type galaxies at
$z=0.1-1.0$, whereas the red curves indicate the distribution of
measurements for the same galaxies but obtained by modelling the total
matter density distributions with power-law profiles (see Sect.\,6.1.3
for details). The solid and the dashed lines give the median and the
90\% boundaries of the simulation distributions. The black squares
indicate measurements for the SLACS and SL2S galaxy samples, where the
error bars show 1\,$\sigma$ error of the data. Specifically, a dark
halo component was modelled by a NFW profile, and a de-projected
best-fitting de Vaucouleur distribution was adopted to model the
stellar distribution. The sum of the two was then used to fit both
strong lensing and kinematics data (Sonnenfeld et al. 2015).


As can be seen from the figure, the distribution of the measurements
derived under the power-law profile assumption (in red) has larger
scatter than the true distribution (in blue). But both distributions
indicate that dark matter on average contributes with $40\%-50\%$ to
the (projected) total matter distributions in centres of early-type
Illustris galaxies. This fraction is higher than suggested by the
observational results. Such tension poses a potential challenge to the
stellar formation and feedback models adopted by the simulation.

Observations also suggest that the projected central dark matter
fraction within the effective radius $f_{\rm dm}(R\leqslant R_{\rm
  eff})$ is mass-dependent: the more massive galaxies are, the larger
is their dark matter fraction. A noticeable positive correlation was
also found between $f_{\rm dm}(R\leqslant R_{\rm eff})$ and $\Reff$,
albeit with large scatter (e.g., \citealt{Tortora2009, Napolitano2010,
  HumphreyBuote2010, GF2010FPDMF}; Auger et al. 2010b;
\citealt{Shu2015SLACSXII}). Using a significantly larger statistical
sample of early-type galaxies from the simulation, we investigated
such dependences.

\begin{figure}
\centering
\includegraphics[width=8cm]{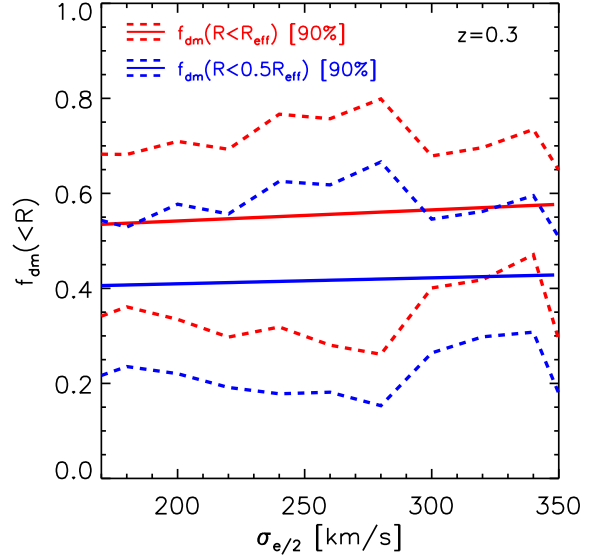}
\includegraphics[width=8cm]{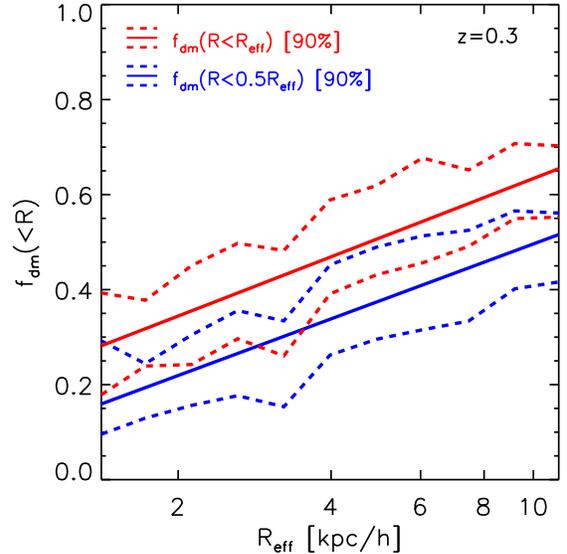}
\caption{The projected central dark matter fraction $f_{\rm
    dm}(\leqslant R)$ as a function of the central velocity dispersion
  $\sigma_{e/2}$ (top panel) and of the effective radius $R_{\rm eff}$
  (bottom panel) for the selected early-type galaxy sample at
  $z=0.3$. Blue and red represent the fractions measured within a
  projected radius of $0.5\,R_{\rm eff}$ and $R_{\rm eff}$,
  respectively. The solid and the dashed lines give the best linear
  fit to the data and the 90\% boundaries of the simulation
  distribution, respectively. Linear regression resulted in $\partial
  f_{\rm dm}(R\leqslant 0.5\,R_{\rm eff}) / \partial \sigma_{e/2} =
  0.0001 \pm 0.0001 $ with a linear correlation coefficient $r=0.05$;
  $\partial f_{\rm dm}(R\leqslant R_{\rm eff}) / \partial \sigma_{e/2}
  = 0.0002 \pm 0.0001 $ with $r=0.09$ and $\partial f_{\rm
    dm}(R\leqslant 0.5\,R_{\rm eff}) / \partial \lg R_{\rm eff} = 0.40
  \pm 0.01 $ with $r=0.89$; $\partial f_{\rm dm}(R\leqslant R_{\rm
    eff}) / \partial \lg R_{\rm eff} = 0.41 \pm 0.01 $ with $r=0.90$.}
\label{fig:Fdmep5dependence}
\end{figure}

Fig.\,\ref{fig:Fdmep5dependence} presents $f_{\rm dm}(\leqslant R)$
versus $\sigma_{e/2}$ (top panel) and versus $R_{\rm eff}$ (bottom
panel) for the early-type galaxy sample selected at $z=0.3$. Blue and
red represent the fractions measured within a radius of $0.5\,R_{\rm
  eff}$ and $R_{\rm eff}$, respectively. The solid and the dashed
lines give the best linear fit to the data and the 90\% boundaries of
the simulation distribution. Linear regression resulted in $\partial
f_{\rm dm}(R\leqslant 0.5\,R_{\rm eff}) / \partial \sigma_{e/2} =
0.0001 \pm 0.0001 $ with a linear correlation coefficient $r=0.05$;
$\partial f_{\rm dm}(R\leqslant R_{\rm eff}) / \partial \sigma_{e/2} =
0.0002 \pm 0.0001 $ with $r=0.09$ and $\partial f_{\rm dm}(R\leqslant
0.5\,R_{\rm eff}) / \partial \lg \Reff = 0.40 \pm 0.01 $ with
$r=0.89$; $\partial f_{\rm dm}(R\leqslant R_{\rm eff}) / \partial \lg
\Reff = 0.41 \pm 0.01 $ with $r=0.90$.


As can be seen, for the early-type galaxy sample, the dependence on
the stellar velocity dispersions is much weaker than on galaxy
sizes. The latter shows a tight and clear positive correlation between
the two quantities. Similar dependences were also found by
\citet{Remus2016}, where early-type galaxies selected from the
Magneticum Pathfinder Simulations (\citealt{Dolag2015MagPathFind})
were studied.

These dependences suggest that the {\it projected} central dark matter
fraction in terms of $f_{\rm dm}(R\leqslant R_{\rm eff})$ [or $f_{\rm
    dm}(R\leqslant 0.5\,R_{\rm eff})$] has a very mild mass dependence
for our early-type galaxy samples. The clear positive correlation
between $f_{\rm dm}(R\leqslant R_{\rm eff})$ [or $f_{\rm
    dm}(R\leqslant 0.5\,R_{\rm eff})$] and $\lg \Reff$ may purely be
an aperture effect: the dark matter fraction drops with decreasing
radius as baryons dominate more and more towards the galactic centre
(also see \citealt{Grillo2010ETGDMF}; and Fig.\,3 of
\citealt{Xu2016IllustrisPL}).

\begin{figure}
\centering
\includegraphics[width=8cm]{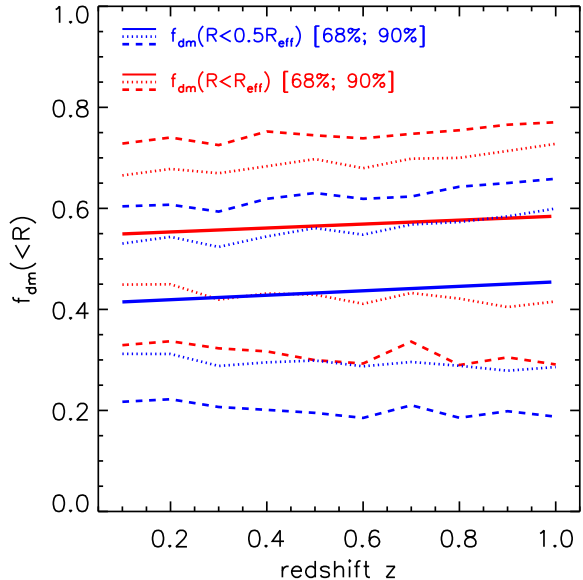}
\caption{The redshift evolution of the central dark matter fractions
  $f_{\rm dm}(R\leqslant 0.5\,R_{\rm eff})$ (blue) and $f_{\rm
    dm}(R\leqslant R_{\rm eff})$ (red) measured for the selected
  early-type galaxy samples at $z=0.1-1.0$. The solid, dotted and
  dashed lines indicate the best linear fit to the data, the 68\% and
  90\% boundaries of either distribution, respectively. For the entire
  sample, both $\partial f_{\rm dm}(R\leqslant 0.5\,R_{\rm eff}) /
  \partial z$ and $\partial f_{\rm dm}(R\leqslant R_{\rm eff}) /
  \partial z$ are equal to $0.04 \pm 0.01$ and both fits have linear
  correlation coefficient $r=0.09$. }
\label{fig:Fdmevolution}
\end{figure}

We also studied the redshift dependence of the projected central dark
matter fraction $f_{\rm dm}(\leqslant R)$ for the selected early-type
galaxy samples in different redshift bins between $z=0.1$ to
$z=1.0$. The result is shown in Fig.\,\ref{fig:Fdmevolution}, where
the solid, dotted and dashed lines indicate the best linear fit to the
data, the 68\% and 90\% boundaries of the distribution,
respectively. Using linear regression approach, we found that both
$\partial f_{\rm dm}(R\leqslant 0.5\,R_{\rm eff}) / \partial z$ and
$\partial f_{\rm dm}(R\leqslant R_{\rm eff}) / \partial z$ are equal
to $0.04 \pm 0.01$ and both fits have linear correlation coefficients
$r=0.09$.

Fair comparisons with observations (or studies using simulations of
the same kind) would require applying identical sample selection
criteria. Dye et al. (2014) reported a similar increasing trend of
$f_{\rm dm}(R\leqslant 0.5\,R_{\rm eff})$ with redshift for a galaxy
sample from the Herschel Astrophysical Terahertz Large Area Survey
(see also Sonnenfeld et al. 2015, Fig.\,6). The result from the
simulation is not in stark contrast with observations. We note that
the early-type galaxy samples at different redshifts were selected
according to their $\sigma_{e/2}$, which also evolve with redshift for
individual galaxies. Therefore, the redshift trends found here hold
for a statistical sample defined as such, but not necessarily for
individual galaxies (also see Remus et al. 2016).

\section{Central matter density profiles}

\begin{figure}
\centering
\includegraphics[width=8cm]{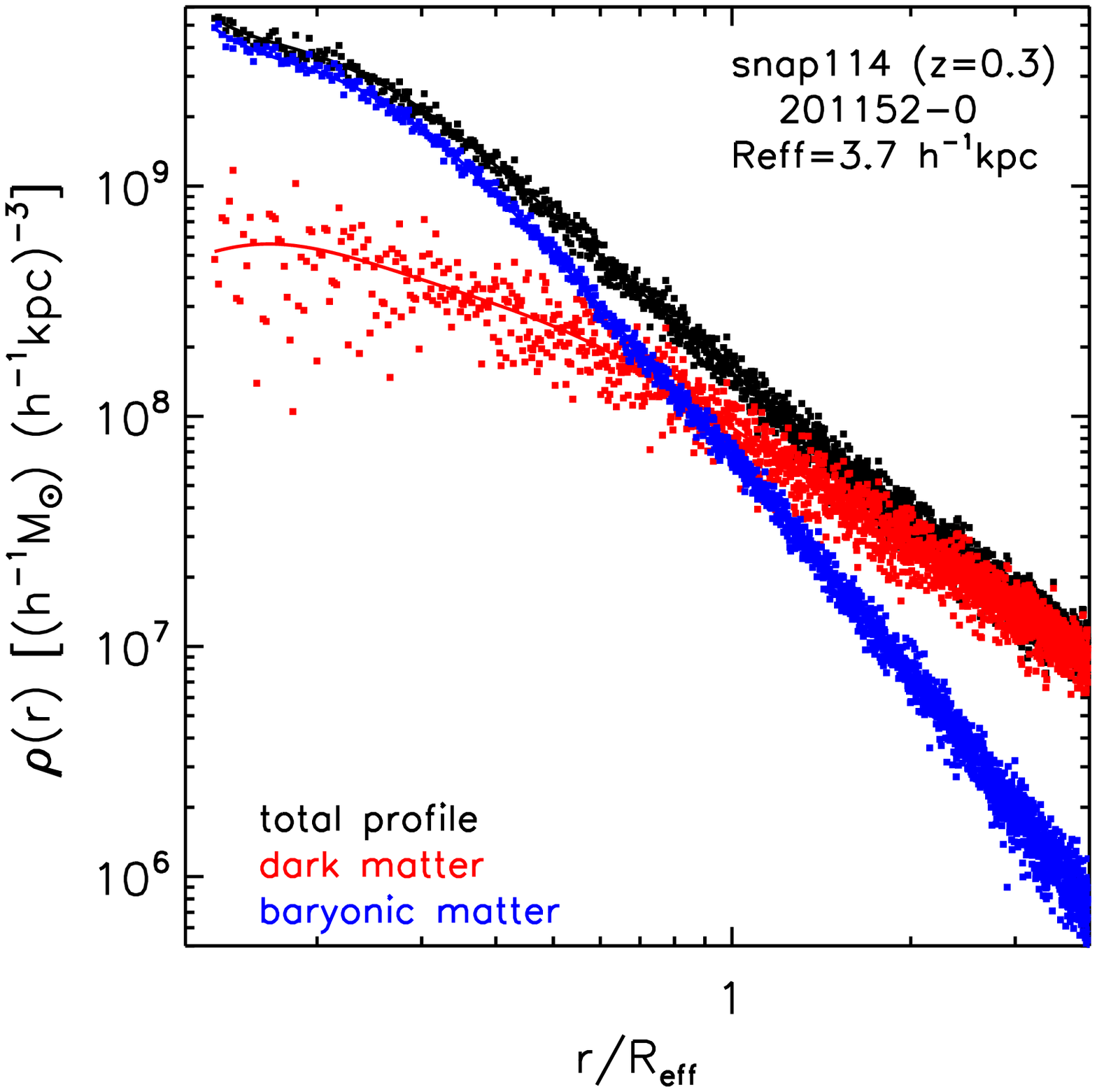}
\caption{Matter density distribution of an early-type galaxy at
  $z=0.3$ (the same galaxy as presented in Fig.\,3). The black, red
  and blue points indicate the density distributions of the total,
  dark matter and baryonic matter, respectively. }
\label{fig:EllDensExample}
\end{figure}

As already shown in Sect.\,5, baryons contribute a large fraction of
the total matter distribution in centres of early-type
galaxies. Fig.\,\ref{fig:EllDensExample} shows the density
distribution of an early-type galaxy at $z=0.3$ (the same galaxy as
presented in Fig.\,3). The black, red and blue symbols indicate the
density distributions of the total, dark matter and baryonic matter,
respectively. The profile of baryons is much steeper than that of dark
matter in the central region. Interestingly, the projected
galactocentric radii where typical strong lensing and stellar
kinematics data are available coincide with the radii where the radial
profiles of dark matter and baryons intercept (e.g., see Fig.\,3 of Xu
et al. 2016). In this radial range (normally $0.5-2.0\,\RE$,
corresponding to the inner few kpc), the slope of the total density
profile depends on both components. Its quantification can be
problematic because the sum of the two components does not necessarily
obey simple global power-law distributions, i.e., no single slope can
describe the overall distribution.

For a simulated galaxy, one can find approximate slope estimators
within given radial ranges. For example, an average slope $\gamma^{\rm
  AV}(r_1,\,r_2)$ of the density profile between two radii $r_1$ and
$r_2$ can be expressed using a power-law interpolation (the
superscript ``AV'' refers to ``average''):
\begin{equation}
\gamma^{\rm
  AV}(r_1,\,r_2)\equiv\frac{\ln[\rho(r_2)/\rho(r_1)]}{\ln(r_1/r_2)}.
\label{eq:slpDirec}
\end{equation}

One can also define $\gamma^{\rm PL}(r_1,\,r_2)$ as the local
logarithmic slope of the power-law profile that best fits the radial
density distribution between $r_1$ and $r_2$ (the superscript ``PL''
refers to ``power-law''). In particular, this definition has been
adopted in many studies on simulated galaxies (e.g.,
\citeauthor{Nipoti2009LD} 2009a; Johansson et al. 2012; Remus et
al. 2013; Li et al. 2016) when comparing to observations.

Another definition is a mass-weighted density slope
$\gamma^{\rm MW}(r)$ (the superscript ``MW'' refers to
``mass-weighted'', see \citealt{Dutton2014}):
\begin{eqnarray}
\gamma^{\rm MW}(r) & \equiv & \frac{1}{M(r)}\int^{r}_0 4\pi x^2
\rho(x)\, \gamma(x) \;{\rm d}x \nonumber \\ & = & 3-\frac{4\pi r^3
  \rho(r)}{M(r)} = 3-\frac{{\rm d}\ln M(r)}{{\rm d}\ln r},
\label{eq:slpMW}
\end{eqnarray}
where the local slope $\gamma(r)$ is given by
\begin{equation}
\gamma(r)\equiv-\frac{{\rm d}\ln\rho(r)}{{\rm d}\ln r},
\label{eq:LocalSlp}
\end{equation}
and the 3-D enclosed total mass $M(r)$ is given by
\begin{equation}
  M(r)\equiv\int^{r}_0 4\pi x^2 \rho(x) \;{\rm d}x.
\label{eq:3DMass}
\end{equation}
Note that for a matter density distribution that follows a perfect
power law, i.e., $\rho(r)\propto r^{-\gamma^{\rm PL}}$,
$\gamma^{\rm MW}(r)=\gamma^{\rm PL}$ at all radii $r$.

Despite different definitions, all the above-mentioned slope
estimators quantify some intrinsic matter density distribution to
first-order. The interpretation of the resulting measurements under
these definitions is straightforward and model-independent. However,
they cannot be applied to observed galaxies, unlike the brightness
distribution, which can be directly measured as long as the galaxy is
spatially resolved. The (total) matter density distribution can only
be determined indirectly by dynamical methods, for example
gravitational lensing or stellar kinematics, and through fitting
parameterized models based on certain assumptions.

For galaxies at lower redshifts where 2-D kinematical data (e.g., the
integral-field spectroscopic data) are available, one can implement
sophisticated dynamical methods (e.g., \citealt{Barnabe2011III};
Cappellari et al. 2015). Assuming parameterized density profiles
within radial ranges (e.g., from $\sim0.1\Reff$ to a few $\Reff$) for
data fitting purposes, the method allows simultaneous fitting to the
matter distribution as well as the stellar orbital anisotropies. In
particular, Li et al. (2016) investigated the validity of such
techniques using the Illustris simulation. They found that although
the orbital anisotropies cannot be accurately recovered and
degeneracies exist between the dark matter and stellar components, the
total mass distributions and their density slopes $\gamma_{\rm
  tot}^{\rm PL}$ within $2.5\,\Reff$ are well recovered with 10\%
accuracy.

For galaxies at higher redshifts where only single-aperture
kinematical data are available, simple approaches that use multiple
mass measurements at different radii to make predictions about matter
density slopes can be adopted. For example, in the SLACS (Auger et
al. 2010b) and SL2S (Sonnenfeld et al. 2013, 2015) surveys, the
central density slopes were derived for the observed lensing galaxies
assuming spherical symmetry, power-law profiles, and isotropic orbital
distributions ($\beta=0$). The derived slopes could therefore suffer
from more systematic biases than those using 2-D kinematics methods.

In order to make fair comparisons, one should apply the observational
estimators also to the simulated samples. In this work, we adopted a
simple approach along these lines that combines strong lensing and
single-aperture kinematics for simulated early-type galaxies. In
Sect.\,6.1, we present the derived slopes, compare them with
observational results, and discuss two associated major systematic
effects. In Sect.\,6.2, different slope estimators (as presented
above) of the total matter density distributions are compared. In
Sect.\,6.3, we present the inner density slopes of the dark matter and
stellar distributions of the early-type Illustris galaxies, and in
Sect.\,6.4, their cosmic evolution is discussed.

\subsection{Total density slopes from combining strong lensing and
  single-aperture kinematics}

For making fair comparisons of the total power-law slopes between the
simulation and, in particular, the SLACS and SL2S survey results, we
adopted a similar practice as used in these studies.  Here we first
briefly describe the main features of the method (details can be found
in e.g., Koopmans et al. 2006; Auger et al. 2010b; Sonnenfeld et
al. 2013).

The total matter distribution of a galaxy is assumed to be spherically
symmetric, with a radial profile that follows a power law, i.e.,
$\rho(r)\propto r^{-\gamma^\prime}$. The stellar distribution is
obtained by de-projecting the Sersic profile that best fits the
surface brightness distribution (see Sect.\,2.3). This latter
component is assumed to be a massless tracer sitting in the
gravitational potential of the former. The stellar orbital anisotropy
$\beta$, defined by Eq.\,(\ref{eq:beta}), is assumed to be zero. Two
``measurements'' are made: (1) the mass $M_{\rm E}$ projected within
the Einstein radius $\RE$ (the strong lensing constraint); (2) the
line-of-sight stellar velocity dispersion measured within a circular
aperture of radius $1.5{\arcsec}$, same as used for the SLACS data
(the stellar kinematics constraint). For any given power-law slope
$\gamma^{\prime}$, measurement (1) constrains the normalization of the
total matter density profile, with which the radial distribution of
the stellar velocity dispersion can be derived by solving the
spherical Jeans equation. The slope $\gamma_0^{\rm LD}$ (where the
superscript ``LD'' stands for ``strong lensing and dynamics'' and the
subscript 0 refers to the assumption of $\beta=0$) that results in the
best fit to measurement (2) is then taken as the power-law slope of
the total density profile. We searched for $\gamma_0^{\rm LD}$ within
[1.2,\,2.8] with a step of 0.02. This leads to the differences between
the best-fitting and the ``observed'' velocity dispersion in most
cases smaller than 1 km/s, and in all cases no larger than 2 km/s,
much smaller than the observational uncertainty (2\%-10\%).

\subsubsection{Comparisons to observations}

\begin{figure}
\centering
\includegraphics[width=8cm]{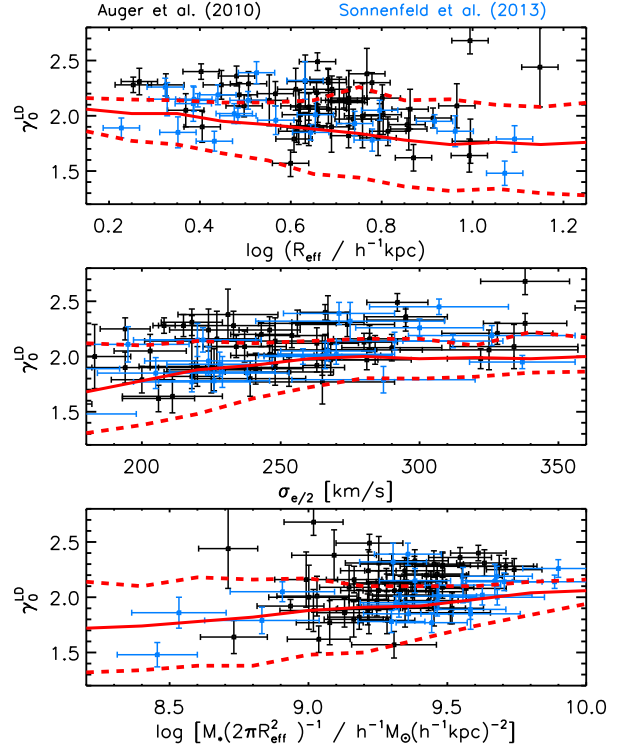}
\caption{The total matter density slope $\gamma_0^{\rm LD}$ (see
  Sect.\,6.2) as a function of $\Reff$ (top panel), $\sigma_{e/2}$
  (middle panel), and $\Sigma_{*}\equiv M_{*}/(2\pi \Reff^2)$ (bottom
  panel). The solid and the dashed lines indicate the median and the
  90\% boundaries of the distributions from the selected early-type
  galaxies at $z=0.1-1.0$. The blue and black squares show the
  observational results for the SLACS (Auger et al. 2010b) and SL2S
  (Sonnenfeld et al. 2013) samples, respectively, where the error bars
  indicate 1\,$\sigma$ error of the data.}
\label{fig:JEslpvsObs}
\end{figure}

Fig.\,\ref{fig:JEslpvsObs} presents the total matter density slope
$\gamma_0^{\rm LD}$ as a function of $\Reff$ (top panel),
$\sigma_{e/2}$ (middle panel), and
$\Sigma_{*}\equiv M_{*}/(2\pi \Reff^2)$ (bottom panel). The solid and
dashed lines indicate the median and the 90\% boundaries of the
distributions for the selected early-type galaxies between $z=0.1$ and
$z=1.0$. The blue and black squares with their error bars indicate
measurements for the SLACS (Auger et al. 2010b) and SL2S (Sonnenfeld
et al. 2013) galaxies, respectively.

The simulation reproduces the general observational trends and
scatter. The dependencies are noticeable: $\gamma_0^{\rm LD}$ on
average decreases with increasing $\Reff$ but increases with
increasing $\sigma_{e/2}$ and $\Sigma_{*}$. In particular, higher-mass
(higher-$\sigma_{e/2}$) galaxies on average have larger $\gamma_0^{\rm
  LD}$ with smaller scatter in comparison to their lower-mass
(lower-$\sigma_{e/2}$) counterparts. We note that, however, such a
mass dependence is in part due to the correlation between the
measurements of the velocity dispersion and the density slope, as well
as other systematic biases (Sect. 6.1.2 and 6.1.3). In Sect.\,6.2,
mass dependences of different slope estimators are presented and
discussed.

The measurements of $\gamma_0^{\rm LD}$ for the simulated sample are
overall shallower than observations. We note that although the
selected Illustris galaxy sample shares a similar range of
$\sigma_{e/2} \in [160,400]\kms$ with the observed sample, they have
rather different probability distributions. The former increases in
number towards lower $\sigma_{e/2}$ while the latter almost has a peak
around $\sim250\kms$. For this reason, we present quantitative
comparisons made at given $\sigma_{e/2}$ (within a small
$\sigma_{e/2}$ range): for galaxies with
$\sigma_{e/2}\in[220,280]\kms$, $\langle\gamma_0^{\rm
  LD}\rangle=1.92\pm0.18$ (rms) for the Illustris galaxy sample, while
$\langle\gamma_0^{\rm LD}\rangle=2.07\pm0.18$ (rms) for the combined
SLACS and SL2S galaxy sample. For galaxies with
$\sigma_{e/2}\in[250,300]\kms$, $\langle\gamma_0^{\rm
  LD}\rangle=1.98\pm0.14$ (rms) for the Illustris sample, and
$\langle\gamma_0^{\rm LD}\rangle=2.13\pm0.19$ (rms) for the
observation sample.

The notably shallower $\gamma_0^{\rm LD}$ for the simulated galaxy
sample is in fact consistent with lower central velocity dispersion
$\sigma_{e/2}$ and higher central dark matter fraction $f_{\rm dm}$
when comparing to the observational sample. This could indicate
potentially inaccurate modelling of the involved baryonic physics
processes in galaxy centres. We must note, however, the ad-hoc choice
of a unique source redshift that was assigned to each lens redshift
resulted in averagely smaller $\RE/\Reff$ distribution of the
simulated galaxy sample compared to the observation. As a mild
increase of $\gamma_0^{\rm LD}$ with increasing $\RE/\Reff$ (i.e.,
steeper slopes at larger radii) was observed, such a systematic
difference in $\RE/\Reff$ would also cause a lower average of
$\gamma_0^{\rm LD}$ for the simulation sample.

\subsubsection{Biases from the isotropic orbital assumption}

\begin{figure}
\centering
\includegraphics[width=8cm]{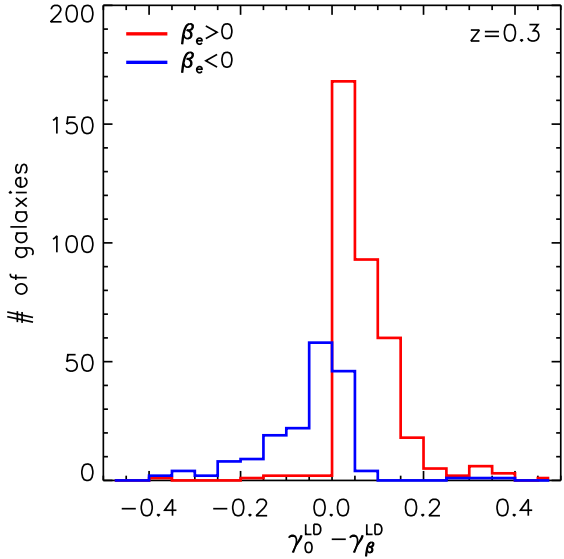}
\caption{Histograms of $\gamma_0^{\rm LD}-\gamma_\beta^{\rm LD}$ for
  the selected early-type galaxy sample at $z=0.3$. Red and blue lines
  indicate galaxies that have $\beta>0$ and $\beta<0$ (measured within
  $\Reff$), respectively.}
\label{fig:JEslpBeta}
\end{figure}

Combining strong lensing and single-aperture stellar kinematics to
derive the total density slopes, $\beta=0$ is commonly assumed due to
a lack of sufficient observational constraints.  To see the effect
of a non-zero anisotropy parameter $\beta$, we also calculated the
total-density power-law slopes $\gamma_\beta^{\rm LD}$ under the same
assumptions as before, but using the true $\beta$ measured for the
simulated galaxies.  Fig.\,\ref{fig:JEslpBeta} shows the histograms of
$\gamma_0^{\rm LD}-\gamma_\beta^{\rm LD}$ for the selected early-type
galaxy sample at $z=0.3$. Red and blue lines indicate galaxies that
have $\beta>0$ and $\beta<0$ (measured within $\Reff$),
respectively. The former distribution (for radially anisotropic
galaxies) peaks at $\gamma_0^{\rm LD}-\gamma_\beta^{\rm LD}>0$, while
the latter (for tangentially anisotropic galaxies) peaks at
$\gamma_0^{\rm LD}-\gamma_\beta^{\rm LD}<0$. This demonstrates that
the true slopes of radially anisotropic systems tend to be
overestimated assuming $\beta=0$, while those of tangential ones tend
to be underestimated (see also Koopmans 2006; 2009).

It is worth noting that, as shown in Fig.~7 and Fig.~9, $\beta$
markedly depends on stellar velocity dispersion and also evolves with
redshift. This leads to possible systematic biases of the
observational measurements. The total density slopes of higher-
(lower-) mass galaxies or galaxies at lower- (higher-) redshifts are
likely to be overestimated (underestimated) by $\gamma_0^{\rm LD}$ when
assuming $\beta=0$.

\subsubsection{Robustness of the power-law assumption}

As shown by Xu et al. (2016), true profiles of realistic galaxies
deviate from power-law distributions. In order to validate the
robustness of the applied power-law assumption despite this fact, we
first defined a curvature parameter for the 3-D enclosed mass
distribution $M(r)$ (defined by Eq.\,(\ref{eq:3DMass})) between $r_1$
and $r_2$:
\begin{equation}
\xi_{M_{\rm 3d}}(r_1,r_2)\equiv\frac{M(\sqrt{r_1
    r_2})}{\sqrt{M(r_1)M(r_2)}}.
\label{eq:xiM}
\end{equation}
This curvature parameter $\xi_{M_{\rm 3d}}(r_1,r_2)$ directly
quantifies the closeness of $M(r)$, and thus $\rho(r)$, to a power-law
distribution between $r_1$ and $r_2$. If the deviation of
$\xi_{M_{\rm 3d}}(r_1,r_2)$ from unity is large, then the power-law
approximation is poor. We set $[r_1,~r_2]$ to be $[0.5,~2.0]\,\RE$,
which is the most relevant radial range for approaches that combine
strong lensing and stellar kinematics.

\begin{figure*}
\centering
\includegraphics[width=5.84cm]{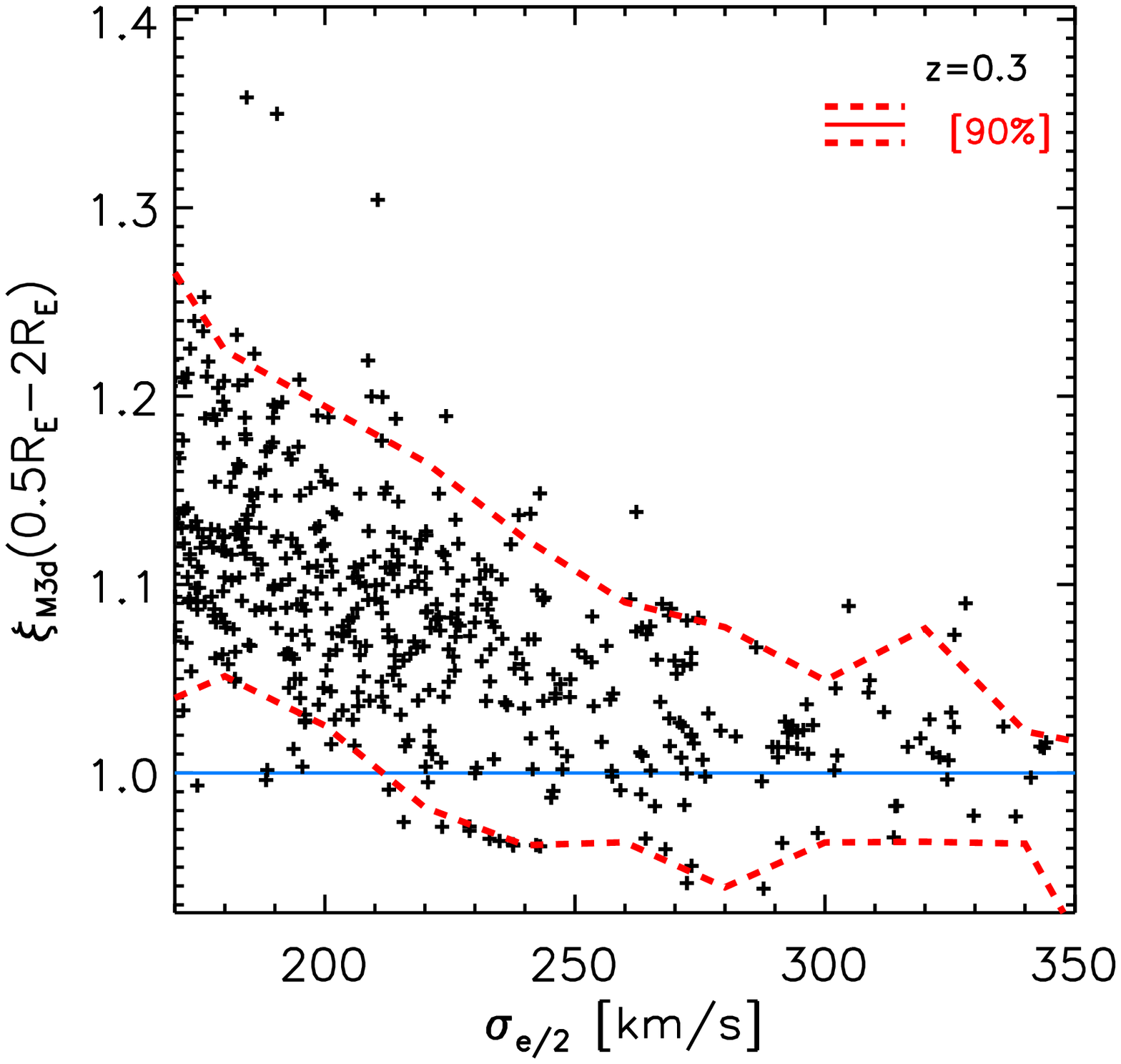}
\includegraphics[width=5.84cm]{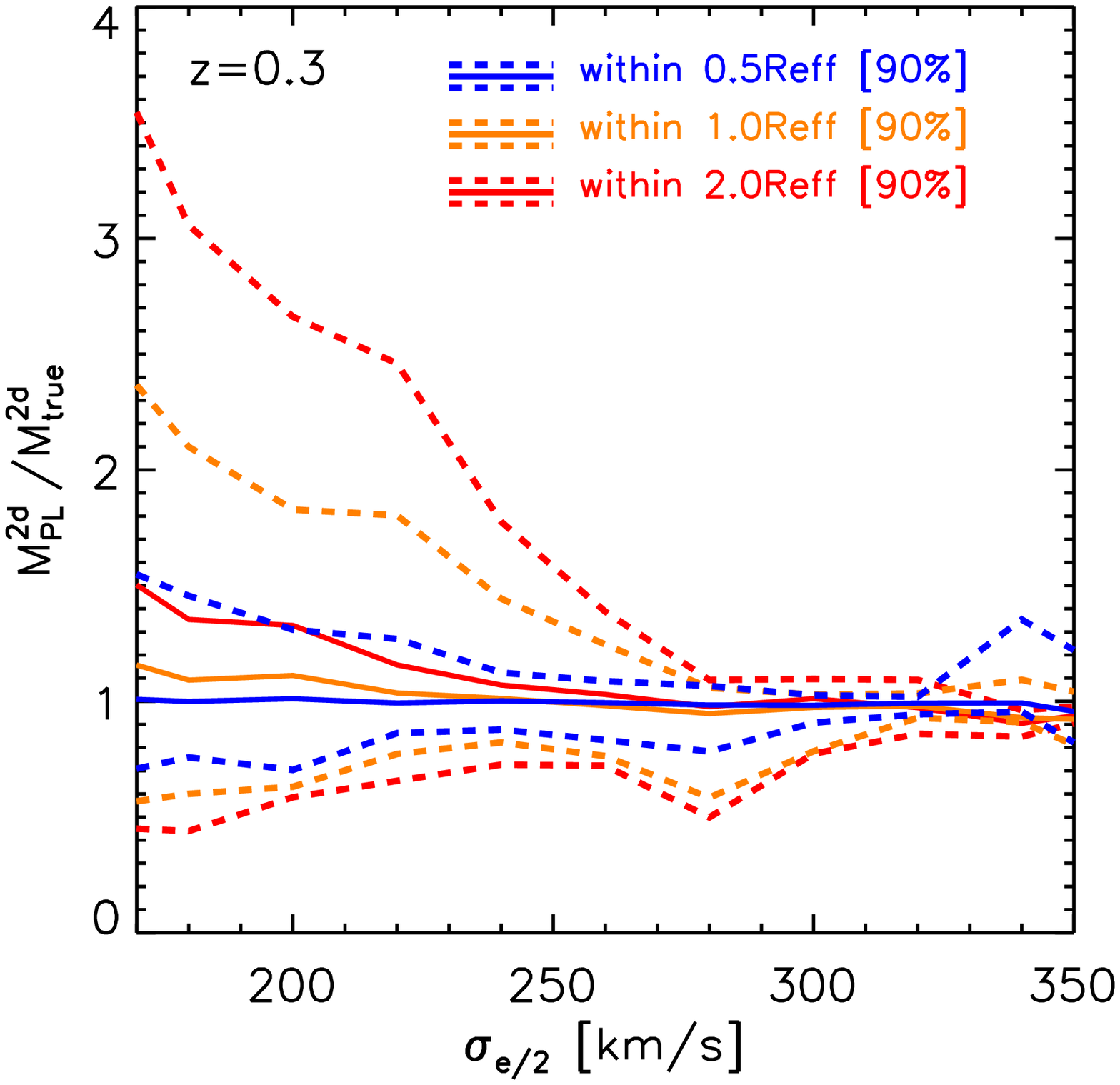}
\includegraphics[width=5.84cm]{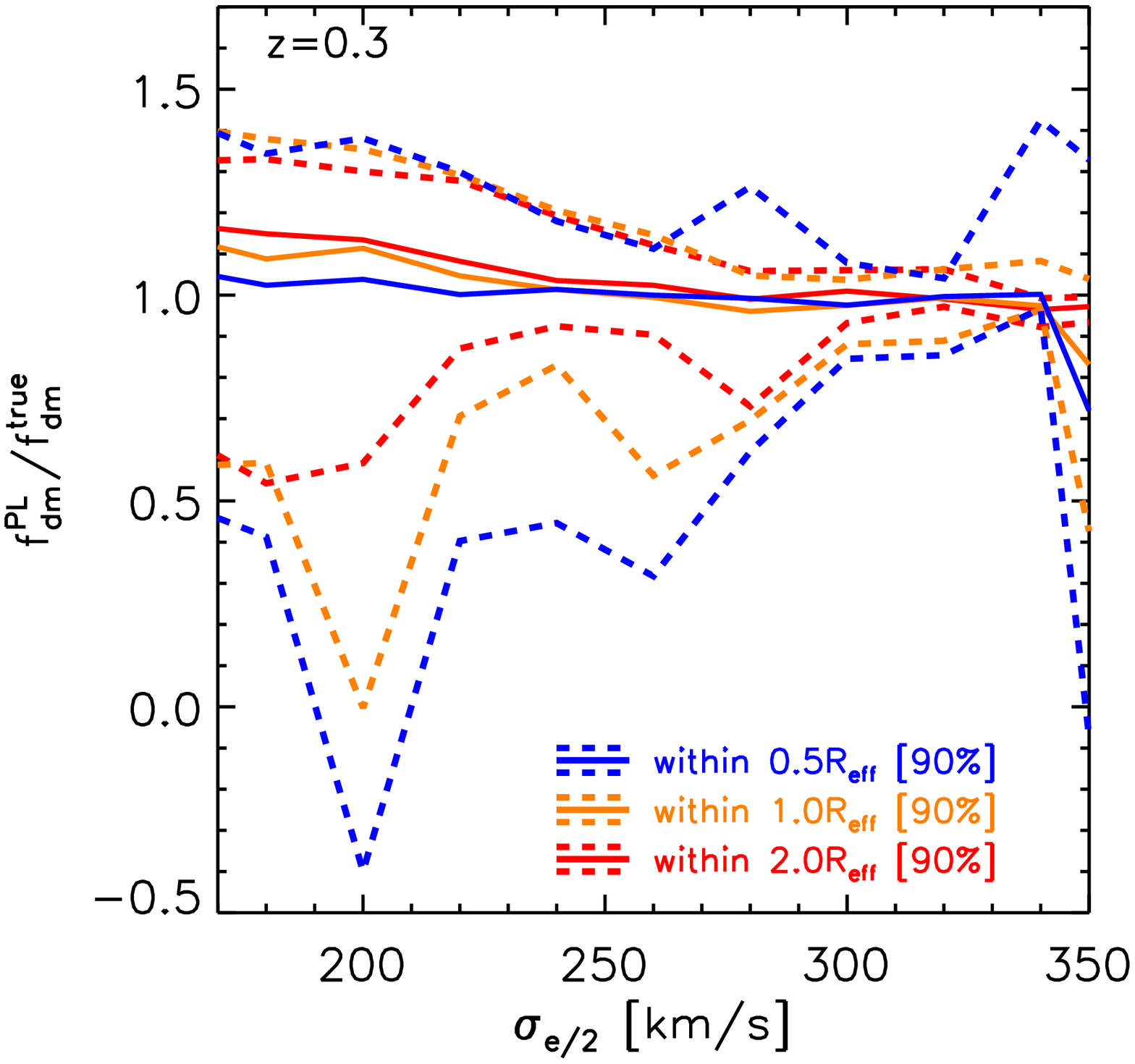}
\caption{The curvature parameter $\xi_{M_{\rm
      3d}}(0.5\,\RE,~2.0\,\RE)$ (left panel), the enclosed mass ratio
  $M_{\rm 2d}^{\rm PL}/M_{\rm 2d}^{\rm true}$ (middle panel) and the
  projected dark matter fraction $f_{\rm dm}^{\rm PL}/f_{\rm dm}^{\rm
    true}$ (right panel), as a function of the central stellar
  velocity dispersion $\sigma_{e/2}$ for the early-type galaxy sample
  at $z=0.3$. In all three panels, the solid and the dashed lines
  indicate the median and the 90\% boundaries of the
  distributions. The blue, red and orange curves present measurements
  made within different aperture sizes as specified in the
  panels. Note that the median values of the ratios measured within
  $0.5\,\Reff$ are close to 1.0, this is because the simulation sample
  has on average $\langle\RE/\Reff\rangle\sim0.5$, and by construction
  $M_{\rm 2d}^{\rm PL}(R\leqslant\RE)=M_{\rm 2d}^{\rm
    true}(R\leqslant\RE)$. In addition, a small fraction of galaxies
  have negative $f_{\rm dm}^{\rm PL}$ (as can be seen from the right
  panel), which originates in the power-law assumption. For these
  galaxies, the power-law model that best fits both the strong lensing
  and kinematical constraints predicts total masses smaller than the
  stellar masses within the studied aperture.}
\label{fig:XiM}
\end{figure*}

We calculated $\xi_{M_{\rm 3d}}(0.5\,\RE,~2.0\,\RE)$ for all the
selected early-type galaxies from the simulation. Its dependence on
$\sigma_{e/2}$ is presented in the left-most panel of
Fig.\,\ref{fig:XiM}. The crosses indicate the simulation results at
$z=0.3$, the solid and the dashed lines indicate the median and the
90\% boundaries of this distribution. In fact, both $\xi_{M_{\rm
    3d}}(0.5\,\RE,~2.0\,\RE)$ and $\xi_{M_{\rm
    3d}}(0.5\,\Reff,~2.0\,\Reff)$ show decreasing trends towards
larger systems. In this case, the power-law approximation becomes
worse for galaxies with central velocity dispersion $\sigma_{e/2}\la
250 \kms$. As shown by Xu et al. (2016), this leads to significantly
biased determinations of the Hubble constant $H_0$ when power-law mass
models are used to describe the lensing galaxies as they artificially
break the so-called ``mass sheet degeneracy'' (e.g.,
\citealt{Falco1985MST, SS13}).

The breakdown of the power-law approximation can also lead to biased
estimates of other derived quantities, e.g., the 2-D enclosed mass
distribution $M_{\rm 2d}(\leqslant R)$ and the projected dark matter
fraction $f_{\rm dm}(\leqslant R)$, if they were derived assuming a
power-law total density profile with slope $\gamma_0^{\rm LD}$ (for
$f_{\rm dm}$, the dark matter mass is obtained by subtracting the
observationally-constrained stellar mass from the total mass projected
within a given aperture). In order to identify potential biases in
these quantities, we further calculate two ratios: (1) between $M_{\rm
  2d}^{\rm PL}(\leqslant R)$ that is derived under the power-law
assumption and the true mass distribution $M_{\rm 2d}^{\rm
  true}(\leqslant R)$; and (2) between $f_{\rm dm}^{\rm PL}(\leqslant
R)$ that is derived under the power-law assumption and the true
fraction $f_{\rm dm}^{\rm true}(\leqslant R)$. The ``true'' quantities
are directly measured for simulated galaxies. Note that by
construction, $M_{\rm 2d}^{\rm PL}(\leqslant\RE)=M_{\rm 2d}^{\rm
  true}(\leqslant\RE)$.  However, this is not necessarily the case at
other radii, unless the true distribution is indeed a power law.

We measured the two ratios within different aperture radii for the
selected early-type Illustris galaxies. In Fig.\,\ref{fig:XiM}, the
middle and right-most panels show the enclosed mass ratio $M_{\rm
  2d}^{\rm PL}/M_{\rm 2d}^{\rm true}$ versus $\sigma_{e/2}$, and the
projected dark matter fraction ratio $f_{\rm dm}^{\rm PL}/f_{\rm
  dm}^{\rm true}$ versus $\sigma_{e/2}$, respectively (for the same
galaxy sample at $z=0.3$). We note that the ratios deviate from unity
due to a combination of two effects: (1) poor approximations of the
power-law models for lower-mass galaxies; and (2) biased (power-law)
slope estimates due to single-aperture kinematics data and the stellar
isotropy assumption. These combined effects could lead to
significantly biased estimates of the enclosed mass and the projected
dark matter fraction, especially when contraints are made for
lower-mass galaxies. The deviations from unity as well as the scatter
also increase with redshift up to $z=1.0$.

\subsection{Different estimators of the total density slopes}

To investigate the variations among different total-slope estimators,
we calculated several sets of slopes for the simulated galaxies within
different radial ranges, namely, the simple power-law dynamical slopes
$\gamma_0^{\rm LD}$ and $\gamma_\beta^{\rm LD}$, the mass-weighted
slopes $\gamma^{\rm MW}(\Reff)$ and $\gamma^{\rm MW}(2.0\,\Reff)$, the
power-law fitted slopes $\gamma^{\rm PL}(0.5\,\Reff,\,\Reff)$ and
$\gamma^{\rm PL}(0.5\,\Reff,\,2.0\,\Reff)$, and finally the average
slopes $\gamma^{\rm AV}(0.5\,\Reff,\,\Reff)$ and $\gamma^{\rm
  AV}(0.5\,\Reff,\,2.0\,\Reff)$.

\begin{table}
\caption{Stellar mass-weighted mean and standard deviation of the
  density slopes for the selected early-type galaxy sample at $z=0.3$}
\begin{minipage}{\textwidth}
\begin{tabular}{l | c | c} \hline
slope estimator & ~~~mean~~~ & ~~~standard deviation~~~ \\\hline
$\gamma_0^{\rm LD}$ & 1.83 & 0.24 \\
$\gamma_\beta^{\rm LD}$ & 1.80 & 0.23 \\
$\gamma^{\rm MW}(\Reff)$ & 1.95 & 0.14 \\
$\gamma^{\rm MW}(2.0\,\Reff)$ & 1.99 & 0.14 \\
$\gamma^{\rm PL}(0.5\,\Reff,\,\Reff)$ & 2.08 & 0.27 \\
$\gamma^{\rm PL}(0.5\,\Reff,\,2.0\,\Reff)$ & 2.07 & 0.26 \\
$\gamma^{\rm AV}(0.5\,\Reff,\,\Reff)$ & 2.08 & 0.26 \\
$\gamma^{\rm AV}(0.5\,\Reff,\,2.0\,\Reff)$ & 2.06 & 0.22 \\ \hline
$\gamma^{\rm PL}_{\rm dm}(0.5\,\Reff,\,\Reff)$ & 1.49 & 0.22 \\
$\gamma^{\rm PL}_{\rm dm}(0.5\,\Reff,\,2.0\,\Reff)$ & 1.58 & 0.17 \\
$\gamma^{\rm PL}_{\rm *}(0.5\,\Reff,\,\Reff)$ & 2.74 & 0.30 \\ 
$\gamma^{\rm PL}_{\rm *}(0.5\,\Reff,\,2.0\,\Reff)$ & 2.87 & 0.26 \\ \hline
\end{tabular}
\end{minipage}
\begin{flushleft}
\end{flushleft}
\label{tab:SlpTable}
\end{table}

\begin{figure}
\centering
\includegraphics[width=8cm]{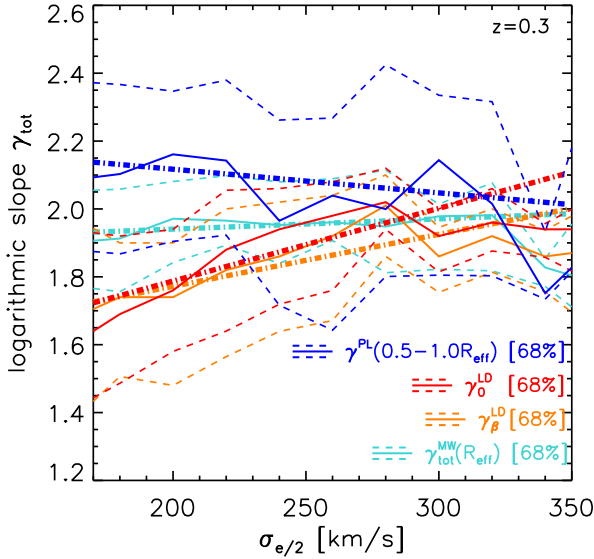}
\caption{Total density slopes measured using different methods, as a
  function of the central stellar velocity dispersion $\sigma_{e/2}$,
  for the selected early-type galaxy sample at $z=0.3$. The red,
  orange, green and blue curves indicate the distributions of
  $\gamma_0^{\rm LD}$, $\gamma_\beta^{\rm LD}$, $\gamma^{\rm
    MW}(\Reff)$ and $\gamma^{\rm PL}(0.5\,\Reff,\,\Reff)$,
  respectively. The solid and the dashed line styles show the median
  and the 68\% boundaries of each distribution, respectively. The
  thicker dashed and dotted lines present the best linear fits to the
  data. Linear regression of $\gamma$ as a function of $\sigma_{e/2}$
  resulted in $\partial \gamma^{\rm LD}_{0} / \partial \sigma_{e/2} =
  0.0021 \pm 0.0002$ with a linear correlation coefficient $r=0.36$;
  $\partial \gamma^{\rm LD}_{\beta} / \partial \sigma_{e/2} = 0.0015
  \pm 0.0002$ with $r=0.27$; $\partial \gamma^{\rm MW}(\Reff) /
  \partial \sigma_{e/2} = 0.0003 \pm 0.0001$ with $r=0.10$ and
  $\partial \gamma^{\rm PL}_{\rm tot}(0.5\,\Reff,\,\Reff) / \partial
  \sigma_{e/2} = -0.0007 \pm 0.0003 $ with $r=-0.11$. }
\label{fig:SlopeTOT}
\end{figure}

Table~\ref{tab:SlpTable} presents the stellar mass-weighted mean and
standard deviation of each above-mentioned slope estimator for the
selected early-type galaxy sample at $z=0.3$.
In Fig.\,\ref{fig:SlopeTOT}, we plot, in particular, four typical
slopes measured around and within $\Reff$, for the same galaxy sample,
as a function of $\sigma_{e/2}$. The red, orange, green and blue
curves indicate the distributions of $\gamma_0^{\rm LD}$,
$\gamma_\beta^{\rm LD}$, $\gamma^{\rm MW}(\Reff)$ and $\gamma^{\rm
  PL}(0.5\,\Reff,\,\Reff)$, respectively. The solid and the dashed
line styles show the median and the 68\% range of each distribution,
respectively. The thicker dashed and dotted lines present the best
linear fits to the data. 

We see that within a large fraction of the velocity range investigated
here, the median slopes are all about isothermal. In particular, the
median and scatter of the intrinsic density slope estimators
$\gamma^{\rm PL}$ and $\gamma^{\rm AV}$ are consistent with the
observational results from 2-D kinematical data (e.g.,
\citealt{Barnabe2011III}; Cappellari et al. 2015).

Interestingly, systematic discrepancies are seen among different slope
estimators. For most galaxies, since the local slope $\gamma(r)$
(defined by Eq.\,(\ref{eq:LocalSlp})) tends to decrease, i.e., the
profile turning flatter, towards smaller $r$ (see also
\citealt{Dutton2014} for more discussion), one finds $\gamma^{\rm
  MW}(\Reff) < \gamma^{\rm PL}(0.5\,\Reff,\,\Reff)$ as expected.
Between the two simple power-law dynamical slopes, the fact that on
average $\gamma^{\rm LD}_0>\gamma_\beta^{\rm LD}$ (within a large span
of $\sigma_{e/2}$) can be explained by the stellar orbital
anisotropies, because the majority of the galaxy sample has radial
anisotropies (as can be seen from Fig.\,\ref{fig:betadependence}). For
galaxies with $\sigma_{e/2}\la 250 \kms$, a moderate disagreement
exists between the simple dynamical slope $\gamma^{\rm LD}$ and the
intrinsic slope $\gamma^{\rm MW}(\Reff)$. This difference increases
with redshift and can be attributed to the poor approximations of the
power-law models for lower-mass galaxies.

Specifically, as can be seen from Fig.\,\ref{fig:XiM}, the curvature
parameter fulfils $\xi_{M_{\rm 3d}}(0.5\,\RE,~2.0\,\RE)>1.0$ for the
majority of lower-mass galaxies. Mathematically, $\xi_{M_{\rm
    3d}}>1.0$ means that, to first-order approximation, $M(r)$ is
concave-upward, lying above the power-law interpolation between the
two radii. The local mass slope $\frac{{\rm d}\ln M(r)}{{\rm d}\ln r}$
therefore decreases with increasing $r$, which according to
Eq.\,(\ref{eq:slpMW}) corresponds to an increase of $\gamma_{\rm
  tot}^{\rm MW}(r)$ within the same radial range. For galaxies with
$\xi_{M_{\rm 3d}}>1.0$ ($\xi_{M_{\rm 3d}}<1.0$), the larger the radius
$r$, the larger (smaller) the slope $\gamma^{\rm MW}(r)$.

As pointed out by Dutton \& Treu (2014), for a perfect power-law
distribution, the density slopes $\gamma^{\rm LD}$ that are derived
under the power-law assumption are essentially the same as
$\gamma^{\rm MW}$. For a realistic galaxy, as the power-law model
could be a poor approximation, this, however, is not necessarily the
case. $\gamma^{\rm LD}$ rather measures some averaged $\gamma_{\rm
  tot}^{\rm MW}(r)$ between $0.5\,\RE$ and $2.0\,\RE$. As $\xi_{M_{\rm
    3d}}>1.0$ holds for the majority of lower-mass galaxies,
$\gamma^{\rm MW}(0.5\,\RE) \la \gamma^{\rm LD} \la
\gamma^{\rm MW}(2.0\,\RE)$ can therefore be in general
expected. For these galaxies, we also found an average radius ratio of
$\langle \Reff/\RE \rangle \sim 2.0$, which eventually led to the
general trend of $\gamma^{\rm LD} \la \gamma^{\rm
  MW}(\Reff)$ seen for galaxies with $\sigma_{e/2}\la 250 \kms$.

It is also interesting to note that different slope estimators have
different mass dependences. As both the stellar anisotropy $\beta$ and
the profile curvature $\xi_{\rm M_{\rm 3d}}$ depend on mass, the
assumptions of isotropic orbits and/or power-law profiles affect the
mass dependence of the lensing and dynamic slope $\gamma^{\rm
  LD}$. Specifically, for the early-type galaxies selected at $z=0.3$,
linear regression of $\gamma^{\rm LD}$ as a function of $\sigma_{e/2}$
resulted in $\partial \gamma^{\rm LD}_{0} / \partial \sigma_{e/2} =
0.0021 \pm 0.0002$ with a linear correlation coefficient $r=0.36$; and
$\partial \gamma^{\rm LD}_{\beta} / \partial \sigma_{e/2} = 0.0015 \pm
0.0002$ with $r=0.27$, indicating positive correlations with
$\sigma_{e/2}$.

On the contrary, the true density slope estimator $\gamma^{\rm PL}$
shows decreasing trends with increasing $\sigma_{e/2}$ (see also,
e.g., Humphrey \& Buote 2010; Remus et al. 2013, 2016). In particular,
for the galaxy sample at $z=0.3$, linear regression of $\gamma^{\rm
  PL}$ as a function of $\sigma_{e/2}$ resulted in $\partial
\gamma^{\rm PL}_{\rm tot}(0.5\,\Reff,\,\Reff) / \partial \sigma_{e/2}
= -0.0007 \pm 0.0003$ with a linear correlation coefficient $r=-0.11$;
and $\partial \gamma^{\rm PL}(0.5\,\Reff,\,2.0\,\Reff) / \partial
\sigma_{e/2} = -0.0010 \pm 0.0002 $ with $r=-0.18$. We also note that
such dependences on $\sigma_{e/2}$ were observed across all the
redshifts studied here.

Relating the anti-correlation between $\gamma^{\rm PL}$ and
$\sigma_{e/2}$ with previously studied $\sigma_{e/2}$ dependences (see
Sect.\,4), we see that lower-mass galaxies have higher central gas
fractions, bluer colours, more tangentially anisotropic orbits and
slightly steeper inner slopes than their more massive counterparts.

\subsection{Inner slope estimates of the dark matter and the stellar distributions}

In centres of galaxies, both dark matter and stellar components shape
the total density profile, which approximately follows an isothermal
distribution. In order to see how much the slopes of individual
components deviate from isothermal, and in particular, how much the
NFW profile is modified due to the presence of baryons, we calculated
$\gamma^{\rm PL}_{\rm dm} (0.5\,\Reff,\,\Reff)$ and $\gamma^{\rm
  PL}_{*} (0.5\,\Reff,\,\Reff)$ for the simulated galaxies. Table
\ref{tab:SlpTable} reports the stellar mass-weighted mean and standard
deviation of the stellar and dark matter slopes for the selected
early-type galaxy sample at $z=0.3$. 

\begin{figure}
\centering
\includegraphics[width=8cm]{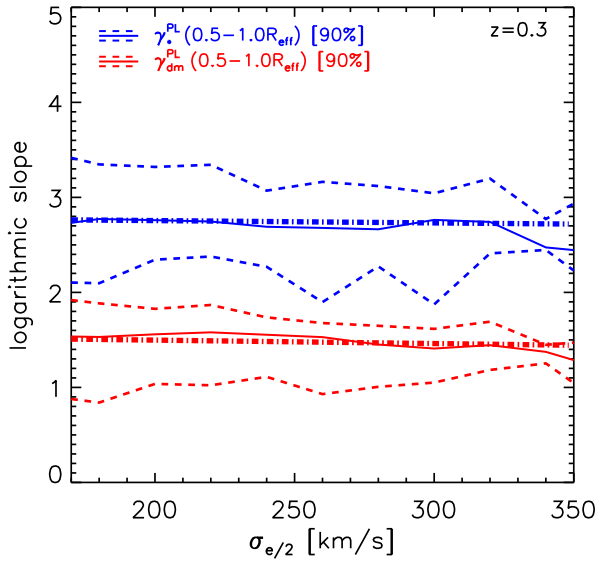}
\caption{The power-law fitted slopes of the inner dark matter
  distribution $\gamma^{\rm PL}_{\rm dm} (0.5\,\Reff,\,\Reff)$ (red)
  and of the inner stellar distribution $\gamma^{\rm PL}_{*}
  (0.5\,\Reff,\,\Reff)$ (blue), as a function of the central stellar
  velocity dispersion $\sigma_{e/2}$, for the selected early-type
  galaxy sample at $z=0.3$. The solid and the dashed line styles
  indicate the median and the 90\% boundaries of either distribution,
  respectively. The thicker dashed and dotted lines present the best
  linear fits to the data. Linear regression of $\gamma_{\rm dm}$ and
  $\gamma_{*}$ as functions of $\sigma_{e/2}$ resulted in $\partial
  \gamma^{\rm PL}_{\rm dm} (0.5\,\Reff,\,\Reff) / \partial
  \sigma_{e/2} = -0.0004 \pm 0.0002$ with a linear correlation
  coefficient $r=-0.06$; $\partial \gamma^{\rm PL}_{\rm dm}
  (0.5\,\Reff,\,2.0\,\Reff) / \partial \sigma_{e/2} = -0.0009 \pm
  0.0002$ with $r=-0.22$ for the dark matter density slopes, and
  $\partial \gamma^{\rm PL}_{*} (0.5\,\Reff,\,\Reff) / \partial
  \sigma_{e/2} = -0.0002 \pm 0.0003$ with a linear correlation
  coefficient $r=-0.03$; $\partial \gamma^{\rm PL}_{*}
  (0.5\,\Reff,\,2.0\,\Reff) / \partial \sigma_{e/2} = -0.0004 \pm
  0.0002$ with $r=-0.07$ for the stellar density slopes.  }
\label{fig:SlopeDMST}
\end{figure}

Their dependences on the central stellar velocity dispersion
$\sigma_{e/2}$ are shown in Fig.\,\ref{fig:SlopeDMST}, where the solid
and the dashed lines indicate the median and the 90\% boundaries of
either distribution, respectively; and the thicker dashed and dotted
lines present the best linear fits to the data. Linear regression of
$\gamma_{\rm dm}$ and $\gamma_{*}$ as functions of $\sigma_{e/2}$
resulted in $\partial \gamma^{\rm PL}_{\rm dm} (0.5\,\Reff,\,\Reff) /
\partial \sigma_{e/2} = -0.0004 \pm 0.0002$ with a linear correlation
coefficient $r=-0.06$; $\partial \gamma^{\rm PL}_{\rm dm}
(0.5\,\Reff,\,2.0\,\Reff) / \partial \sigma_{e/2} = -0.0009 \pm
0.0002$ with $r=-0.22$ for the dark matter density slopes, and
$\partial \gamma^{\rm PL}_{*} (0.5\,\Reff,\,\Reff) / \partial
\sigma_{e/2} = -0.0002 \pm 0.0003$ with a linear correlation
coefficient $r=-0.03$; $\partial \gamma^{\rm PL}_{*}
(0.5\,\Reff,\,2.0\,\Reff) / \partial \sigma_{e/2} = -0.0004 \pm
0.0002$ with $r=-0.07$ for the stellar density slopes. 

In addition, the stellar component, with a mean slope of $\langle
\gamma^{\rm PL}_{*} (0.5\,\Reff,\,\Reff) \rangle \sim 2.74$, is
significantly steeper than the total matter distribution (for which
$\langle \gamma^{\rm PL} (0.5\,\Reff,\,\Reff) \rangle \sim
2.08$).  The inner dark matter slope is notably steeper than the
predicted logarithmic slope of unity for an NFW profile at small
radii. A mean slope of $\langle \gamma^{\rm PL}_{\rm dm}
(0.5\,\Reff,\,\Reff) \rangle \sim 1.49$ is well consistent with both
observations (e.g., Sonnenfeld et al. 2012; Cappellari et al. 2013;
Oguri et al. 2014; Bruderer et al. 2016) and published simulations
(e.g., Johansson et al. 2012; Remus et al. 2013).

\subsection{Evolutionary trends}

\begin{figure}
\centering
\includegraphics[width=8cm]{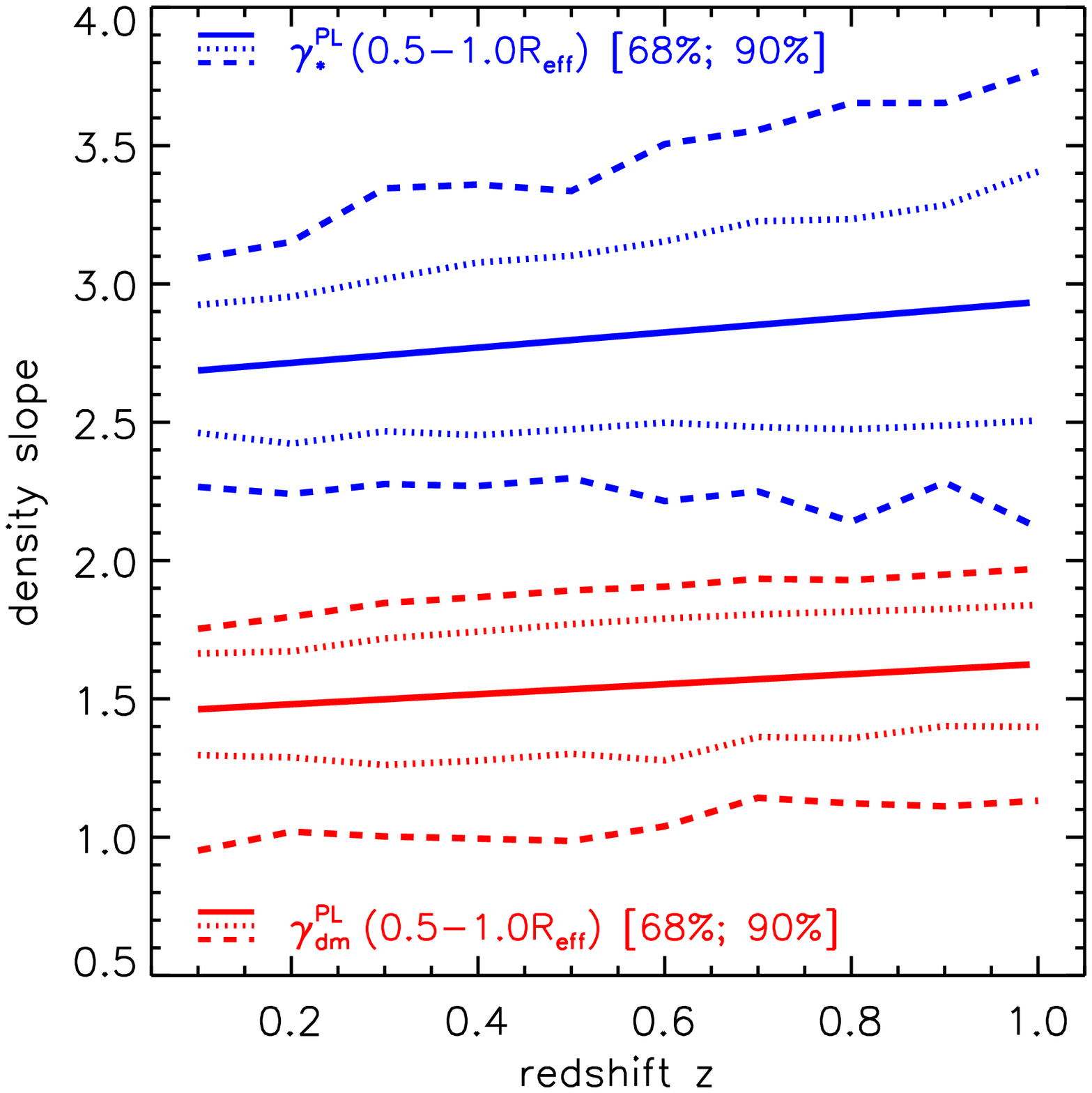}
\includegraphics[width=8cm]{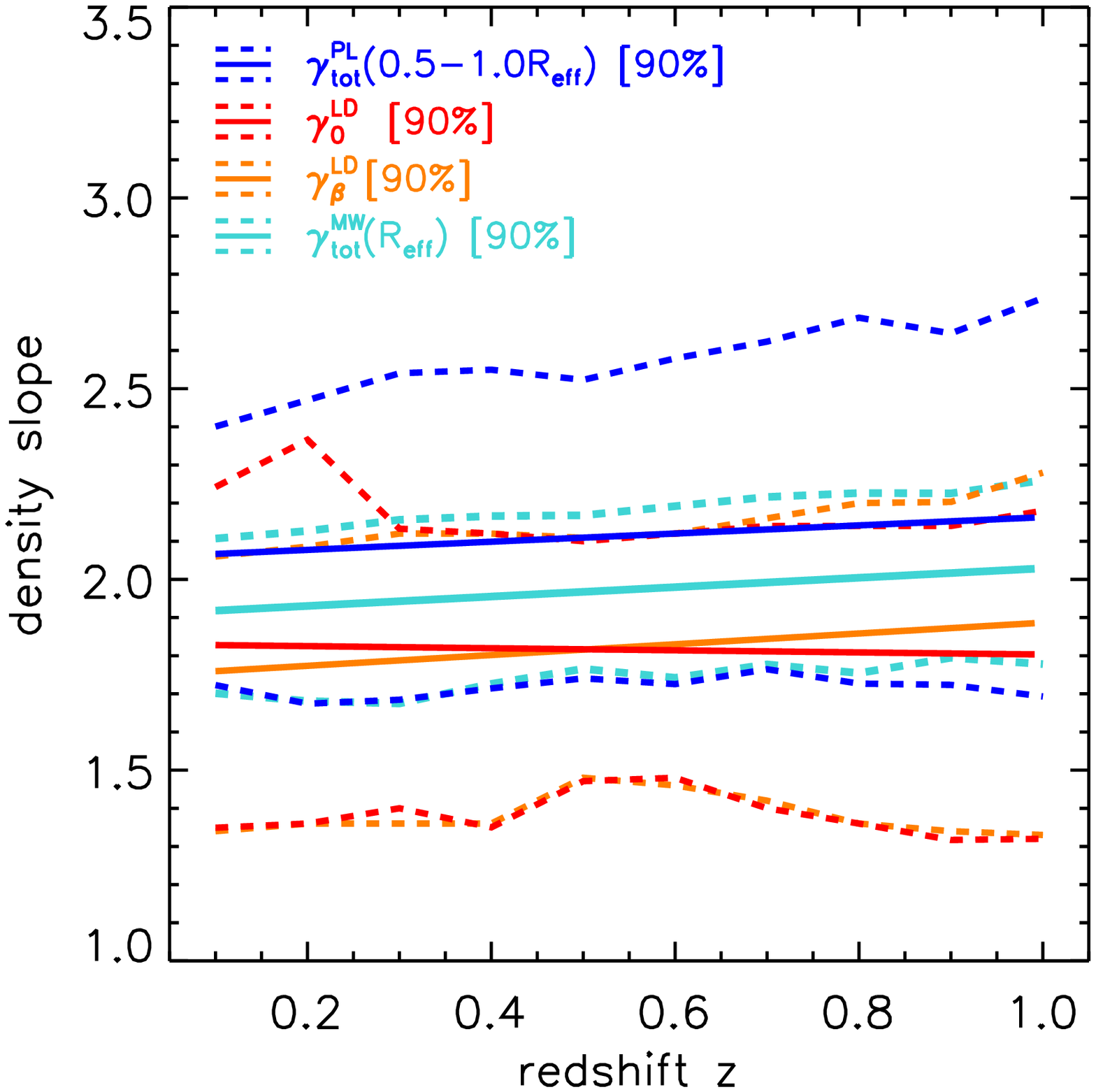}
\caption{The cosmic evolution of the matter density slopes measured
  for the selected early-type galaxies from the Illustris
  simulation. {\em Top panel:} the red and blue curves present the
  slopes, $\gamma_{\rm dm}^{\rm PL}(0.5\,\Reff,\,\Reff)$ and
  $\gamma_{*}^{\rm PL}(0.5\,\Reff,\,\Reff)$, of the dark matter and
  stellar distributions, respectively. {\em Bottom panel:} the red,
  orange, green and blue curves give the slopes of $\gamma_0^{\rm
    LD}$, $\gamma_\beta^{\rm LD}$, $\gamma^{\rm MW}(\Reff)$ and
  $\gamma^{\rm PL}(0.5\,\Reff,\,\Reff)$, respectively. The solid,
  dotted and dashed lines show the best linear fit to the data, the
  68\% and 90\% boundaries of the distributions, respectively. Note
  that linear regression resulted in $\partial \gamma_{\rm dm}^{\rm
    PL}(0.5\,\Reff,\,\Reff) / \partial z = 0.18 \pm 0.01$ with a
  correlation coefficient $r=0.20$ for the dark matter slope
  evolution; $\partial \gamma_{*}^{\rm PL}(0.5\,\Reff,\,\Reff) /
  \partial z = 0.28 \pm 0.02$ with $r=0.20$ for the stellar slope
  evolution. For the total matter density slopes, $\partial
  \gamma_0^{\rm LD} / \partial z = -0.03 \pm 0.01$ with $r=-0.03$;
  $\partial \gamma_\beta^{\rm LD} / \partial z=0.14 \pm 0.01$ with
  $r=0.15$; $\partial \gamma^{\rm MW}(\Reff) / \partial z = 0.12 \pm
  0.01$ with $r=0.25$ and $\partial \gamma^{\rm
    PL}(0.5\,\Reff,\,\Reff) / \partial z =0.11 \pm 0.01$ with
  $r=0.11$.}
\label{fig:slpZev}
\end{figure}

Fig.\,\ref{fig:slpZev} shows the cosmic evolution of the matter
density slopes measured for the selected early-type galaxies from the
Illustris simulation. Again we applied linear regression to the slopes
studied above as functions of redshift $z$. The best linear fit
resulted in $\partial \gamma_{\rm dm}^{\rm PL}(0.5\,\Reff,\,\Reff) /
\partial z = 0.18 \pm 0.01$ with a correlation coefficient $r=0.20$
for the dark matter slope evolution; $\partial \gamma_{*}^{\rm
  PL}(0.5\,\Reff,\,\Reff) / \partial z = 0.28 \pm 0.02$ with $r=0.20$
for the stellar slope evolution. Both become shallower with decreasing
redshift.

For the total matter density slopes, $\partial \gamma_0^{\rm LD} /
\partial z = -0.03 \pm 0.01$ with $r=-0.03$; $\partial
\gamma_\beta^{\rm LD} / \partial z=0.14 \pm 0.01$ with $r=0.15$;
$\partial \gamma^{\rm MW}(\Reff) / \partial z = 0.12 \pm 0.01$ with
$r=0.25$ and $\partial \gamma^{\rm PL}(0.5\,\Reff,\,\Reff) / \partial
z =0.11 \pm 0.01$ with $r=0.11$. As can be seen, most slope estimators
also indicate a shallower trend towards lower redshifts. In
particular, the mean magnitude and evolutionary trend of $\gamma^{\rm
  PL}$ found for the early-type Illustris galaxies are consistent with
results from previous cosmological simulations and theoretical studies
(e.g., Johansson et al. 2012; Remus et al. 2013, 2016; Sonnenfeld et
al. 2014).

It is noteworthy to point out that unlike the redshift evolution of
the intrinsic slope estimators, a mild increase of $\gamma_0^{\rm LD}$
was seen at lower redshifts for the early-type Illustris
galaxies. Interestingly, this is roughly consistent with strong
lensing observations. Currently, the only observationally constrained
redshift evolution (up to $z=1.0$) comes from combining strong lensing
techniques with single-aperture stellar kinematics. The derived slopes
$\gamma_0^{\rm LD}$ for the SLACS, SL2S and BELLS galaxy samples
consistently show a mild steepening towards lower redshifts (e.g.,
Koopmans et al. 2006; Auger et al. 2010b; Ruff et al. 2011; Bolton et
al. 2012; Sonnenfeld et al. 2013, 2015). We stress that such an
apparent evolutionary trend of $\gamma_0^{\rm LD}$ may not represent
the true evolution of the total central density slopes of early-type
galaxies, because the derivation involves various model assumptions
(e.g., spherical symmetry, isotropic stellar orbits and power-law
distributions) and thus suffers from systematic biases (see
Sect.\,6.1.2, 6.1.3 and Fig.\,\ref{fig:slpZev}).

On the other hand, we also caution that $\gamma_0^{\rm LD}$ derived
for the simulated sample cannot be directly compared to the
observational results due to sampling bias. In particular, the
simulation sample and the observational sample have different
probability distributions for $\sigma_{e/2}$ and (normalized) $\RE$ in
such a way that the former sample would experience a relatively lower
mean of $\gamma_0^{\rm LD}$ because of the existence of a larger
fraction of lower-mass systems and/or systems with smaller
(normalized) $\RE$. A fair comparison between the simulation and
observations would require (1) strictly adopting observational
criteria to select simulation samples, and (2) 2-D kinematics data and
high-resolution imaging data for a large number of galaxies out to
high redshifts. This could be facilitated by future
Integral-Field-Unit deep surveys.

\section{Conclusions and Discussion}

In this work, we used the highest resolution run of the Illustris
simulation project (\citeauthor{Illustris2014MN} 2014a,\,b;
\citealt{Genel2014Illustris, Sijacki2015Illustris,
  Nelson2015IllustrisDataRelease}) to study the stellar orbital
anisotropies, the central dark matter fractions and the central radial
density slopes of early-type galaxies from $z=1.0$ to the present.
The early-type galaxies were identified according to their radial
surface brightness distributions (in the SDSS $g$, $r$ and $i$ bands)
as those systems which can be better fit by de Vaucouleurs profiles
than by exponential profiles (see Sect.\,2.4).

In particular, we selected central early-types galaxies with stellar
velocity dispersions within an observationally motivated range equal
to $\sigma_{e/2} \in[160,400]\kms$ (e.g., Bolton et al. 2008a). The
resulting galaxy sample roughly reproduces the observed
mass-size-velocity dispersion relations (see Sect.\,3.2). A variety of
galaxy properties of this simulation sample were compared to observed
early-type galaxies from existing galaxy surveys and strong lensing
surveys. Below we summarize the main findings of our analysis:

\begin{itemize}
\item{Towards galaxy centres, baryons dominate the shape of the total
    matter distribution. The ratio between the ellipticities of the
    luminous and the total distributions is on average
    $(b/a)_{\rm gal} / (b/a)_{\rm tot}\sim0.9$, measured within half
    of the effective radius. As the radius increases, the total matter
    distribution becomes rounder and this ratio becomes smaller. The
    misalignment angle $\Delta\phi_{\rm RA}$ between the luminous and
    the total matter distributions is consistent with zero and shows a
    standard deviation of $\la 10$ degree. In general, the inner
    regions of low-redshift galaxies show better alignment between
    the luminous and the total matter distributions than
    higher-redshift galaxies (see Sect.\,3.1).}

\item{We measured the velocity anisotropy parameter $\beta$ as defined
    in Eq.\,(\ref{eq:beta}) for the simulated early-type galaxies,
    assuming spherically symmetric density distributions. We found
    that higher-mass galaxies and galaxies at lower redshifts tend to
    have more radially anisotropic stellar orbits ($\beta > 0$)
    compared to their lower-mass and higher-redshift counterparts,
    consistent with their redder $B-V$ colours and smaller central
    (cold) gas fractions, which indicates the existence of relatively
    old stellar populations that are primarily passively evolving (see
    Sect.\,4). }

\item{We measured the projected dark matter fraction $f_{\rm dm}$
  within 5 kpc from the galaxy centres to be $40\%-50\%$ for the
  Illustris early-type galaxies. This range is noticeably higher than
  suggested by measurements from the SLACS and SL2S surveys. The
  projected dark matter fractions measured within the effective radius
  $\Reff$ show a very mild dependence on the central velocity
  dispersion $\sigma_{e/2}$, but a clear positive correlation with
  $\lg \Reff$. The latter could be an aperture effect due to
  increasing $f_{\rm dm}$ at larger radii (see Sect.\,5).}

\item{We applied a common technique used observationally that combines
  strong lensing with single-aperture kinematics to constrain the
  total (central) density slopes $\gamma_0^{\rm LD}$ of the simulated
  galaxies. The inferred $\gamma_0^{\rm LD}$ is on average shallower
  than in observations, for which the mean distribution is much closer
  to isothermal (see Auger et al. 2010b, Sonnenfeld et
  al. 2013). However, the simulation roughly reproduces the observed
  dependences of $\gamma_0^{\rm LD}$ on other galaxy properties (e.g.,
  effective radius, stellar velocity dispersion, and surface density,
  see Sect.\,6.1.1). }

\item{The slope $\gamma_0^{\rm LD}$ derived using the above-mentioned
  simple practical technique suffers from systematic biases due to two
  major assumptions, i.e., the isotropic stellar orbits and the
  power-law density model. As a result, (1) radially (tangentially)
  anisotropic orbits cause $\gamma_0^{\rm LD}$ to overestimate
  (underestimate) the true density slopes (see Sect.\,6.1.2); and (2)
  the poor approximation of the power-law assumption causes
  $\gamma_0^{\rm LD}$ to diverge from the true density slope, in
  particular for lower-mass galaxies with $\sigma_{e/2}\la 250\kms$
  (see Sect.\,6.1.3). These model assmuptions could have also
  introduced biased mass dependence and biased redshift evolution as
  probed by $\gamma_0^{\rm LD}$ (see Sect.\,6.4).}

\item{We compared slope $\gamma_0^{\rm LD}$ to several different slope
  estimators ($\gamma^{\rm MW}$, $\gamma^{\rm PL}$ and $\gamma^{\rm
    AV}$), which measure some intrinsic slopes of the central density
  profiles. In particular, the power-law fitted slope $\gamma^{\rm
    PL}(\Reff)$ decreases with increasing $\sigma_{e/2}$ and is on
  average (slightly) steeper than isothermal, consistent with the
  observational results from 2-D kinematical data (e.g.,
  \citealt{Barnabe2011III}; Cappellari et al. 2015). In comparison,
  both $\gamma_{\rm tot\ }^{\rm MW}(\Reff)$ and $\gamma_0^{\rm LD}$
  are shallower than isothermal; and $\gamma_{\rm tot\ }^{\rm
    MW}(\Reff)>\gamma_0^{\rm LD}$ was observed especially for
  lower-mass galaxies ($\sigma_{e/2}\la 250 \kms$). The difference is
  due to a combination of two effects, which hold for the majority of
  the selected galaxy samples: (1) the power-law assumption breaks
  down and, in particular, the curvature parameter exceeds unity,
  i.e., $\xi_{\rm M_{\rm 3d}}>1.0$, essentially resulting in steeper
  (local) density slopes at larger radii; and (2) on average, the
  effective radii $\Reff$ are larger than the Einstein radii $\RE$,
  which eventually makes $\gamma^{\rm MW}(\Reff)$ larger than
  $\gamma_0^{\rm LD}$, as the latter quantity is essentially
  normalized at $\RE$ (see Sect.\,6.2). }

\item{The baryonic component is much more centrally concentrated than
    the total matter distribution. Due to the existence of baryons,
    the inner dark matter slope is notably steeper than the NFW
    prediction (see Sect.\,6.3). For the selected early-type galaxy
    samples from the simulation, the density slopes, either of the
    individual dark matter and baryonic components, or of the sum of
    the two, become shallower with cosmic time (see Sect.\,6.4). }

\item {Several disagreements between the simulation and observational
  results that we found in this work seem to be related: the
  simulation predicted some higher central dark matter fractions,
  which would have suppressed the dominanting role of baryons and thus
  led to somehow shallower total density profiles in the inner regions
  of galaxies. The tension pose a potential challenge to the stellar
  formation and feedback models adopted by the simulation.}

\end{itemize}

It is worth noting that for the selected early-type Illustris
galaxies, we obtained self-consistent findings regarding the mass
dependences and redshift evolutions of their colours, central cold gas
(HI) fractions, stellar orbital anisotropies and central total density
slopes. The early-type galaxies at higher redshifts are seen to be
bluer and to contain a higher fraction of cold gas in their central
regions. These features indicate star formation activity in their
recent histories, which also leads to steeper central density profiles
and more tangentially anisotropic stellar orbits. In comparison, their
lower-redshift counterparts host much older stellar populations with
redder colours and smaller cold gas fractions. In particular, as a
consequence of another 7 Gyrs of passive evolution, their central
density profiles become shallower and develop more radially
anisotropic orbits. To test these theoretical predictions, 2-D
kinematical data and high-resolution multi-band imaging data are
required for a large number of galaxies, and out to high redshifts.

\section*{ACKNOWLEDGEMENTS}

We would like to thank Stefan Hilbert, Hongyu Li, Yiping Shu, Adam
Bolton and Matthias Bartelmann for various useful discussions. We
would like to acknowledge the constructive discussion held at the TAP
group meeting, in particular we thank Robert Grand, Ruediger Pakmor,
Christine Simpson and Thomas Guillet. We would also like to thank the
anonymous referee for very insightful and useful comments to improve
the quality of the paper. DDX and VS would like to thank the Klaus
Tschira Foundation. VS acknowledges support through the European
Research Council under ERC-StG grant EXAGAL-308037. DS acknowledges
funding support from a {\it {Back to Belgium}} grant from the Belgian
Federal Science Policy (BELSPO). MV gratefully acknowledges support of
the Alfred P. Sloan Foundation and the MIT RSC Reed fund.

\onecolumn
\appendix
\section{Summary of the catalogue fields}

We calculated a variety of properties (including galaxy types, sizes,
morphologies, photometries, matter contents and distributions, and
measurements on strong lensing and stellar kinematics) for all the
Illustris galaxies that have stellar masses $M_{*}\ga10^{10}M_{\odot}$
at redshift $z=[0.1,~0.2,~0.3, ~0.4, ~0.5, ~0.6, ~0.7, ~0.8, ~0.9,
  ~1.0]$. Accordingly we defined artificial source redshifts at
$z_{\rm s}=[0.5,~0.6,~0.7,~0.9,~2.0,~2.0,~2.0,~2.0,~2.0,~2.0]$ in
order to calculate the expected Einstein radii $\RE$ for the
galaxies. These choices of $z_{\rm s}$ were motived by the lens-source
redshift distributions of the SLACS and SL2S surveys. All these data
are publicly available at the Illustris website
(www.illustris-project.org). In this Appendix, we give an overview of
the catalogued quantities in Table~\ref{tab:size} to
Table~\ref{tab:slp}, where we only provide descriptions of the
projection-dependent quantities in their X-projection (those for the
Y- and Z-projections follow the same fashion) for the sake of
simplicity.

In the catalogue, all radii involved are centred on the galactic
centre of light (see Table.\,\ref{tab:Morphology}), which is also
referred to as the ``galaxy centre''. All properties that were
evaluated at a given radius were obtained through interpolation using
the polynomial functions that were fitted to the corresponding radial
distributions (assuming a circular/spherical symmetry) within a given
radial range, i.e., from $R_{\rm promin}$ to $R_{\rm promax}$ (see
Table~\ref{tab:size} for their definitions). If the evaluation radius
falls beyond this range, then the evaluation was actually made at the
closer boundary radius of either $R_{\rm promin}$ or $R_{\rm
  promax}$. For density slopes that were measured between a given
radial range of ($r_1$, $r_2$), if $r_2<R_{\rm promin}$ or $r_1>R_{\rm
  promax}$ was true, then the slope was set to be the default value of
1E10. In addition, any fractional quantity should be a number between
0.0 and 1.0; however the interpolated value could go beyond this range
if the interpolation radius is close to the boundaries where the
polynomial fit becomes divergent. In this case, the interpolated
fractions would be set to be the default value of -1.0. Tables given
below further clarify any other specific cases where the default
values would be reached.

We note that apart from the online catalogue, galaxy surface
brightness distribution, as well as the best-fitted Sersic, de
Vaucouleurs and exponential profiles (see Sect.2 for details) can be
available by sending e-mail request to the author. In addition, the
following radial profiles (and the polynomial fitting functions) of
the projected/3d matter distribution that were extracted under the
assumption of a circular/spherical symmetry (in regardless of galaxy
types) within a radial range between $R_{\rm promin}$ and $R_{\rm
  promax}$ can be also available by sending e-mail request to the
author. These profiles include: (1) the convergence (surface density)
distribution $\kappa(R)$ at a given projected radius of $R$, (2) the
mean convergence distribution $\overline{\kappa}(\leqslant R)\equiv
2\int_0^{R} {R^{\prime} \kappa(R^{\prime}) {\rm d}R{^{\prime}}}/R^2$,
(3) the projected local dark matter fraction $f^{2D}_{\rm dm}(R)$, (4)
the projected cumulative dark matter fraction $\overline{f^{2D}_{\rm
    dm}}(\leqslant R)$, (5) the total matter density distribution
$\rho(r)$ at a 3D radius of $r$, (6) the local dark matter fraction
$f^{3D}_{\rm dm}(r)$, (7) the cumulative dark matter fraction
$\overline{f^{3D}_{\rm dm}}(\leqslant r)$, (8) the cumulative gas
fraction $\overline{f^{3D}_{\rm gas}}(\leqslant r)$, (9) the
cumulative cold gas (HI) fraction $\overline{f^{3D}_{\rm
    cgs}}(\leqslant r)$. Note that all the mass fraction distributions
are with respect to the total matter distribution. 

\begin{table*}
\caption{Galaxy size measurements. Note that all quantities below are
  calculated from the galactic centre of light (see
  Table.\,\ref{tab:Morphology}) and are in unit of arcsec. For more
  details, see Sect.\,2.3 and Sect.\,2.4. }
\begin{minipage}{\textwidth}
\begin{tabular}{l | l } \hline\hline
Group Name  &  Description \\\hline
R\_promin & The minimum of the radial range, within which radial distributions of relevant quantities were measured \\
& (and used for interpolation); set to be the angular size (at the snapshot redshift) corresponding to a physical \\
& scale of 0.7 kpc, which is the softening length of the simulation   \\
\hline
R\_promax & The maximum of the radial range, within which radial distributions of relevant quantities were measured (and \\
& used for interpolation); set to be the angular size (at the snapshot redshift) corresponding to min(5$\times$hsmr, 30 kpc), \\
& where hsmr is the half-stellar-mass radius of the galaxy subhalo as calculated by {\sc subfind} \\ 
\hline
Rein\_x  & The Einstein radius in X-projection; set to be 0.0 if Rein\_x $<$ R\_promin \\
\hline
Rc50\_x  & The radius within which the projected cumulative dark matter fraction is 50\% in X-projection; \\
& set to be 0.0 if Rc50\_x $<$ R\_promin; or set to be 1E10 if Rc50\_x $>$ R\_promax \\
\hline
Rl50\_x  & The radius at which the projected local dark matter fraction is 50\% in X-projection; \\
& set to be 0.0 if Rl50\_x $<$ R\_promin; or set to be 1E10 if Rl50\_x $>$ R\_promax \\
\hline
Reff\_ser\_x  & The effective radius by fitting Sersic profile in rest-frame SDSS r-band in X-projection \\
\hline
Reff\_dev\_x  & The effective radius by fitting de Vaucouleurs profile in rest-frame SDSS r-band in X-projection \\
\hline
Reff\_exp\_x  & The effective radius by fitting exponential profile in rest-frame SDSS r-band in X-projection \\
\hline
Reff\_of\_sdss\_umod\_x  & The Sersic-fitted effective radius in observer-frame SDSS u-band in X-projection \\ 
\hline
Reff\_of\_sdss\_u\_x  & The radius which encloses half of the total luminosity measured in observer-frame SDSS u-band within \\
& a projected radius of 30  kpc from galaxy centre in X-projection \\
\hline
Reff\_of\_sdss\_gmod\_x  & The Sersic-fitted effective radius in observer-frame SDSS g-band in X-projection \\
\hline
Reff\_of\_sdss\_g\_x  & The radius which encloses half of the total luminosity measured in observer-frame SDSS g-band within \\
& a projected radius of 30  kpc from galaxy centre in X-projection \\
\hline
Reff\_of\_sdss\_rmod\_x  & The Sersic-fitted effective radius in observer-frame SDSS r-band in X-projection \\
\hline
Reff\_of\_sdss\_r\_x  & The radius which encloses half of the total luminosity measured in observer-frame SDSS r-band within \\
& a projected radius of 30  kpc from galaxy centre in X-projection \\
\hline
Reff\_of\_sdss\_imod\_x  & The Sersic-fitted effective radius in observer-frame SDSS i-band in X-projection \\
\hline
Reff\_of\_sdss\_i\_x  & The radius which encloses half of the total luminosity measured in observer-frame SDSS i-band within \\
& a projected radius of 30  kpc from galaxy centre in X-projection \\
\hline
Reff\_of\_sdss\_zmod\_x  & The Sersic-fitted effective radius in observer-frame SDSS z-band in X-projection \\
\hline
Reff\_of\_sdss\_z\_x  & The radius which encloses half of the total luminosity measured in observer-frame SDSS z-band within \\
& a projected radius of 30  kpc from galaxy centre in X-projection \\
\hline
Reff\_rf\_sdss\_gmod\_x  & The Sersic-fitted effective radius in rest-frame SDSS g-band in X-projection \\
\hline
Reff\_rf\_sdss\_g\_x  & The radius which encloses half of the total luminosity measured in rest-frame SDSS g-band within \\
& a projected radius of 30  kpc from galaxy centre in X-projection \\
\hline
Reff\_rf\_sdss\_rmod\_x  & The Sersic-fitted effective radius in rest-frame SDSS r-band in X-projection \\
\hline
Reff\_rf\_sdss\_r\_x  & The radius which encloses half of the total luminosity measured in rest-frame SDSS r-band within \\
& a projected radius of 30  kpc from galaxy centre in X-projection \\
\hline
Reff\_rf\_sdss\_imod\_x  & The Sersic-fitted effective radius in rest-frame SDSS i-band in X-projection \\
\hline
Reff\_rf\_sdss\_i\_x  & The radius which encloses half of the total luminosity measured in rest-frame SDSS i-band within \\
& a projected radius of 30  kpc from galaxy centre in X-projection \\
\hline
Reff\_of\_hst\_bmod\_x  & The Sersic-fitted effective radius in observer-frame HST B-F435w in X-projection \\
\hline
Reff\_of\_hst\_b\_x  & The radius which encloses half of the total luminosity measured in observer-frame HST B-F435w within \\
& a projected radius of 30  kpc from galaxy centre in X-projection \\
\hline
Reff\_of\_hst\_vmod\_x  & The Sersic-fitted effective radius in observer-frame HST V-F606w in X-projection \\
\hline
Reff\_of\_hst\_v\_x  & The radius which encloses half of the total luminosity measured in observer-frame HST V-F606w within \\
& a projected radius of 30  kpc from galaxy centre in X-projection \\
\hline
Reff\_of\_hst\_imod\_x  & The Sersic-fitted effective radius in observer-frame HST I-F814w in X-projection \\
\hline
Reff\_of\_hst\_i\_x  & The radius which encloses half of the total luminosity measured in observer-frame HST I-F814w within \\
& a projected radius of 30  kpc from galaxy centre in X-projection \\
\hline
Reff\_rf\_john\_bmod\_x  & The Sersic-fitted effective radius in rest-frame Johnson B-band in X-projection \\
\hline
Reff\_rf\_john\_b\_x  & The radius which encloses half of the total luminosity measured in rest-frame Johnson B-band within \\
& a projected radius of 30  kpc from galaxy centre in X-projection \\
\hline
Reff\_rf\_john\_vmod\_x  & The Sersic-fitted effective radius in rest-frame Johnson V-band in X-projection \\
\hline
Reff\_rf\_john\_v\_x  & The radius which encloses half of the total luminosity measured in rest-frame Johnson V-band within \\
& a projected radius of 30  kpc from galaxy centre in X-projection \\
\hline
\end{tabular}
\end{minipage}
\label{tab:size}
\end{table*}

\begin{table*}
\caption{Galaxy photometry. Note that all magnitudes below are
  absolute AB magnitudes. The surface brightnesses are in unit of mag
  arcsec$^{-2}$. For more details, see Sect.\,2.3 and Sect.\,2.4 for
  details.}
\begin{minipage}{\textwidth}
\begin{tabular}{l | l } \hline\hline
Group Name & Description \\\hline
Sersic\_m\_x  & The Sersic index of the best-fitted Sersic profile in rest-frame SDSS r-band in X-projection \\
\hline
IRe\_ser\_x  & The surface brightness at Reff\_ser\_x in rest-frame SDSS r-band in X-projection \\
\hline
IRe\_dev\_x  & The surface brightness at Reff\_dev\_x in rest-frame SDSS r-band in X-projection \\
\hline
IRe\_exp\_x  & The surface brightness at Reff\_exp\_x in rest-frame SDSS r-band in X-projection \\
\hline
Mag\_of\_sdss\_umod\_x  & The magnitude in observer-frame SDSS u-band derived from best-fitted Sersic model in X-projection \\ 
\hline
Mag\_of\_sdss\_u\_x  & The magnitude in observer-frame SDSS u-band derived from direct measurement \\
& within a projected radius of 30  kpc from galaxy centre in X-projection \\
\hline
Mag\_of\_sdss\_gmod\_x  & The magnitude in observer-frame SDSS g-band derived from best-fitted Sersic model in X-projection \\
\hline
Mag\_of\_sdss\_g\_x  & The magnitude in observer-frame SDSS g-band derived from direct measurement \\
& within a projected radius of 30  kpc from galaxy centre in X-projection \\
\hline
Mag\_of\_sdss\_rmod\_x  & The magnitude in observer-frame SDSS r-band derived from best-fitted Sersic model in X-projection \\
\hline
Mag\_of\_sdss\_r\_x  & The magnitude in observer-frame SDSS r-band derived from direct measurement \\
& within a projected radius of 30  kpc from galaxy centre in X-projection \\
\hline
Mag\_of\_sdss\_imod\_x  & The magnitude in observer-frame SDSS i-band derived from best-fitted Sersic model in X-projection \\
\hline
Mag\_of\_sdss\_i\_x  & The magnitude in observer-frame SDSS i-band derived from direct measurement \\
& within a projected radius of 30  kpc from galaxy centre in X-projection \\
\hline
Mag\_of\_sdss\_zmod\_x  & The magnitude in observer-frame SDSS z-band derived from best-fitted Sersic model in X-projection \\
\hline
Mag\_of\_sdss\_z\_x  & The magnitude in observer-frame SDSS z-band derived from direct measurement \\
& within a projected radius of 30  kpc from galaxy centre in X-projection \\
\hline
Mag\_rf\_sdss\_gmod\_x  & The magnitude in rest-frame SDSS g-band derived from best-fitted Sersic model in X-projection \\
\hline
Mag\_rf\_sdss\_g\_x  & The magnitude in rest-frame SDSS g-band derived from direct measurement \\
& within a projected radius of 30  kpc from galaxy centre in X-projection \\
\hline
Mag\_rf\_sdss\_rmod\_x  & The magnitude in rest-frame SDSS r-band derived from best-fitted Sersic model in X-projection \\
\hline
Mag\_rf\_sdss\_r\_x  & The magnitude in rest-frame SDSS r-band derived from direct measurement \\
& within a projected radius of 30  kpc from galaxy centre in X-projection \\
\hline
Mag\_rf\_sdss\_imod\_x  & The magnitude in rest-frame SDSS i-band derived from best-fitted Sersic model in X-projection \\
\hline
Mag\_rf\_sdss\_i\_x  & The magnitude in rest-frame SDSS i-band derived from direct measurement \\
& within a projected radius of 30  kpc from galaxy centre in X-projection \\
\hline
Mag\_of\_hst\_bmod\_x  & The magnitude in observer-frame HST B-F435w derived from best-fitted Sersic model in X-projection \\
\hline
Mag\_of\_hst\_b\_x  & The magnitude in observer-frame HST B-F435w derived from direct measurement \\
& within a projected radius of 30  kpc from galaxy centre in X-projection \\
\hline
Mag\_of\_hst\_vmod\_x  & The magnitude in observer-frame HST V-F606w derived from best-fitted Sersic model in X-projection \\
\hline
Mag\_of\_hst\_v\_x  & The magnitude in observer-frame HST V-F606w derived from direct measurement \\
& within a projected radius of 30  kpc from galaxy centre in X-projection \\
\hline
Mag\_of\_hst\_imod\_x  & The magnitude in observer-frame HST I-F814w derived from best-fitted Sersic model in X-projection \\
\hline
Mag\_of\_hst\_i\_x  & The magnitude in observer-frame HST I-F814w derived from direct measurement \\
& within a projected radius of 30  kpc from galaxy centre in X-projection \\
\hline
Mag\_rf\_john\_bmod\_x  & The magnitude in rest-frame Johnson B-band derived from best-fitted Sersic model in X-projection \\
\hline
Mag\_rf\_john\_b\_x  & The magnitude in rest-frame Johnson B-band derived from direct measurement \\
& within a projected radius of 30  kpc from galaxy centre in X-projection \\
\hline
Mag\_rf\_john\_vmod\_x  & The magnitude in rest-frame Johnson V-band derived from best-fitted Sersic model in X-projection \\
\hline
Mag\_rf\_john\_v\_x  & The magnitude in rest-frame Johnson V-band derived from direct measurement \\
& within a projected radius of 30  kpc from galaxy centre in X-projection \\
\hline
MagB\_ep5\_x  & The rest-frame Johnson-B magnitude measured within a projected radius of $0.5\times$Reff\_rf\_john\_bmod\_x \\
& from galaxy centre in X-projection \\
\hline
MagB\_e1\_x  & The rest-frame Johnson-B magnitude measured within a projected radius of Reff\_rf\_john\_bmod\_x from \\
& galaxy centre in X-projection \\
\hline
MagB\_e2\_x  & The rest-frame Johnson-B magnitude measured within a projected radius of $2.0\times$Reff\_rf\_john\_bmod\_x \\
& from galaxy centre in X-projection \\
\hline
MagV\_ep5\_x  & The rest-frame Johnson-V magnitude measured within a projected radius of $0.5\times$Reff\_rf\_john\_vmod\_x \\
& from galaxy centre in X-projection \\
\hline  
MagV\_e1\_x  & The rest-frame Johnson-V magnitude measured within a projected radius of Reff\_rf\_john\_vmod\_x from \\
& galaxy centre in X-projection \\
\hline
MagV\_e2\_x  & The rest-frame Johnson-V magnitude measured within a projected radius of $2.0\times$Reff\_rf\_john\_vmod\_x \\
& from galaxy centre in X-projection \\
\hline
\end{tabular}
\end{minipage}
\label{tab:Photometry}
\end{table*}

\begin{table*}
\caption{Galaxy morphologies. For more details, see Sect.\,2.2.}
\begin{minipage}{\textwidth}
\begin{tabular}{l | l } \hline\hline
Group Name &  Description \\\hline 
Galxc\_x  & The x-coordinate (in the plane of projection) of the light centre in the rest-frame Johnson-V band in X-projection, \\
& measured within a projected radius of 3.0$\times$hsmr from and with respect to the centre of subhalo as calculated by {\sc subfind} \\
\hline
Galyc\_x  & The y-coordinate (in the plane of projection) of the light centre in the rest-frame Johnson-V band in X-projection, \\
& measured within a projected radius of 3.0$\times$hsmr from and with respect to the centre of subhalo as calculated by {\sc subfind} \\
\hline
Galb2a\_ep5\_x  & The axial ratio of the projected (rest-frame Johnson-V band) light distribution measured within \\
 &  a radius of $0.5\times$Reff\_rf\_john\_vmod\_x from galaxy centre in X-projection \\
\hline
Galb2a\_e1\_x  & The axial ratio of the projected (rest-frame Johnson-V band) light distribution measured within \\
 &  a radius of Reff\_rf\_john\_vmod\_x from galaxy centre in X-projection \\
\hline
Galb2a\_e2\_x  & The axial ratio of the projected (rest-frame Johnson-V band) light distribution measured within \\
 &  a radius of $2.0\times$Reff\_rf\_john\_vmod\_x from galaxy centre in X-projection \\
\hline
Subb2a\_ep5\_x  & The axial ratio of the projected total matter distribution measured within a radius of \\
 &  $0.5\times$Reff\_rf\_john\_vmod\_x from galaxy centre  in X-projection \\
\hline
Subb2a\_e1\_x  & The axial ratio of the projected total matter distribution measured within a radius of \\
 &  Reff\_rf\_john\_vmod\_x from galaxy centre in X-projection \\
\hline
Subb2a\_e2\_x  & The axial ratio of the projected total matter distribution measured within a radius of \\
 &  $2.0\times$Reff\_rf\_john\_vmod\_x from galaxy centre in X-projection \\
\hline
GalRA\_ep5\_x  & The orientation angle of the projected (rest-frame Johnson-V band) light distribution measured \\
 &  within a radius of $0.5\times$Reff\_rf\_john\_vmod\_x from galaxy centre in X-projection \\
\hline
GalRA\_e1\_x  & The orientation angle of the projected (rest-frame Johnson-V band) light distribution measured \\
 &  within a radius of Reff\_rf\_john\_vmod\_x from galaxy centre in X-projection \\
\hline
GalRA\_e2\_x  & The orientation angle of the projected (rest-frame Johnson-V band) light distribution measured \\
 &  within a radius of $2.0\times$Reff\_rf\_john\_vmod\_x from galaxy centre in X-projection \\
\hline
SubRA\_ep5\_x  & The orientation angle of the projected total matter distribution measured within a radius of \\
 &  $0.5\times$Reff\_rf\_john\_vmod\_x from galaxy centre in X-projection \\
\hline
SubRA\_e1\_x  & The orientation angle of the projected total matter distribution measured within a radius of \\
 &  Reff\_rf\_john\_vmod\_x from galaxy centre in X-projection \\
\hline
SubRA\_e2\_x  & The orientation angle of the projected total matter distribution measured within a radius of \\
 &  $2.0\times$Reff\_rf\_john\_vmod\_x from galaxy centre in X-projection \\
\hline
TypeDec\_x   &  Galaxy Type in X-projection: 1 for early-type; 0 for late-type; -1 if lack of resolution for surface brightness fitting \\
\hline
\end{tabular}
\end{minipage}
\label{tab:Morphology}
\end{table*}

\begin{table*}
\caption{Matter contents and fractions. Note that all masses below are
  in unit of $h^{-1}M_{\odot}$; all fractions are with respect to the
  total matter. To derive the percentage of HI out of the total gas
  content, one needs to take, e.g., Fcgs3\_e1\_x/Fgas3\_e1\_x. }
\begin{minipage}{\textwidth}
\begin{tabular}{l | l} \hline\hline
Group Name &  Description \\\hline
Mrein\_x   &  The mass projected within a radius of Rein\_x from galaxy centre in X-projection; set to be 0.0 if Rein\_x $<$ R\_promin \\
\hline
MstarB\_ep5\_x   &  The stellar mass projected within a radius of $0.5\times$Reff\_rf\_john\_bmod\_x from galaxy centre in X-projection \\
\hline
MstarB\_e1\_x   &  The stellar mass projected within a radius of Reff\_rf\_john\_bmod\_x from galaxy centre in X-projection  \\
\hline
MstarB\_e2\_x   &  The stellar mass projected within a radius of $2.0\times$Reff\_rf\_john\_bmod\_x from galaxy centre in X-projection  \\
\hline
MstarV\_ep5\_x   &  The stellar mass projected within a radius of $0.5\times$Reff\_rf\_john\_vmod\_x from galaxy centre in X-projection  \\
\hline
MstarV\_e1\_x   &  The stellar mass projected within a radius of Reff\_rf\_john\_vmod\_x from galaxy centre in X-projection  \\
\hline
MstarV\_e2\_x   &  The stellar mass projected within a radius of $2.0\times$Reff\_rf\_john\_vmod\_x from galaxy centre in X-projection  \\
\hline
MtotB\_ep5\_x   &  The total mass projected within a radius of $0.5\times$Reff\_rf\_john\_bmod\_x from galaxy centre in X-projection  \\
\hline
MtotB\_e1\_x   &  The total mass projected within a radius of Reff\_rf\_john\_bmod\_x from galaxy centre in X-projection  \\
\hline
MtotB\_e2\_x   &  The total mass projected within a radius of $2.0\times$Reff\_rf\_john\_bmod\_x from galaxy centre in X-projection  \\
\hline
MtotV\_ep5\_x   &  The total mass projected within a radius of $0.5\times$Reff\_rf\_john\_vmod\_x from galaxy centre in X-projection  \\
\hline
MtotV\_e1\_x   &  The total mass projected within a radius of Reff\_rf\_john\_vmod\_x from galaxy centre in X-projection  \\
\hline
MtotV\_e2\_x   &  The total mass projected within a radius of $2.0\times$Reff\_rf\_john\_vmod\_x from galaxy centre in X-projection  \\
\hline
Fdm2in5kpc\_x   &  The cumulative dark matter fraction within a projected radius of 5 kpc from galaxy centre in X-projection; \\
& set to be -1.0 if the angular scale of 5 kpc is larger than R\_promax \\
\hline
Fdm2inRein\_x   &  The cumulative dark matter fraction within a projected radius of Rein\_x from galaxy centre in X-projection; \\
& set to be -1.0 if Rein\_x $<$ R\_promin \\
\hline
Fdm2\_ep5\_x   &  The cumulative dark matter fraction within a projected radius of $0.5\times$Reff\_rf\_john\_vmod\_x \\
& from galaxy centre in X-projection \\
\hline
Fdm2\_e1\_x   &  The cumulative dark matter fraction within a projected radius of Reff\_rf\_john\_vmod\_x  \\
 &  from galaxy centre in X-projection \\
\hline
Fdm2\_e2\_x   &  The cumulative dark matter fraction within a projected radius of $2.0\times$Reff\_rf\_john\_vmod\_x  \\
 &  from galaxy centre in X-projection \\
\hline
Fdm3\_ep5\_x   &  The cumulative dark matter fraction within a 3D radius of $0.5\times$Reff\_rf\_john\_vmod\_x from galaxy centre \\
\hline
Fdm3\_e1\_x   &  The cumulative dark matter fraction within a 3D radius of Reff\_rf\_john\_vmod\_x from galaxy centre \\
\hline
Fdm3\_e2\_x   &  The cumulative dark matter fraction within a 3D radius of $2.0\times$Reff\_rf\_john\_vmod\_x from galaxy centre \\
\hline
Fgas3\_ep5\_x   &  The cumulative gas fraction within a 3D radius of $0.5\times$Reff\_rf\_john\_vmod\_x from galaxy centre \\
\hline
Fgas3\_e1\_x   &  The cumulative gas fraction within a 3D radius of Reff\_rf\_john\_vmod\_x from galaxy centre \\
\hline
Fgas3\_e2\_x   &  The cumulative gas fraction within a 3D radius of $2.0\times$Reff\_rf\_john\_vmod\_x from galaxy centre \\
\hline
Fcgs3\_ep5\_x   &  The cumulative cold gas (HI) fraction within a 3D radius of $0.5\times$Reff\_rf\_john\_vmod\_x from galaxy centre \\
\hline
Fcgs3\_e1\_x   &  The cumulative cold gas (HI) fraction within a 3D radius of Reff\_rf\_john\_vmod\_x from galaxy centre \\
\hline
Fcgs3\_e2\_x   &  The cumulative cold gas (HI)fraction within a 3D radius of $2.0\times$Reff\_rf\_john\_vmod\_x from galaxy centre \\
\hline
\end{tabular}
\end{minipage}
\begin{flushleft}
\end{flushleft}
\end{table*}

\begin{table*}
\caption{Stellar kinematics measurements. Note that all velocities
  below are in unit of $\kms$ and with respect to the centre-of-mass
  velocity of its host dark matter subhalo.}
\begin{minipage}{\textwidth}
\begin{tabular}{l | l } \hline\hline
Group Name & Description \\\hline
Vmean\_mav\_x  & The stellar-mass-weighted stellar line-of-sight mean velocity measured within a projected radius \\
& of 1.5 arcsec from galaxy centre in X-projection \\ 
\hline
Vmean\_lav\_x  & The (rest-frame SDSS-r band) luminosity-weighted stellar line-of-sight mean velocity measured \\
& within a projected radius of 1.5 arcsec from galaxy centre in X-projection \\
\hline
Vmean\_ep5\_x  & The (rest-frame SDSS-r band) luminosity-weighted stellar line-of-sight mean velocity measured \\
& within a projected radius of $0.5\times$Reff\_rf\_john\_vmod\_x from galaxy centre in X-projection \\
\hline
Vmean\_e1\_x  & The (rest-frame SDSS-r band) luminosity-weighted stellar line-of-sight mean velocity measured \\
& within a projected radius of Reff\_rf\_john\_vmod\_x from galaxy centre in X-projection \\
\hline
Vmean\_e2\_x  & The (rest-frame SDSS-r band) luminosity-weighted stellar line-of-sight mean velocity measured \\
& within a projected radius of $2.0\times$Reff\_rf\_john\_vmod\_x from galaxy centre in X-projection \\
\hline
Vsigma\_mav\_x & The stellar-mass-weighted stellar line-of-sight velocity dispersion measured within a projected \\
& radius of 1.5 arcsec from galaxy centre in X-projection \\
\hline
Vsigma\_lav\_x & The (rest-frame SDSS-r band) luminosity-weighted stellar line-of-sight velocity dispersion measured \\
& within a projected radius of 1.5 arcsec from galaxy centre in X-projection \\
\hline
Vsigma\_ep5\_x  & The (rest-frame SDSS-r band) luminosity-weighted stellar line-of-sight velocity dispersion measured \\
& within a projected radius of $0.5\times$Reff\_rf\_john\_vmod\_x from galaxy centre in X-projection \\
\hline
Vsigma\_e1\_x  & The (rest-frame SDSS-r band) luminosity-weighted stellar line-of-sight velocity dispersion measured \\
& within a projected radius of Reff\_rf\_john\_vmod\_x from galaxy centre in X-projection \\
\hline
Vsigma\_e2\_x  & The (rest-frame SDSS-r band) luminosity-weighted stellar line-of-sight velocity dispersion measured \\
& within a projected radius of $2.0\times$Reff\_rf\_john\_vmod\_x from galaxy centre in X-projection \\
\hline
Beta\_mav\_x & The stellar-mass-weighted stellar orbital anisotropy parameter measured within a 3D radius of \\
& 1.5 arcsec from galaxy centre \\
\hline
Beta\_lav\_x & The (rest-frame SDSS-r band) luminosity-weighted stellar orbital anisotropy parameter measured \\
& within a 3D radius of 1.5 arcsec from galaxy centre \\
\hline
Beta\_ep5\_x  & The (rest-frame SDSS-r band) luminosity-weighted stellar orbital anisotropy parameter measured \\
& within a 3D radius of $0.5\times$Reff\_rf\_john\_vmod\_x from galaxy centre \\
\hline
Beta\_e1\_x  & The (rest-frame SDSS-r band) luminosity-weighted stellar orbital anisotropy parameter measured \\
& within a 3D radius of Reff\_rf\_john\_vmod\_x from galaxy centre \\
\hline
Beta\_e2\_x  & The (rest-frame SDSS-r band) luminosity-weighted stellar orbital anisotropy parameter measured \\
& within a 3D radius of $2.0\times$Reff\_rf\_john\_vmod\_x from galaxy centre \\
\hline
\end{tabular}
\end{minipage}
\label{tab:LensingKinematics}
\end{table*}

\begin{table*}
\caption{Matter density slopes. For more details, see Sect.\,6.}
\begin{minipage}{\textwidth}
\begin{tabular}{l | l } \hline\hline
Group Name  & Description \\\hline
slpMWtot\_ep5\_x & The mass-weighted total density slope calculated using Eq.\,1 of Dutton \& Treu 2014, evaluated \\
& at a radius of $0.5\times$Reff\_rf\_john\_vmod\_x from galaxy centre \\
\hline
slpMWtot\_e1\_x & The mass-weighted total density slope calculated using Eq.\,1 of Dutton \& Treu 2014, evaluated \\
& at a radius of Reff\_rf\_john\_vmod\_x from galaxy centre \\
\hline
slpMWtot\_e2\_x & The mass-weighted total density slope calculated using Eq.\,1 of Dutton \& Treu 2014, evaluated \\
& at a radius of $2.0\times$Reff\_rf\_john\_vmod\_x from galaxy centre \\
\hline
slp3tot\_ep5\_x & The average total density slope calculated using Eq. (15) of the paper, between 0.2$-$0.5 times \\
& Reff\_rf\_john\_vmod\_x  \\
\hline
slp3tot\_e1\_x & The average total density slope calculated using Eq. (15) of the paper, between 0.5$-$1.0 times \\
& Reff\_rf\_john\_vmod\_x \\
\hline
slp3tot\_e2\_x & The average total density slope calculated using Eq. (15) of the paper, between 0.5$-$2.0 times \\
& Reff\_rf\_john\_vmod\_x \\
\hline
slp3totPLf\_ep5\_x & The fitted power-law slope of the total density distribution between 0.2$-$0.5 times Reff\_rf\_john\_vmod\_x \\
\hline
slp3totPLf\_e1\_x & The fitted power-law slope of the total density distribution between 0.5$-$1.0 times Reff\_rf\_john\_vmod\_x \\
\hline
slp3totPLf\_e2\_x & The fitted power-law slope of the total density distribution between 0.5$-$2.0 times Reff\_rf\_john\_vmod\_x \\
\hline
slpJESER\_x & The total density slope derived by combining strong lensing measurement of Mrein\_x and single-aperture \\
& stellar kinematics data of Vsigma\_lav\_x, assuming the stellar orbital anisotropy is given by Beta\_lav\_x; \\
& set to be 1E10 if Rein\_x $<$ R\_promin \\
\hline
slpJEbeta0\_x & The total density slope derived by combining strong lensing measurement of Mrein\_x and \\
& single-aperture stellar kinematics data of Vsigma\_lav\_x, assuming isotropic stellar orbital distribution; \\
& set to be 1E10 if Rein\_x $<$ R\_promin \\
\hline
slp3dm\_ep5\_x & The average dark matter density slope calculated using Eq. (15) of the paper, between 0.2$-$0.5 \\
& times Reff\_rf\_john\_vmod\_x \\
\hline
slp3dm\_e1\_x & The average dark matter density slope calculated using Eq. (15) of the paper, between 0.5$-$1.0 \\
& times Reff\_rf\_john\_vmod\_x \\
\hline
slp3dm\_e2\_x & The average dark matter density slope calculated using Eq. (15) of the paper, between 0.5$-$2.0 \\
& times Reff\_rf\_john\_vmod\_x \\
\hline
slp3dmPLf\_ep5\_x & The fitted power-law slope of the dark matter density distribution between 0.2$-$0.5 times \\
& Reff\_rf\_john\_vmod\_x \\
\hline
slp3dmPLf\_e1\_x & The fitted power-law slope of the dark matter density distribution between 0.5$-$1.0 times \\
& Reff\_rf\_john\_vmod\_x \\
\hline
slp3dmPLf\_e2\_x & The fitted power-law slope of the dark matter density distribution between 0.5$-$2.0 times \\
& Reff\_rf\_john\_vmod\_x \\
\hline
slp3st\_ep5\_x & The average stellar density slope calculated using Eq. (15) of the paper, \\
& between 0.2$-$0.5 times Reff\_rf\_john\_vmod\_x \\
\hline
slp3st\_e1\_x & The average stellar density slope calculated using Eq. (15) of the paper, \\
& between 0.5$-$1.0 times Reff\_rf\_john\_vmod\_x \\
\hline
slp3st\_e2\_x & The average stellar density slope calculated using Eq. (15) of the paper, \\
& between 0.5$-$2.0 times Reff\_rf\_john\_vmod\_x \\
\hline
slp3stPLf\_ep5\_x & The fitted power-law slope of the stellar density distribution between 0.2$-$0.5 times Reff\_rf\_john\_vmod\_x \\
\hline
slp3stPLf\_e1\_x & The fitted power-law slope of the stellar density distribution between 0.5$-$1.0 times Reff\_rf\_john\_vmod\_x \\
\hline
slp3stPLf\_e2\_x & The fitted power-law slope of the stellar density distribution between 0.5$-$2.0 times Reff\_rf\_john\_vmod\_x \\
\hline

\end{tabular}
\end{minipage}
\label{tab:slp}
\end{table*}

\twocolumn
\bibliographystyle{mn2e}
\bibliography{ms_xudd}
\label{lastpage}
\end{document}